\documentclass[aps,prb,10pt,superscriptaddress,twocolumn,longbibliography]{revtex4-2}

\usepackage[utf8]{inputenc}
\usepackage{amssymb}
\usepackage{amsmath} 
\usepackage{graphicx}
\usepackage{amsmath} 
\usepackage{mathtools}
\usepackage{array,multirow}
\usepackage{longtable}
\newcommand{\PreserveBackslash}[1]{\let\temp=\\#1\let\\=\temp}
\usepackage{physics}
\usepackage{tikz}
\usepackage{xcolor}
\usepackage{hyperref}
\usepackage{float}
\advance\textwidth by .5cm
\advance\evensidemargin by -.33cm
\advance\oddsidemargin by -.33cm

\begin{document}

\title{Scaling of many-body localization transitions:\\Quantum dynamics in Fock space and real space}

\author{Thibault Scoquart}
\email{thibault.scoquart@kit.edu}

\affiliation{\mbox{Institute for Quantum Materials and Technologies, Karlsruhe Institute of Technology, 76131 Karlsruhe, Germany}}
\affiliation{\mbox{Institut f\"ur Theorie der Kondensierten Materie, Karlsruhe Institute of Technology, 76131 Karlsruhe, Germany}}

\author{Igor V.~Gornyi}

\affiliation{\mbox{Institute for Quantum Materials and Technologies, Karlsruhe Institute of Technology, 76131 Karlsruhe, Germany}}
\affiliation{\mbox{Institut f\"ur Theorie der Kondensierten Materie, Karlsruhe Institute of Technology, 76131 Karlsruhe, Germany}}

\author{Alexander D.~Mirlin}

\affiliation{\mbox{Institute for Quantum Materials and Technologies, Karlsruhe Institute of Technology, 76131 Karlsruhe, Germany}}
\affiliation{\mbox{Institut f\"ur Theorie der Kondensierten Materie, Karlsruhe Institute of Technology, 76131 Karlsruhe, Germany}}

\begin{abstract}
Many-body-localization (MBL) transitions are studied in a family of single-spin-flip spin-$\frac12$ models, including the one-dimensional (1D) chain with nearest-neighbor interactions, the quantum dot (QD) model with all-to-all pair interactions, and the quantum random energy model (QREM). We investigate the generalized imbalance that characterizes propagation in Fock space out of an initial basis state and, at the same time, can be efficiently probed by real-space measurements. For all models considered, the average imbalance and its quantum and mesoscopic fluctuations provide excellent indicators for the position of the MBL transition $W_c(n)$, where $n$ is the number of spins. Combining these findings with earlier results on level statistics, we determine phase diagrams of the MBL transitions in the $n$-$W$ plane. Our results provide evidence for a direct transition between the ergodic and MBL phases for each of the models, without any intermediate phase. For QREM and QD model, $W_c(n)$ grows as a power law of $n$ (with logarithmic corrections), in agreement with analytical predictions $W_c^{\rm QREM}(n) \sim n^{1/2} \ln n$ and 
$W_c^{\rm QD}(n) \gtrsim n^{3/4} \ln^{1/2} n$. This growth is in stark contrast to the 1D model, where $W_c(n)$ is essentially independent of $n$, consistent with the analytic expectation $W_c^{\rm 1D}(n\to \infty)= {\rm const}$. We also determine the scaling of the transition width $\Delta W (n) / W_c(n)$ and estimate the system size $n$ needed to study the asymptotic scaling behavior. Although the corresponding values of $n$ are larger than those accessible to exact simulations on a classical computer, they are within the reach of quantum simulators. Our results indicate feasibility of experimental studies of $n$-$W$ phase diagrams and scaling properties of MBL transitions in models of 1D and QD type and in their extensions to other spatial geometry or distance-dependent interactions.
\end{abstract} 

\maketitle

\section{Introduction}
\label{sec:intro}

The field of many-body localization
(MBL) deals with a fundamentally important problem of ergodicity and its violation in disordered interacting many-body problems at a finite energy density \cite{gornyi2005interacting, basko2006metal}. 
The MBL transitions, i.e., transitions between ergodic phases and non-ergodic MBL phases characterized by local integrals of motion, can be viewed as an extension of Anderson-localization transitions \cite{anderson58, evers08} to a (much more complex) setting of highly excited states of a many-body problem. In the past years, the MBL problem has been addressed for a rich variety of models, including interacting spins, fermions, and bosons in various spatial dimensionalities, with short-range and long-range interactions; see reviews  \cite{nandkishore15, Alet2018a, abanin2019colloquium, gopalakrishnan2020dynamics, tikhonov2021from, doggen2021many, Sierant2024}.

Despite remarkable progress in investigations of MBL, many questions of key importance remain (at least partly) open, in particular: What is the asymptotics of critical disorder $W_c(n)$ of the MBL transition and of the transition width $\Delta W(n) / W_c(n)$  for large system size (say, number of spins) $n$? What is the scaling behavior of various observables around the transition?  
What is the physics of a broad intermediate regime observed in numerical simulations in relatively small systems? Does its width shrink asymptotically to zero, as expected for a single transition?
Or are there actually two transitions asymptotically, with an intermediate phase in the thermodynamic limit ($n\to \infty$)?

One-dimensional (1D) models have attracted particular interest in the MBL context.  Most analytical works favor a single transition with $W_c(n) = \text{const }$ asymptotically~\cite{gornyi2005interacting, basko2006metal, Ros2015a, imbrie16a, Imbrie2016JSP, roeck17, Thiery2017a, goremykina2019analytically, morningstar2019renormalization, morningstar2020a}. 
On the numerical side, the model of a spin-$\frac12$ random-field Heisenberg chain attracted particular attention. Exact-diagonalization investigations of this model for $n$ in the range from 8 to 22 revealed an MBL transition at disorder $W_c(n) \simeq 3 - 4$, with a drift towards larger $W_c(n)$ at increasing $n$ \cite{Arijeet2010, Luitz2015}. Matrix-product-state study of longer chains yielded an estimate $W_c(n) \simeq 5.5$ for $n=50$ and $n=100$ \cite{Doggen2018a}. Very close values were obtained for the estimated thermodynamic-limit critical disorder, $W_c(n\to\infty) \simeq 5 - 6$, by phenomenological extrapolation~\cite{Sierant2020b, colbois2024interaction-driven} of exact-diagonalization data to $n\to \infty$. 
Several works have hypothesized (with arguments based mainly on numerical results) 
alternative 
possibilities, 
such as  $W_c(n)$ diverging at $n \to \infty$ \cite{Suntajs2019a, vsuntajs2020ergodicity, Sels2021dynamical, sels2022bath-induced, sels2023thermalization} or a scenario of two transitions, with an intermediate non-ergodic phase (between the ergodic and the ``conventional'' MBL phases) in the asymptotic $n \to \infty$ limit \cite{Weiner2019slow, biroli2024largedeviation, colbois2024statistics}.

One can view a disordered interacting many-body system as a tight-binding model on a graph representing the many-body Hilbert space (``Fock space''). In this representation, energies of many-body basis states are associated with sites of the graph, and interaction-induced transition matrix elements between these states are associated with links of the graph. 
Such a Fock-space view on the MBL problem suggests the use of observables akin to those known from the Anderson localization problem, including level statistics, eigenfunction statistics, the time dependence of propagation on the graph, etc. In particular, the Wigner-Dyson level statistics serves as a hallmark of the ergodic phase, with Poisson statistics signaling the ergodicity breaking. It is worth emphasizing that, from the Fock-space point of view, the concept of a transition between ergodic and MBL phases, with the questions formulated above, is applicable also to models with all-to-all interactions and without real-space localization (``quantum-dot models'')  \cite{altshuler1997quasiparticle, jacquod1997emergence, mirlin1997localization, silvestrov1997decay, silvestrov1998chaos, gornyi2016many, gornyi2017spectral, jacquod1997emergence, georgeot1997breit, leyronas2000scaling, shepelyansky2001quantum,  PhysRevE.62.R7575, jacquod2001duality, rivas2002numerical, bulchandani2022onset, garcia-garcia2018chaotic, micklitz2019nonergodic, monteiro2020minimal, Monteiro2021, nandy2022delayed, larzul2022quenches, Herre2023, dieplinger2023finite-size}. 
Recent theoretical works addressed various facets of the Fock-space physics around MBL transitions
\cite{serbyn2017thouless, tikhonov18, mace19multifractal, tikhonov2021from, tikhonov2021eigenstate, nag2019many-body, tarzia2020many, roy2021fock, crowley2022constructive, bulchandani2022onset, long2023phenomenology, creed2023probability, Roy2023, Ghosh2024, biroli2024largedeviation}.
Recent years witnessed also major progress in experimental investigations of the MBL transitions 
in the framework of Fock-space approaches	\cite{smith2016many-body, roushan2017spectroscopic, xu2018emulating, lukin2019probing, Yao2023, Wang2025}.

\begin{figure}[H]
    \centering
  \includegraphics[width=\columnwidth]{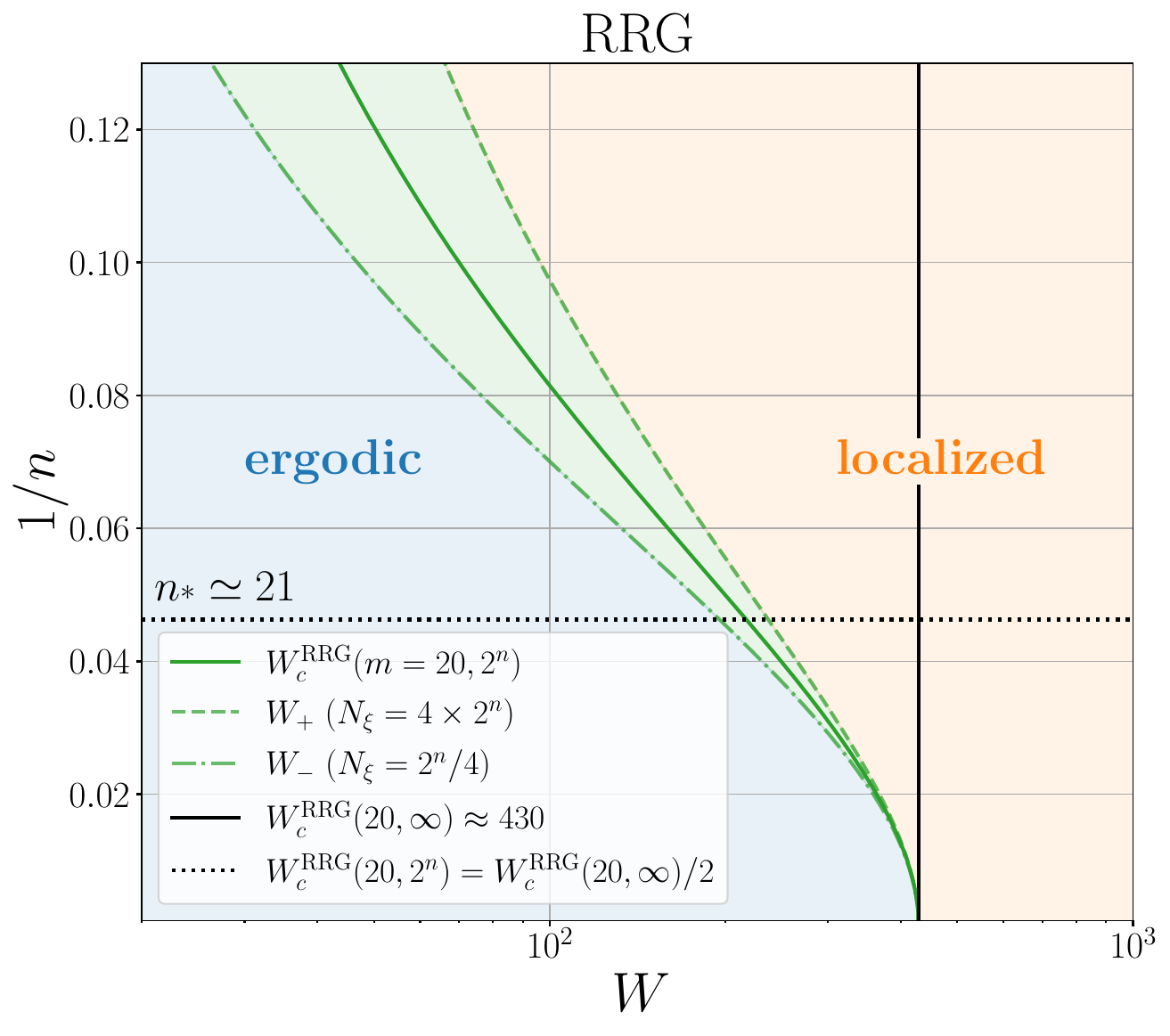}
    \caption{Finite-size ``phase diagram" of the Anderson transition on RRG with connectivity ${m+1}$, where ${m=20}$, in the plane $(W, 1/n)$, where $W$ is disorder and $N = 2^n$ is the number of vertices. The diagram is obtained analytically following Refs.~\cite{Herre2023, Scoquart2024}, see Appendix \ref{app:RRG} for details. The vertical line marks the thermodynamic-limit critical value $W^{\rm RRG}_c(m, \infty)$, while the full green line is the effective finite-size location of the transition $W^{\rm RRG}_c(m, 2^n)$, with the transition region shown around it. In order to probe the critical behavior of the correlation length $\xi(W)$ at $W \to W^{\rm RRG}_c(m, \infty)$, system sizes $n > n_*$ are needed, with $n_*$ marked by the horizontal dotted line. 
   }
\label{fig:analytics_RRG_Gaussian_m20}
\end{figure}

The Fock-space view on MBL provides a connection with the problem of 
Anderson localization on random regular graphs (RRG). An RRG is a random graph with a fixed (hence ``regular'') coordination number $m+1$ (where $m \ge 2$) for all vertices. The link to MBL has driven a considerable interest to Anderson localization on RRG in recent years \cite{biroli2012difference, de2014anderson, tikhonov2016anderson, garcia-mata17, metz2017level, Biroli2017, kravtsov2018non, biroli2018, PhysRevB.98.134205, tikhonov19statistics, tikhonov19critical, PhysRevResearch.2.012020, tikhonov2021eigenstate, garcia-mata2022critical, colmenarez2022subdiffusive, sierant2023universality, valba2022mobility, Herre2023, biroli2024critical, rizzo2024localized, kochergin2024robust}.
It is worth mentioning that a related sparse random matrix model, which differs from RRG by fluctuations of the coordination number, was studied earlier \cite{mirlin1991universality, fyodorov1991localization}. We refer the reader to a recent review \cite{tikhonov2021from} of localization on RRG and its relations to MBL. Importantly, key observables for the RRG model can be determined analytically, thus providing an excellent benchmark for numerical simulations \cite{tikhonov2016anderson, metz2017level, tikhonov19statistics, tikhonov19critical, biroli2024critical, rizzo2024localized}. 
While the RRG model is simpler than genuine MBL models, there are many similarities between them, making the understanding of Anderson localization on RRG very instructive for exploring MBL physics. 

In view of this, we show in Fig.~\ref{fig:analytics_RRG_Gaussian_m20} the analytically determined finite-size ``phase diagram'' of the Anderson transition on RRG in the parameter plane $(W, 1/n)$, where $W$ is the disorder strength and $N = 2^n$ is the Hilbert-space volume (number of vertices of the graph). The parameter $n$, rescaled by a factor $1/ \log_2 m$, is the diameter of the graph (i.e., the maximal distance between two vertices); it is analogous to the number of spins (i.e., to the real-space volume) in MBL models of interacting spins.  The boundary between the ergodic and localized phases is at the critical disorder strength $W^{\rm RRG}_c(m, 2^n)$; it is broadened by the finite-size transition width (green region).

A pronounced drift of the finite-size critical disorder towards its thermodynamic-limit value 
$W^{\rm RRG}_c(m, \infty)$ (shown by the vertical line) is obvious. The magnitude of this drift increases logarithmically with $m$; we have chosen a moderately large $m=20$ for this plot. Another noticeable property of the transition clearly seen in the figure is that its finite-size width is much smaller than the distance to the thermodynamic-limit value 
$W^{\rm RRG}_c(m, \infty)$. Let us note that the localized region in Fig.~\ref{fig:analytics_RRG_Gaussian_m20} includes both the thermodynamic-limit localized phase, $W>W^{\rm RRG}_c(m, \infty)$, and the regime to the left of the $n\to\infty$ critical disorder, where the system exhibits localization properties because its size is not sufficiently large.

Thus, for the RRG model, answers to the questions formulated at the beginning of this section are known. At the same time, several key properties of the RRG model (such as an exponential increase of Hilbert space with $n$ and the character of the critical point $W^{\rm RRG}_c(m, \infty)$, which is a continuation of the localized phase) hold also for MBL problems. This suggests that at least some of important features of the transition between ergodic and localized phases on RRG are also relevant to MBL. However, as the MBL physics is more complex than that of the RRG model, more work is needed to better understand phase diagrams of the MBL transitions that are counterparts of the RRG phase diagram of Fig.~\ref{fig:analytics_RRG_Gaussian_m20}. This is the challenge addressed by the present paper.

To answer the questions posed above for MBL, we study quantum dynamics around the MBL transition in several single-spin-flip models (introduced in Ref.~\cite{Scoquart2024}), at infinite temperature. This includes a 1D model (a spin chain with only nearest-neighbor interaction) and models with all-to-all interactions. 
We focus on single-spin-flip models for the following reason. The Fock space of a many-body spin model can be viewed as a set of vertices of a hypercube, and a single-spin-flip matrix element corresponds to an edge of this hypercube. Consequently, all our models ``live'' on the same hypercube graph in the Fock-space representation, which simplifies the analytical study of the models and their comparison.

In Ref.~\cite{Scoquart2024}, we demonstrated the role of Fock-space correlations between the matrix elements of the many-body Hamiltonian for the scaling of critical disorder $W_c(n)$ and the transition width $\Delta W / W_c(n)$ in these single-spin-flip models. 
Numerics in Ref.~\cite{Scoquart2024} was based on statistics of the many-body energy levels and eigenfunctions. Measuring such observables experimentally requires full tomography of many-body eigenstates and energies, which quickly becomes extremely difficult or even impossible with increasing system size. 

Here, we consider different observables that characterize the dynamics of the system initially prepared in one of the basis states (i.e., eigenstates of all $\hat{S}_i^z$ operators, where $i=1,\ldots,n$ labels spins). Importantly, while the observables that we consider---the generalized imbalance and its fluctuations---characterize the dynamics in Fock space, they can also be measured by real-space measurements and are thus experimentally accessible also for large systems.  Furthermore, comparing the results for these observables with those for level statistics provides information on the phase diagram of the problem. Indeed, imagine that there would be two distinct transitions in the $n \to \infty$ limit, at disorder strengths $W_{c1}$ and $W_{c2}$, such that the system is asymptotically in the ergodic phase for $W<W_{c1}$, in the MBL phase for $W>W_{c2}$, and in an intermediate phase for $W_{c1}<W<W_{c2}$, which is non-ergodic but at the same time exhibits hybridization of states over the maximal distance in Fock space. In this case, the level statistics (which is the indicator of ergodicity) would show a transition at $W_{c1}$, while the imbalance (which is the measure of propagation in Fock space) would exhibit a transition at $W_{c2}$.

Importantly, one of the models that we study is the quantum random energy model (QREM), which allows for an accurate analytical study due to its relation to the RRG model, thus serving as a benchmark for numerical investigations. The analytical findings include the phase diagram (a single transition from ergodicity to MBL), as well as results for the critical disorder $W_c(n)$ (which scales asymptotically as $n^{1/2}\ln n$) and for the transition width $\Delta W(n) / W_c(n)$. Our numerical results are in full agreement with these analytical predictions.

\begin{figure}[H]
    \centering   \includegraphics[width=0.50\textwidth]{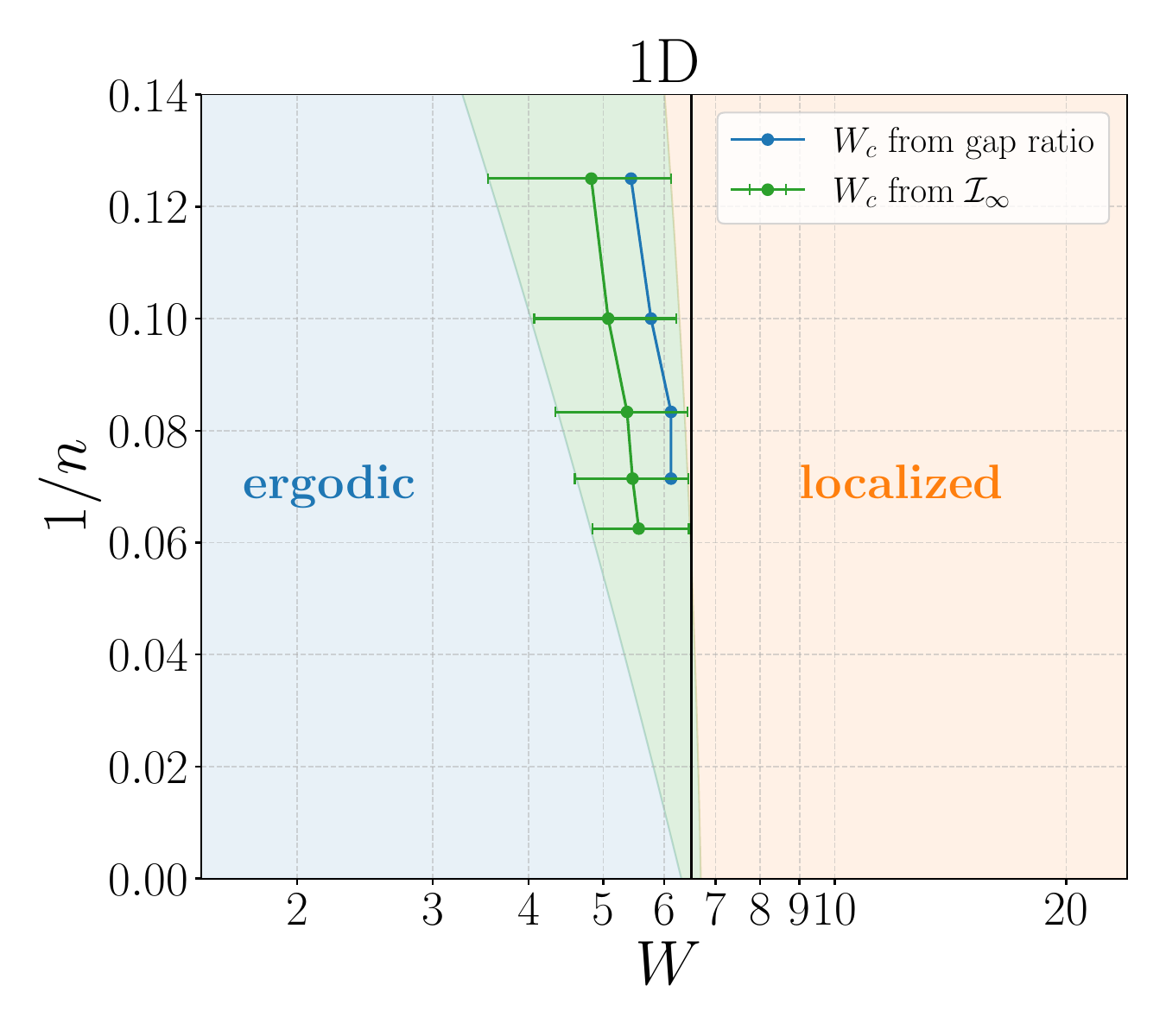}
    \caption{ Finite-size ``phase diagram'' for the 1D model studied in this work. The position of the transition and its width, as obtained from results for the average generalized imbalance, are shown, with extrapolation towards $1/n \to 0 $.
    In addition, the level-statistics indicator of the transition \cite{Scoquart2024} is included. 
    This figure is an alternative representation of the data
    presented and discussed in Sec.~\ref{sec:transition}, see Fig.~\ref{fig:Wc_vs_n_1D_QD_QREM}. 
The phase diagram in this figure bears analogy to that in Fig.~\ref{fig:analytics_RRG_Gaussian_m20} for the RRG model but the finite-size drift of $W_c^{\rm 1D}(n)$ here is much weaker.}
    \label{fig:1D-phase-diagram-W-1_over_n}
    \end{figure}

Our results for a 1D model are of particular interest. Crucially, they support a single ergodicity-to-MBL transition with a $W_c(n) = \text{const }$ behavior of the critical disorder at large $n$. Specifically, the values of $W_c(n)$ that we obtain from the analysis of generalized imbalance and its fluctuations exhibit only a weak drift with $n$ and are in accordance with the finite-size critical disorder found in Ref.~\cite{Scoquart2024} by studying statistics of many-body energy levels and eigenfunctions. 
These results are illustrated in Fig.~\ref{fig:1D-phase-diagram-W-1_over_n}. It is seen that 
 the finite-size drift of $W_c^{\rm 1D}(n)$ is very weak in comparison to that in the RRG model, the phase diagram for which is shown in Fig.~\ref{fig:analytics_RRG_Gaussian_m20}.

Further, we study a quantum-dot (QD) model, which is analogous to the 1D model in that the spin interactions are only of two-spin type. The crucial difference is that, in the QD model, all pairs of spin interact, in contrast to the 1D model where interaction is restricted to pairs of nearest-neighbor spins in a chain. Like for 1D model and for QREM, our results for the QD model support a single transition between ergodic and MBL phases. The numerically obtained scaling of $W_c(n)$ for the QD model is consistent with the analytic lower bound $W_c(n) \gtrsim n^{3/4}\ln^{1/2}n$. Our analytical and numerical results show that the physics of the QD model is in many respects similar to that of QREM. 

The structure of the paper is as follows. 
In Sec.~\ref{sec:models_def}, we introduce the spin models---1D, QD, and QREM---studied in this work. In Sec.~\ref{sec:dynamics_imbalance}, we define the generalized imbalance and present first numerical results for the dynamics of the average imbalance, which show that the imbalance is an appropriate indicator of the MBL transition in all models under consideration. We then turn to an analytical study  of the imbalance and its fluctuations in Sec.~\ref{sec:real-Fock} and \ref{sec:analytics}. Specifically, 
Sec.~\ref{sec:real-Fock} contains a general analytical discussion of the imbalance and its quantum and mesoscopic fluctuations, which in particular emphasizes relations between real-space and Fock-space observables. Section \ref{sec:analytics} contains an analysis of the average imbalance and its fluctuations
in the ergodic and MBL phases as well as in the transition region for each of three considered models. Armed with analytical predictions, we return to a detailed numerical study of the imbalance and its fluctuations in Sec.~\ref{sec:numerics}. 
In Sec.~\ref{sec:transition}, we analyze phase diagrams of the models, using numerical values for $W_c(n)$ and for the transition width $\Delta W(n) / W_c(n)$ obtained in Sec.~\ref{sec:numerics}. Key results of the work are summarized in Sec.~\ref{sec:summary}, which also contains a discussion of prospective research directions.
Technical details are presented in several Appendices.

\section{Models}
\label{sec:models_def}

We consider a family of  spin-$\frac12$ models 
\cite{Scoquart2024},
 specified below. Hamiltonians of these models have following properties (which motivate their choice):
\begin{itemize}
\item[(i)] the models are of single-spin-flip character, implying that
the Fock-space graph for all of them is the same and is formed by vertices and edges of a hypercube;
\item[(ii)] the models are characterized by 
random interactions and random Zeeman field, which makes all basis states of the Fock space statistically equivalent;
\item[(iii)] in the Fock-space representation, fluctuations of energies and hopping matrix elements are the same for all the models;  the difference between the models is only in energy and hopping correlations; 
\item[(iv)]  the only conserved quantity is energy.
\end{itemize}

These properties common to all of the models are greatly advantageous for a direct comparison of observables between the models. Also, the single-spin-flip terms in the Hamiltonian connect Fock-space basis states separated by the minimal Hamming distance (unity). It is plausible that this 
reduces finite-size effects, such as the drift of $W_c(n)$ in 1D models, in comparison with frequently studied XXZ models
involving two-spin-flip terms that connect basis states separated by Hamming distance two.

\subsection{1D model}
\label{sec:1D-model}

The 1D model considered here is a spin chain of length $n$ with periodic boundary conditions and the following Hamiltonian involving nearest-neighbor interactions \footnote{The 1D model defined by Eqs.~\eqref{eq:1D-splittting}--\eqref{eq:1D-Vxy-variance} is exactly the model that was studied numerically in Ref.~\cite{Scoquart2024}. At the same time, there was a slight inaccuracy in the definition of this model in Ref.~\cite{Scoquart2024}: in Eq.~(27) of Ref.~\cite{Scoquart2024}, the second term of Eq.~\eqref{eq:1D-1} of the present work was missing, which was compensated by an additional factor two in Eq.~(30) of Ref.~\cite{Scoquart2024} compared to Eq.~\eqref{eq:1D-Vxy-variance} of the present work. This did not influence the variances of Fock-space energies $E_\alpha$ and hoppings $T_{\alpha\beta}$, Eq.~\eqref{eq:distrib_energies_hoppings_1D}, as well as the energy covariance matrix 
${(C_E^{\rm 1D})}_{\alpha\beta}$,
Eq.~\eqref{eq:1D-correl_energies_intermediate}. However, this led to a small modification in the covariance matrix ${(C_T^{\rm 1D})}_{\alpha\beta\mu\nu}$: the second term of Eq.~\eqref{eq:1D-CT} was missing in Eq.~(37) of Ref.~\cite{Scoquart2024}, compensated by an additional factor of two. This modified 1D model has the same properties as that defined by Eqs.~\eqref{eq:1D-splittting}--\eqref{eq:1D-Vxy-variance}, with a slight shift of numerical values of $W_c(n)$. Since 
the model that was studied numerically in Ref.~\cite{Scoquart2024} was exactly the model explored in the present work, we will directly compare the numerical values.}:
\begin{align}
&\hat{H}^\text{1D} =  \hat{H}^\text{1D}_0 + \hat{H}^\text{1D}_1 \,, 
\label{eq:1D-splittting} \\
&\hat{H}^\text{1D}_0  = \sum\limits_{i = 1}^{n} \epsilon_i\hat{S}_i^z + 2 \sum\limits_{i = 1}^{n} V_{i,i+1}^{z}\hat{S}_i^z \hat{S}_{i+1}^z \,, 
\label{eq:1D-0} \\
&\hat{H}^\text{1D}_1 = 2\smashoperator{\sum}\limits_{\substack{i = 1\\a \in \{x,y\}}}^{n}  \left[V_{i, i+1}^{a}\hat{S}_i^z \hat{S}_{i+1}^a +V_{i, i-1}^{a}\hat{S}_i^z \hat{S}_{i-1}^a \right],
\label{eq:1D-1}
\end{align}
Here $\smash{\hat{S}_i^\alpha}$ are the spin-$\smash{\frac{1}{2}}$ operators, $\smash{\hat{S}_i^\alpha = \frac{1}{2} \sigma_i^a}$ with Pauli matrices $\sigma_i^a$. The energies $\epsilon_i$ are uncorrelated random variables uniformly distributed in $\left[-W, W\right]$. Further, the interaction couplings $\smash{V_{i,i+1}^{z}}$, $\smash{V_{i,i+1}^{a}}$, and $\smash{V_{i,i-1}^{a}}$ (with $a=x,y$) are uncorrelated real Gaussian random variables with zero mean and variances
\begin{align}
&\langle V^z_{i,i+1} V^z_{k,k+1} \rangle = \delta_{ik}\,, 
\label{eq:1D-Vz-variance}
\\
&\langle V^a_{i,i+1} V^a_{k,k+1} \rangle = \langle V^a_{i,i-1} V^a_{k,k-1} \rangle =  \delta_{ik}\,, \ \ \ \ a=x,y.
\label{eq:1D-Vxy-variance}
\end{align}
Throughout the paper, we denote by a bar $\overline{\vphantom{A}\dots}$ the statistical 
(``mesoscopic") average  over the distribution of both the on-site energies $\epsilon_i$ and the random couplings $V^a_{i,j}$.

The Hamiltonian $\hat{H}^\text{1D}$, as well as any other spin Hamiltonian $\hat{H}$, can be straightforwardly mapped to the Fock-space representation:
\begin{align}
\hat{H} &= \hat{H}_0 + \hat{H}_1\nonumber\\
&=  \sum\limits_{\alpha = 1}^{2^n} E_\alpha \ket{\alpha}\bra{\alpha}  + \sum\limits_{\substack{\alpha,\beta = 1\\ \alpha\neq\beta}}^{2^n} T_{\alpha\beta} \ket{\alpha}\bra{\beta} \,,
\label{eq:Fock_space_rep}
\end{align}
where $\ket{\alpha}$, $\ket{\beta}$ are basis states given by eigenstates of all $\smash{\hat{S}_i^z}$ operators (i.e., by strings of up or down spins). Equations \eqref{eq:1D-splittting}--\eqref{eq:1D-1} split the Hamiltonian
$\hat{H}^\text{1D}$ into diagonal  ($\hat{H}^\text{1D}_0$) and off-diagonal ($\hat{H}^\text{1D}_1$) parts with respect to this basis, which correspond to two terms in Eq.~\eqref{eq:Fock_space_rep}.  Equation \eqref{eq:Fock_space_rep} represents each of our many-body spin Hamiltonians $\hat{H}$ in the form of a tight-binding model
defined on the Fock-space graph, which is an $n$-dimensional hypercube with  $2^n$ nodes and connectivity $n$, with 
energies $E_\alpha$ and hopping matrix elements $T_{\alpha\beta}$.

 For the $\hat{H}^\text{1D}$ Hamiltonian, the many-body energy $E_\alpha \equiv \bra{\alpha}\hat{H}^\text{1D}\ket{\alpha}$ associated with a given basis state $\ket{\alpha}$  is 
\begin{align}
E_\alpha  =  \frac{1}{2}\sum\limits_{i=1}^n s_i^{(\alpha)}\epsilon_i + \frac{1}{2}\sum\limits_{i=1}^n s_{i, i+1}^{(\alpha)} V_{i,i+1}^z \,,
\label{eq:1D_on_site_energies}
\end{align}
with $s_i^{(\alpha)} \equiv \bra{\alpha}\sigma_i^z\ket{\alpha} = \pm 1$ and $s_{i,i+1}^{(\alpha)} \equiv 
s_i^{(\alpha)} s_{i+1}^{(\alpha)} = \pm 1$. The ``hopping'' matrix elements $T_{\alpha\beta} \equiv \bra{\alpha}\hat{H}^\text{1D}\ket{\beta}$ are non-zero if and only if $\ket{\alpha}$ and $\ket{\beta}$ are connected by a single spin flip. For a pair of states $\ket{\alpha}$ and  $\ket{\beta}=\ket{\bar{\alpha}_k}\equiv\sigma_k^x\ket{\alpha}$ that differ only by the sign of the spin $s_k$ in a position $k$, 
we have
\begin{align}
T_{\alpha\beta} &=  2\bra{\alpha}\left[ (V_{k-1,k}^x\hat{S}_{k-1}^z+ V_{k+1,k}^x\hat{S}_{k+1}^z)\hat{S}_k^x\right.\nonumber\\
&\left.+ (V_{k-1,k}^y\hat{S}_{k-1}^z+ V_{k+1,k}^y\hat{S}_{k+1}^z)\hat{S}_k^y \right] \ket{\bar{\alpha}_k} \nonumber \\
&= \frac{1}{2} \left( V_{k-1,k}^x s_{k-1}^{(\alpha)} + \mathrm{i}\, V_{k-1,k}^y s_{k-1,k}^{(\alpha)}\right.\nonumber\\
&\left.+ V_{k+1,k}^x s_{k+1}^{(\alpha)} + \mathrm{i}\, V_{k+1,k}^y s_{k+1,k}^{(\alpha)}  \right),
\label{eq:1D_matrix_elements}
\end{align}
whereas for any $\ket{\beta}$ differing from $\ket{\alpha}$ by more than one spin flip, we have $T_{\alpha\beta}=0$. 

By virtue of the central limit theorem, for $n \gg 1$, the many-body energies $E_\alpha$, given by Eq.~\eqref{eq:1D_on_site_energies}, obey a multivariate Gaussian distribution. Likewise,  the hopping matrix elements $T_{\alpha\beta}$, Eq.~\eqref{eq:1D_matrix_elements}, are  multivariate Gaussian random variables. The individual distributions of each of $E_\alpha$ and $T_{\alpha\beta}$ are [with $\mathcal{N}(\mu, \sigma^2)$ denoting the normal distribution of a real variable with mean $\mu$ and variance $\sigma^2$]: 
\begin{align}
&E_\alpha \sim  \mathcal{N}\left(0,  \frac{nW^2+ 3n}{12}\right),\nonumber\\
&T_{\alpha\beta} \sim \mathcal{N}\left(0, \frac{1}{2} \right) + \mathrm{i}\,\mathcal{N}\left(0,\frac{1}{2}\right),
\label{eq:distrib_energies_hoppings_1D}
\end{align}
and their correlations are governed by off-diagonal elements of covariance matrices $(C_E)_{\alpha \beta} = \langle E_\alpha E_\beta \rangle$ and $(C_T)_{\alpha\beta \mu\nu} = \langle T_{\alpha\beta}^*T_{\mu\nu} \rangle$. The energy covariance matrix 
 $(C_E^{\rm 1D})_{\alpha\beta} = \left\langle E_\alpha E_\beta \right\rangle$ reads:
\begin{align}
(C_E^{\rm 1D})_{\alpha\beta} & = \frac{W^2}{12} \sum\limits_{i=1}^n  s_i^{(\alpha)} s_i^{(\beta)} + \frac{1}{4}\sum\limits_{i=1}^n s_{i,i+1}^{(\alpha)}s_{i,i+1}^{(\beta)} \nonumber \\
& = \frac{W^2}{12} (n - 2 r_{\alpha\beta}) + 
\frac{1}{4} (n - 2 q_{\alpha\beta}).
\label{eq:1D-correl_energies_intermediate}
\end{align}
Here, $r_{\alpha\beta}$ is the Hamming distance
between the basis states $\ket{\alpha}$ and $\ket{\beta}$, i.e., the number of spins to be flipped to transform $\ket{\alpha}$ into $\ket{\beta}$, or, equivalently, the length of the shortest paths connecting these states on the Fock-space hypercube graph. Further, $q_{\alpha\beta}$ is the number of sites $i$ such that $s_{i,i+1}^{(\alpha)} = - s_{i,i+1}^{(\beta)}$.

Crucially, the covariance \eqref{eq:1D-correl_energies_intermediate} depends on the pair of states $\alpha$ and $\beta$ {\it not} solely via the Hamming distance $r_{\alpha\beta}$. The emergence of $q_{\alpha\beta}$ in Fock-space correlations, Eq.~\eqref{eq:1D-correl_energies_intermediate}, reflects the 1D real-space geometry of the model \cite{Scoquart2024} and is of crucial importance for special properties of the 1D model.
An element  $(C_{T}^{\rm 1D})_{\alpha\beta\mu\nu} \equiv  \langle T_{\alpha\beta}^* T_{\mu\nu} \rangle$ of the covariance matrix is different from zero only if the states $\{\alpha, \beta\}$ are connected by a single flip of spin $s_k$ and the states $\{\mu,\nu\}$ are connected by a single flip of the same spin $s_k$, with $s_k^{(\alpha)} = s_k^{(\mu)}$. In this situation, 
\begin{equation}
(C_{T}^{\rm 1D})_{\alpha\beta\mu\nu} = \frac{1}{2}\left(
s_{k-1}^{(\alpha)}  s_{k-1}^{(\mu)} +s_{k+1}^{(\alpha)}  s_{k+1}^{(\mu)} \right).
\label{eq:1D-CT}
\end{equation}
As in the case of energy correlations, 
 $(C_{T}^{\rm 1D})_{\alpha\beta\mu\nu}$ is not a function of Hamming distance, which is again a manifestation of the spatial geometry of the 1D model. 

\subsection{Quantum dot (QD) model}

Like the 1D model, the quantum dot (QD) model involves interactions of pairs of spins. The difference is that, in the QD model, these are all-to-all pair interactions. The Hamiltonian of the QD model reads
\begin{align}
&\hat{H}^\text{QD} =  \hat{H}^\text{QD}_0 + \hat{H}^\text{QD}_1 \,, 
\label{eq:QD}
\\
&\hat{H}^\text{QD}_0= \sum\limits_{i = 1}^{n} \epsilon_i\hat{S}_i^z + \frac{2}{\sqrt{n}}\sum\limits_{i,j = 1}^{n} V_{ij}^{z}\hat{S}_i^z \hat{S}_j^z \,,
\label{eq:QD-H0}
\\
&\hat{H}^\text{QD}_1 = \frac{1}{\sqrt{n}}\smashoperator{\sum}\limits_{\substack{i,j = 1\\ a \in \{x,y\}}}^{n}  V_{ij}^{a} \left( \hat{S}_i^z \hat{S}_j^a + \text{H.c.}\right),
\label{eq:QD-H1}
\end{align}
The single-particle energies $\epsilon_i$ are uncorrelated random variables uniformly distributed in $\left[-W, W\right]$, as in the 1D model. The interaction couplings $\smash{V_{ij}^{a}}$ (with $a\in\{x,y,z\}$) are uncorrelated real Gaussian random variables with zero mean and variances
\begin{align}
&\langle V^z_{ij} V^z_{kl} \rangle = \delta_{ik}\delta_{jl} \,, \label{eq:Vz_variance}\\
&\langle V^x_{ij} V^x_{kl} \rangle = \langle V^y_{ij} V^y_{kl} \rangle = 2 \delta_{ik}\delta_{jl} \,. \label{eq:Vxy_variance}
\end{align}
In the Fock-space representation \eqref{eq:Fock_space_rep}, the many-body energies $E_\alpha \equiv \bra{\alpha}\hat{H}^\text{QD}\ket{\alpha}$ are
\begin{align}
E_\alpha  =  \frac{1}{2}\sum\limits_{i=1}^n \epsilon_i s_i^{(\alpha)} + \frac{1}{2\sqrt{n}}\sum\limits_{i,j=1}^n V_{ij}^z s_{ij}^{(\alpha)}\,.
\label{eq:QD_on_site_energies}
\end{align}
As for the 1D model, the Fock-space matrix element  $T_{\alpha\beta}$ is non-zero if  $\ket{\alpha}$ and $\ket{\beta}$ are connected by a single spin flip, i.e., $\ket{\beta}=\ket{\bar{\alpha}_k}\equiv\sigma_k^x\ket{\alpha}$. It reads then
\begin{align}
T_{\alpha\beta} &=
\bra{\alpha}\hat{H}^\text{QD}_1\ket{\bar{\alpha}_k}=
\frac{1}{2\sqrt{n}} \sum\limits_{\substack{i = 1 \\ i\neq k}}^{n}\left( V_{ik}^x s_{i}^{(\alpha)} + \mathrm{i}\, V_{ik}^y s_{ik}^{(\alpha)} \right).
\label{eq:QD_matrix_elements}
\end{align}
The individual distributions of each of $E_\alpha$ and $T_{\alpha\beta}$ are exactly the same as in the 1D model, Eq.~\eqref{eq:distrib_energies_hoppings_1D}.
The difference between the models is encoded in correlations. The energy covariance matrix for the QD model is
\begin{align}
(C_E^{\rm QD})_{\alpha\beta} &=  n\left[\frac{W^2}{12} \left(1- \frac{2r_{\alpha\beta}}{n}\right) + \frac{1}{4}\left(1-\frac{2r_{\alpha\beta}}{n}\right)^2\right].
\label{eq:correl_energies_final}
\end{align}
Similarly to the 1D model, an element 
$(C_{T}^{\rm QD})_{\alpha\beta\mu\nu} \equiv  \langle T_{\alpha\beta}^* T_{\mu\nu} \rangle$ 
of the hopping covariance matrix is non-zero 
only if the states $\{\alpha, \beta\}$ are connected by a single flip of spin $s_k$ and the states $\{\mu,\nu\}$ are connected by a single flip of the same spin $s_k$, with $s_k^{(\alpha)} = s_k^{(\mu)}$. It is then given by
\begin{equation}
(C_{T}^{\rm QD})_{\alpha\beta\mu\nu} = \frac{1}{n}
\sum\limits_{\substack{i = 1 \\ i \neq k}}^{n}
s_i^{(\alpha)} s_i^{(\mu)} = \frac{n-1-2r_{\alpha\mu}}{n} \simeq 1 - \frac{2r_{\alpha\mu}}{n} \,.
\label{eq:QD-CT-full}
\end{equation}
Importantly, $(C_E^{\rm QD})_{\alpha\beta}$ and non-zero elements of $(C_{T}^{\rm QD})_{\alpha\beta\mu\nu}$ depend on the states involved via Hamming distance only ($r_{\alpha\beta}$ and $r_{\alpha\mu}$, respectively). 

\subsection{QREM}

The 1D and QD model defined above are ``conventional'' interacting spin models, in the sense that their Hamiltonians include only single-spin terms and two-spin interaction terms. In addition to these models, we consider the QREM. 
 The Fock-space representation of this model is obtained from those of the 1D and QD models by removing all energy and hopping correlations. Writing the Hamiltonian of QREM as a sum of terms representing products of spin operators would require terms involving arbitrarily many spin operators (up to the maximal number $n$). Thus, this model is rather far from those that can be realistically implemented in an experiment. The idea behind considering it is that, due to its close relation to the RRG model,  this model can be efficiently studied analytically (for $n \gg 1)$, thus allowing us to test the agreement between analytics and numerics.  Let us emphasize that the Hilbert space for the QREM is the same as for the ``conventional'' (1D, QD) spin models. This permits to study for the QREM the same observables (in particular, the generalized imbalance and its fluctuations in this paper, see below) as for the 1D and QD models. 

We provide now a formal definition of the model. Individual distributions of each of the Fock-space matrix elements $E_\alpha$ and $T_{\alpha\beta}$ are given in QREM by the same
Eq.~\eqref{eq:distrib_energies_hoppings_1D}
as for the 1D and QD models. At the same time, at variance with the 1D and QD models, the energies $E_\alpha$ and hopping matrix elements $T_{\alpha\beta}$ in QREM are independent (i.e., uncorrelated) random variables. We note that our version of the QREM slightly differs from that studied in Refs.~\cite{laumann2014many-body,baldwin2016the_many-body}: the hopping matrix elements 
$T_{\alpha\beta}$ were constant in Refs.~\cite{laumann2014many-body,baldwin2016the_many-body} (see also an earlier work \cite{Goldschmidt1990}) and are uncorrelated Gaussian random variables in our case.

\section{Dynamics of the generalized imbalance}
\label{sec:dynamics_imbalance}

\subsection{Generalized imbalance: definition}
\label{subsec:defs_imbalance}

A natural observable characterizing quantum dynamics in interacting spin systems is a spin autocorrelation function
\begin{equation}
\mathcal{I}^{(\alpha)}(t) = \bra{\alpha} \frac{1}{n} \sum_{j=1}^n \sigma_j^z(t)\sigma_j^z \ket{\alpha},
\label{eq:imbalance-1}
\end{equation}
where $\ket{\alpha}$ is a basis state, 
$\sigma_j^z(t) = \hat{U}^\dagger(t) \sigma_j^z
\hat{U}(t)$ , and $\hat{U}(t) = \exp(-i\hat{H}t)$ is the evolution operator. Since the basis states are eigenstates of $z$ components of all spins, $\sigma_j^z \ket{\alpha} = s_j^{(\alpha)}\ket{\alpha}$, 
Eq.~\eqref{eq:imbalance-1} can be equivalently rewritten in the form
\begin{equation}
\mathcal{I}^{(\alpha)}(t) = \bra{\psi(t)} \hat{\mathcal{I}}^{(\alpha)} \ket{\psi(t)} \,,
\label{eq:imbalance-2}
\end{equation}
where $\ket{\psi(t)} = \hat{U}(t) \ket{\alpha}$ is the result of the unitary evolution of state $\ket{\alpha}$ up to time $t$ and 
\begin{equation}
\hat{\mathcal{I}}^{(\alpha)} = \frac{1}{n} \sum_{j=1}^n s_j^{(\alpha)} \sigma_j^z \,.
\label{eq:imbalance-3}
\end{equation}

For 1D models, the
observable $\mathcal{I}^{(\alpha)}(t)$ was used [mainly for the N\'eel initial state,  $\ket{\alpha}\equiv\ket{\uparrow,\downarrow,\uparrow,\dots,\downarrow}$ with $s_j^{(\alpha)}= (-1)^{j-1}$] in many experimental \cite{schreiber2015observation,choi2016exploring,Wang2025} and theoretical (see, e.g., Refs.~\cite{Luitz2016a,Doggen2018a,Doggen2019a,Doggen2021stark, Sierant2022challenges}) papers in the context of MBL transitions. It was termed ``imbalance'' since for the N\'eel initial state, it characterizes a difference in spin density 
(or in particle density in the language of fermion or hard-core-boson models) between even and odd sites. The above definition \eqref{eq:imbalance-1} extends this notion to any spin system and any initial basis state $\ket{\alpha}$. Hence, $\mathcal{I}^{(\alpha)}(t)$ can be termed ``generalized imbalance''.  

For all models that we consider in this paper, properties of $\mathcal{I}^{(\alpha)}(t)$ are independent of the choice of the initial state $\ket{\alpha}$. Indeed, consider any two basis states $\ket{\alpha}$ and $\ket{\beta}$. They differ by flipping some subset of spins $j$. A unitary transformation by a product of matrices $\sigma_j^x$ over this subset will map $\ket{\alpha} \mapsto \ket{\beta}$. It is easy to see that, for each of our models, the ensemble of Hamiltonians is invariant under this unitary transformation, which proves that properties of $\mathcal{I}^{(\alpha)}(t)$ and $\mathcal{I}^{(\beta)}(t)$ are statistically the same. In particular, the ensemble-averaged generalized imbalance does not depend on the initial state,
$\overline{\mathcal{I}^{(\alpha)}(t)} = \overline{\mathcal{I}^{(\beta)}(t)}$. 
In view of the above invariance, one can choose the initial state to be, e.g., the N\'eel state, $\ket{\alpha}\equiv\ket{\uparrow,\downarrow,\uparrow,\dots,\downarrow}$, without restricting generality.
Below, we will frequently omit the superscript $(\alpha)$ indicating the initial state in $\overline{\mathcal{I}^{(\alpha)}(t)}$. 
 For brevity, we will call $\overline{\mathcal{I}^{(\alpha)}(t)}$ the imbalance, omitting the word ``generalized''. 

 \begin{figure*}[ht!]
    \centering
   \includegraphics[width=0.315\textwidth]{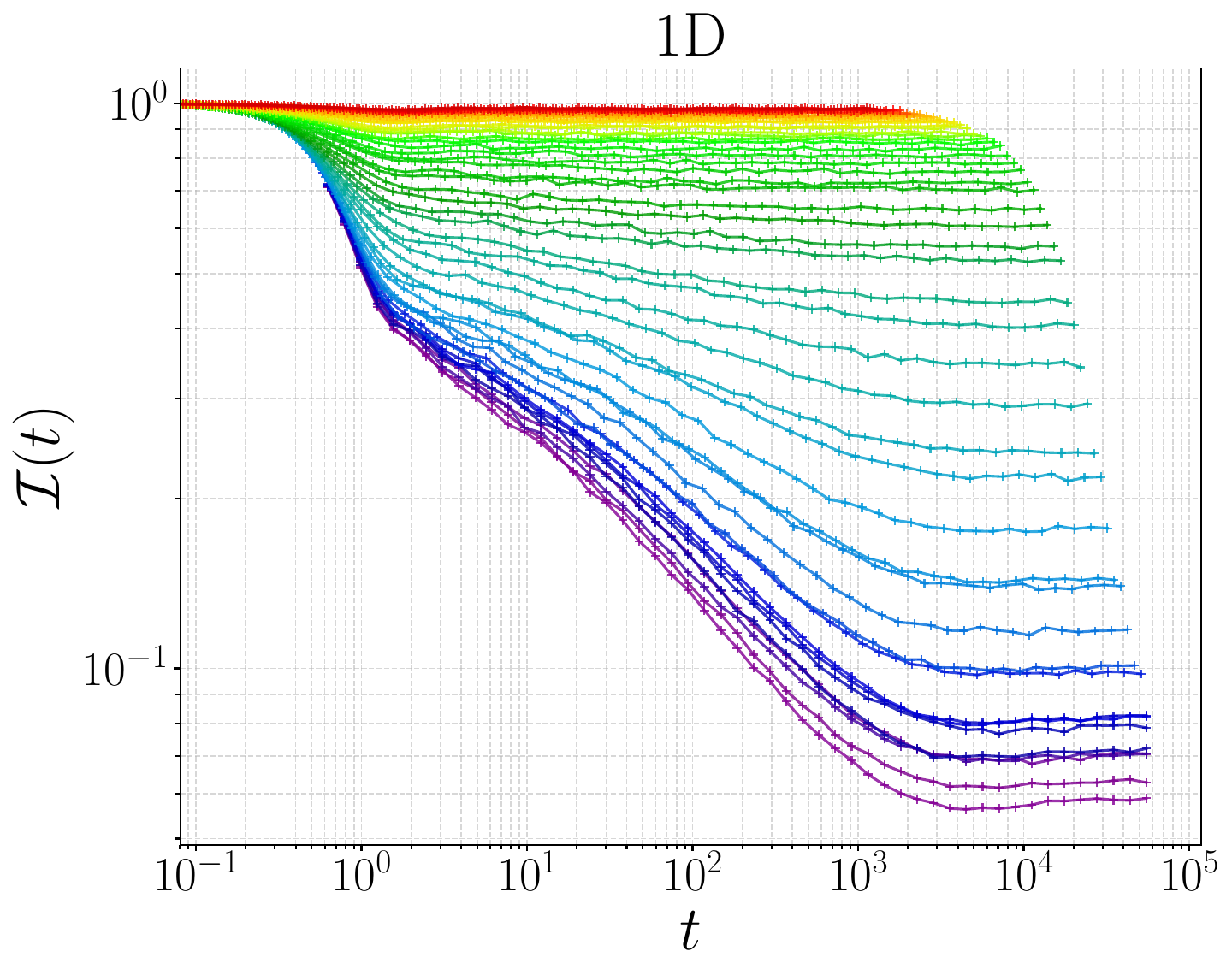}
  \includegraphics[width=0.315\textwidth]{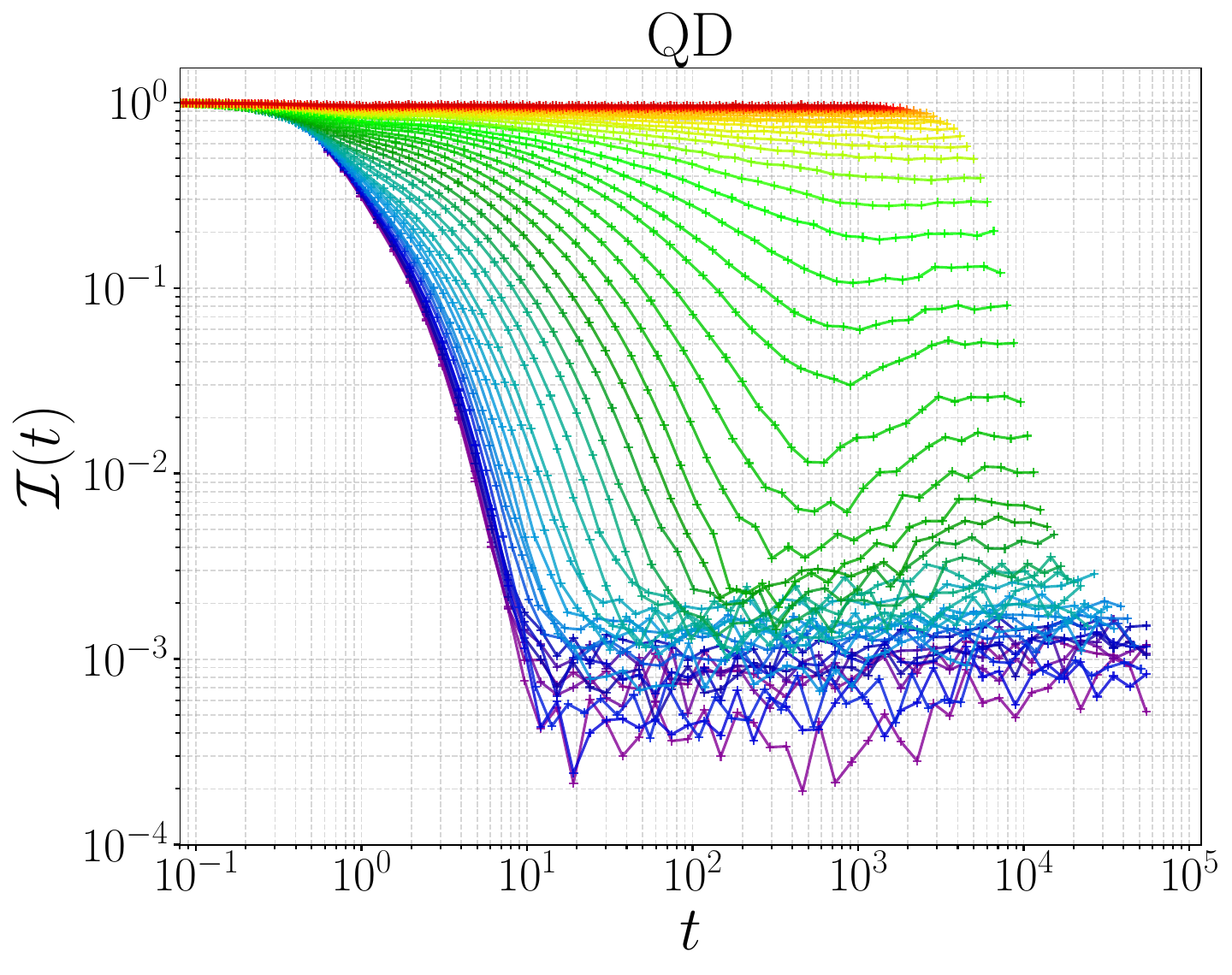}
    \includegraphics[width=0.35\textwidth]{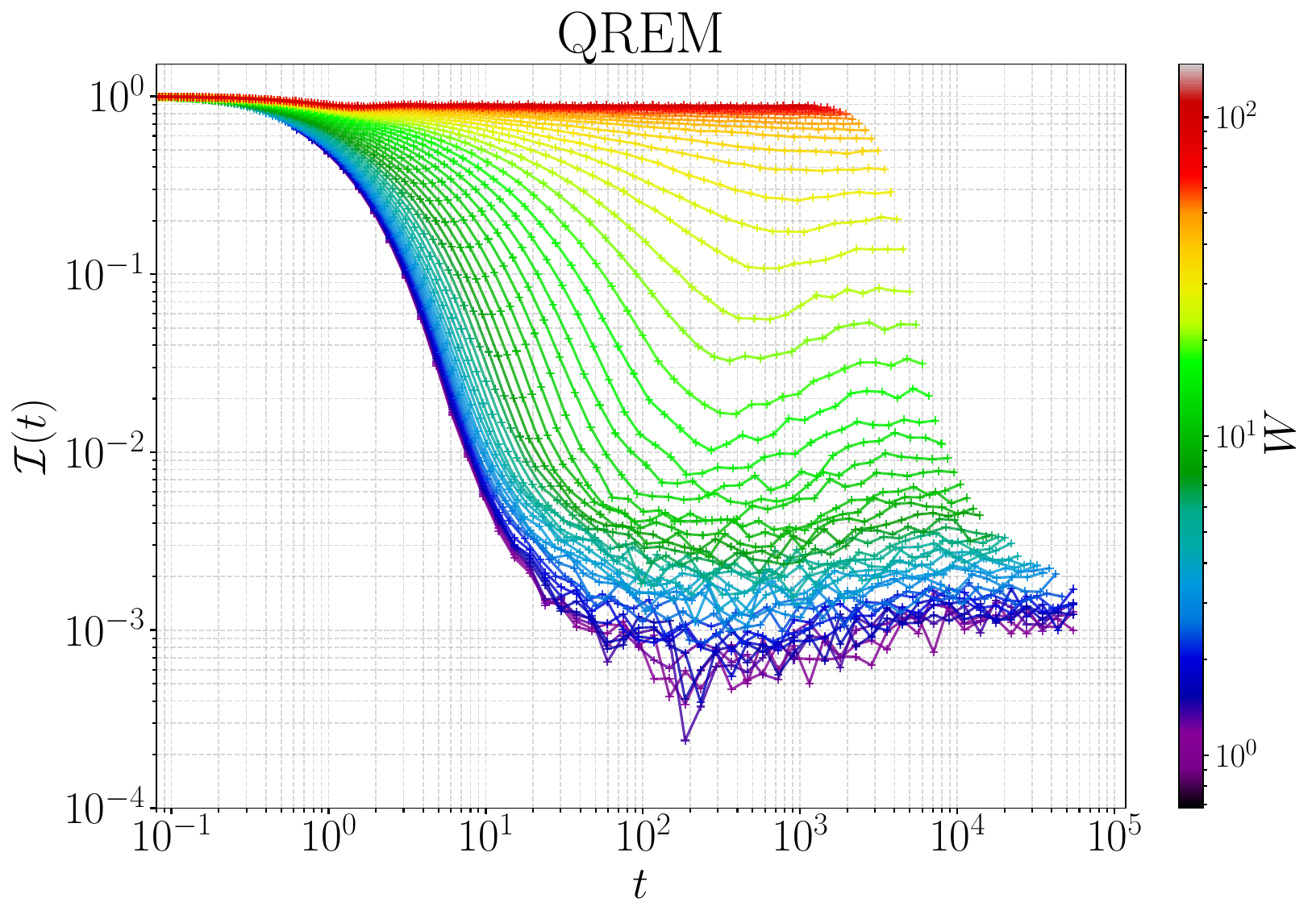}
    \caption{Time evolution of the average imbalance $\overline{\mathcal{I}(t)}$ for the 1D model (left), QD model (center) and QREM (right) with $n=14$ spins, following an interaction quench from a basis state $\ket{\alpha}$. The disorder ranges from $W\simeq 1$ to $W\simeq 100$ for all three models. The maximum times exceed the estimated Heisenberg time $t_H$, Eq.~\eqref{eq:tH}, for all values of $W$. 
    }
    \label{fig:time-evol1D_QD_QREM}
\end{figure*}

The imbalance $\mathcal{I}(t)$ measures how much of the initial order is left at time $t$, as $\mathcal{I}(0)=1$ by construction. For an ergodic system, $\smash{\mathcal{I}(t)}$ tends to zero as $t \to \infty$, up to a small finite-size correction. In the MBL phase (or regime), the system fails to thermalize and retains memory of the initial state, which translates into a finite value of the imbalance at long times: 
\begin{align}
\mathcal{I}(t) \xrightarrow[]{t \to \infty} \mathcal{I}_\infty, \quad\quad 0<\mathcal{I}_\infty\leq 1 \,.
\end{align}    

\subsection{Numerical results for imbalance evolution}
\label{sec:imbalance-numerics}

To illustrate dynamical properties of the imbalance $\mathcal{I}^{(\alpha)}(t)$ across the MBL transition for our models, we present in Fig.~\ref{fig:time-evol1D_QD_QREM} the time-resolved dynamics 
of the average imbalance in a broad range of disorder for the 1D and QD models, as well as the QREM, with $n=14$ spins.  
We take the state with the energy $E_\alpha$ closest to zero as the initial state $\alpha$, in order to additionally minimize the effect of the tails of the spectrum, see a discussion in
Sec.~\ref{sec:analytics-QREM-erg}.
The time evolution is performed numerically using advanced matrix exponentiation algorithms for large sparse matrices and adaptive time step  and averaged over many disorder realizations (see Appendix~\ref{app:numerics_details} for detail).

The maximum time of the dynamics in these plots is larger than the estimated Heisenberg time $t_H$ that sets the largest characteristic time scale for the quantum dynamics,
\begin{equation}
t_H= \frac{2\pi}{\Delta} \simeq 2^n\frac{4\pi}{W \sqrt{n}}. 
\label{eq:tH}
\end{equation}
Here $\Delta$ is the level spacing (around zero energy), and we assumed $W > 1$, which is the range of disorder at which the physics of our interest develops. 
For all three models, we clearly observe an evolution of the asymptotic imbalance 
$\mathcal{I}_\infty^{(\alpha)}$ from values close to zero to values close to unity with increasing $W$, which is a manifestation of the MBL transition.
In all the plots, the range of disorder is from $W\simeq 1$ to $W\simeq 100$, and the same color code is used.  

In the transition region (light blue to dark green in the 1D model, light green in the QD model, and yellow in the QREM), the imbalance exhibits a slow decay on all times up to $t \sim t_H$. This transition region clearly separates the ergodic regime, with strong decay of the imbalance, from the localized regime, for which the imbalance saturates to values close to unity.

Despite these similarities between all the models, clear differences between the 1D model, on one hand, and QD model and QREM, on the other hand, are seen in Fig.~\ref{fig:time-evol1D_QD_QREM}. In particular, 
the imbalance dynamics for the 1D model displays, at times $t \gtrsim 1$, a clear power-law decay on the ergodic side of the transition. This is in contrast with the behavior found for the QD model and QREM, where the ergodic side of the transition reveals a much faster, approximately exponential decay of the imbalance, indicating a very different character of the underlying dynamics.
Also, the value of the disorder in the transition region is a few times larger for the QD model and QREM than for the 1D model. 

The numerics presented in Fig.~\ref{fig:time-evol1D_QD_QREM} thus demonstrates that, for all the models, the imbalance is an appropriate indicator of the MBL transition and, at the same time, there are essential differences between the models. We turn now to an analytical study of the imbalance and its fluctuations, Sec.~\ref{sec:real-Fock} and  \ref{sec:analytics}. After this, 
in Sec.~\ref{sec:numerics} and \ref{sec:transition},
we will present more detailed numerical data and carry out their analysis.

\section{From real space to Fock space: General relations}
\label{sec:real-Fock}

\subsection{Dynamics of generalized imbalance}
\label{sec:dynamics-gener-imbalance}

To determine the imbalance \eqref{eq:imbalance-2}, it is sufficient to measure spins $\hat{S}_j = \frac{1}{2}\sigma_j$ at all sites $j$ after evolving the system for a time $t$. In this sense, the imbalance is a real-space observable: measuring it does not require finding many-body eigenstates or energy levels. This makes the imbalance (including not only its average but also fluctuations discussed below) particularly suitable for experimental studies. 

At the same time, to better understand physical meaning of the imbalance and to study analytically its properties in various models, it is highly instructive to express it in terms of Fock-space observables. The imbalance operator  \eqref{eq:imbalance-3} is diagonal in the eigenbasis of $\hat{H}_0$:
\begin{align}
\bra{\beta}\hat{\mathcal{I}}^{(\alpha)}\ket{\gamma} = \delta_{\beta,\gamma} \:
\frac{1}{n} \sum_{j=1}^n s_j^{(\alpha)} s_j^{(\beta)} =
\delta_{\beta,\gamma}\left(1-\frac{2 r(\alpha,\beta)}{n}\right) ,
\label{eq:imbalance_hamming-1}
\end{align}
where $r(\alpha,\beta)$ denotes the Hamming distance between the basis states $\ket{\alpha}$ and $\ket{\beta}$. Substitution of Eq.~\eqref{eq:imbalance_hamming-1} into Eq.~\eqref{eq:imbalance-2} yields
\begin{align}
\mathcal{I}^{(\alpha)}(t) 
&= \sum\limits_{\beta,\gamma}^{2^n} \bra{\psi(t)}\ket{\beta}\bra{\beta} \hat{\mathcal{I}} \ket{\gamma}\bra{\gamma}\ket{\psi(t)}\nonumber\\
&= \sum\limits_{\beta}^{2^n} |\bra{\beta}\ket{\psi(t)}|^2\left(1-\frac{2r(\alpha,\beta)}{n}\right).
\label{eq:imbalance-4}
\end{align}
Thus, in the Fock-space representation, the time evolution of $\mathcal{I}^{(\alpha)}(t)$ corresponds to the propagation of a state on the Fock-space graph measured by the Hamming distance between the initial state $\ket{\alpha}$ and basis components $\ket{\beta}$ of the evolved state $\ket{\psi(t)}$.

Using the decomposition of the basis states $\ket{\alpha}$,  $\ket{\beta}$ over the eigenbasis $\{\ket{I}\}$ of the full Hamiltonian $\hat{H}$, we can rewrite Eq.~\eqref{eq:imbalance-4} as
\begin{align}
\mathcal{I}^{(\alpha)}(t) &= \sum\limits_{\beta}^{2^n} |\bra{\beta}e^{-i\hat{H}t}\ket{\alpha}|^2\left(1-\frac{2r(\alpha,\beta)}{n}\right)\nonumber\\
&=  \sum\limits_{I,J,\beta}^{2^n}  e^{i(\epsilon_{I}-\epsilon_{J})t} \braket{\alpha}{I}\bra{I}\ket{\beta}\bra{\beta}\ket{J}\braket{J}{\alpha} \nonumber \\
& \times
\left(1-\frac{2 r(\alpha,\beta)}{n}\right).
\end{align}
The matrix elements $\braket{\alpha}{I}$ have a meaning of wave-function amplitudes of exact eigenstates $\ket{I}$ in the Fock-space basis.
In the large-$t$ limit, oscillations originating from terms with $I \ne J$ effectively average out, yielding $\mathcal{I}^{(\alpha)}(t) \to 
\mathcal{I}^{(\alpha)}_\infty$ with 
\begin{align}
\mathcal{I}_\infty^{(\alpha)}
&=\sum\limits_{I,\beta}^{2^n}  |\braket{I}{\alpha}|^2 |\bra{I}\ket{\beta}|^2 \left(1-\frac{2 r(\alpha,\beta)}{n}\right).
\label{eq:I_inf-1}
\end{align}
Equation \eqref{eq:I_inf-1} expresses the saturation value of the imbalance in terms of the exact many-body eigenstates of the Hamiltonian. 

In Ref.~\cite{roy2021fock}, a quantity called ``eigenstate polarization'' that is a close relative of the imbalance \eqref{eq:imbalance-2}  was considered and can be presented in a form similar to Eq.~\eqref{eq:I_inf-1}. An essential difference is that, in order to determine the eigenstate polarization experimentally, one would need to perform the full eigenstate tomography. 

\subsection{Imbalance fluctuations}

When one studies the imbalance, the simplest observable is the average imbalance,
\begin{equation}
\overline{\mathcal{I}^{(\alpha)}(t)} 
= \overline{\bra{\psi(t)} \hat{\mathcal{I}}^{(\alpha)} \ket{\psi(t)}} \equiv \overline{\langle \hat{\mathcal{I}}^{(\alpha)}(t) \rangle} \,.
\label{eq:imbalance-5}
\end{equation}
It includes the quantum averaging over the state $\ket{\psi(t)} = \hat{U}(t) \ket{\alpha}$, which we abbreviate as $\langle \ldots (t) \rangle$, and the statistical (mesoscopic) average over random realizations of the Hamiltonian (denoted by an overbar).

As was appreciated in earlier works \cite{DeTomasi2021, roy2021fock,Yao2023, Sun2024} (for other models), studying fluctuations of the imbalance and related observables is also very useful in the context of MBL transitions. It will be convenient in this context to introduce an operator $\hat{x}$ of Hamming distance (with respect to the basis state $\alpha$),
\begin{align}
\hat{x} &= \frac{n}{2}\left(1 - \hat{\mathcal{I}}^{(\alpha)}\right) = \frac{1}{2}\sum_{i=1}^n\left(1 - s_i^{(\alpha)} \sigma^z_i\right)
\label{eq:x-operator-def}
\end{align}
Then, for a given realization of disorder
\begin{align}
\mathcal{I}^{(\alpha)}(t) =
\langle \hat{\mathcal{I}}^{(\alpha)}(t) \rangle  = 1- \frac{2\langle x(t) \rangle }{n},    
\label{eq:imbalance-x-relation}
\end{align}
 where $\langle x(t) \rangle \equiv \bra{\psi(t)}\hat{x}\ket{\psi(t)}$.
We define, again for a given realization of disorder, the following probability distribution (with the variable $x$ taking integer values from 0 to $n$)
\begin{align}
P^{(\alpha)}(x,t) = \sum_\beta |\bra{\beta}\ket{\psi(t)}|^2 \delta_{r(\alpha,\beta), x} \,,
\label{eq:radial_distrib}
\end{align}
which has a meaning of a ``radial''  (Hamming-distance) probability distribution of the time-evolved state $\ket{\psi(t)}$ on the Fock-space graph.
Then $\langle x(t) \rangle = \sum_x x P^{(\alpha)}(x,t)$. 

We turn now to fluctuations of the imbalance.
 Since Eq.~\eqref{eq:imbalance-5} involves two types of averaging (quantum and mesoscopic), one can define two types of imbalance fluctuations.  The quantum variance $v_q$ of the imbalance reads
\begin{align}
v_q(t) &= \overline{\langle [\hat{\mathcal{I}}^{(\alpha)}]^2(t) \rangle - \langle \hat{\mathcal{I}}^{(\alpha)}(t) \rangle^2} \nonumber\\
& = \left(\frac{2}{n}\right)^{\!2}\,\left[ \overline{\langle x^2(t) \rangle - \langle x(t) \rangle^2}\right] \,,
\label{eq:imbalance-vq}
\end{align}
where $\langle x^2(t) \rangle = \sum_x x^2 P^{(\alpha)}(x,t)$. To determine experimentally $v_q(t)$, one should perform multiple quantum measurements of the imbalance for the same disorder realization, to find the variance of the distribution of results, and then to average this variance over disorder realizations. The mesoscopic variance $v_m(t)$  of the imbalance is defined as 
\begin{align}
v_m(t) &= \overline{\langle \hat{\mathcal{I}}^{(\alpha)}(t) \rangle^2} - \overline{\langle \hat{\mathcal{I}}^{(\alpha)}(t) \rangle}^{\,2} \nonumber \\
&= \left(\frac{2}{n}\right)^{\!2}\left[\,\overline{\langle x(t) \rangle^2} -\overline{\langle x(t) \rangle}^{\,2}\right] \,;
\label{eq:imbalance-vm}
\end{align}
it measures fluctuations of the quantum average of the imbalance from one disorder realization to another one.

\section{Analytical considerations for specific models}
\label{sec:analytics}

The subject of this section is the  analytical study of the generalized imbalance and its fluctuations in models considered in this paper. 

\subsection{QREM}
\label{sec:analytics-QREM}

The localization transition in the QREM can be explored by using its connection with the RRG model, see Appendix \ref{app:QREM}. In the large-$n$ limit, the scaling of the QREM critical disorder is
\begin{equation}
W_c^{\rm QREM}(n) \sim n^{1/2} \ln n \,, \qquad n \to \infty \,,
\end{equation}
see Eqs.~\eqref{eq:RRG-maintext-Wc-rescaled-equation}, \eqref{eq:RRG-Wc-asympt} for more accurate formulas with numerical coefficients.
Below we discuss the imbalance and its fluctuations in QREM, first in the ergodic phase, then in the localized phase, and finally at the localization transition, $W \approx W_c^{\rm QREM}(n)$.

\subsubsection{Ergodic phase}
\label{sec:analytics-QREM-erg}

On the ergodic side of the transition, i.e., for $W$ well below $W_c^{\rm QREM}(n)$, the dynamics of a system initially placed at a given site $\alpha$ of the QREM Fock space is closely related to that on RRG \cite{tikhonov19statistics} or on a Cayley tree \cite{monthus1996random,efetov1987density,mirlin1991localization}. Specifically, the fictitious particle representing the system exhibits diffusion on the Fock-space graph. This diffusion is characterized by a diffusion coefficient $D(W)$. Diffusion on a Cayley tree is very peculiar in the sense that the average distance $r(t)$ of a particle from the origin grows linearly with time, $r(t) = (m-1)Dt$, where $m+1$ is the coordination number. This is because, in a small time interval $\Delta t$, there is a probability $mD \Delta t$ to move one step away from the origin and a probability $D \Delta t$ to move one step closer to the origin. Here we use the normalization of $D$ as in Ref.~\cite{efetov1987density}, such that the total rate $\Gamma$ of escape from the given site is $\Gamma = (m+1)D$. 

On the QREM graph (hypercube), when a particle is at a Hamming distance $x$ from the initial vertex, there are $x$ links back (reducing Hamming distance by unity) and $n-x$ links forward (increasing Hamming distance by unity). Thus, the ``diffusion'' equation for the averaged distance $x(t)$ reads
\begin{align}
\frac{d}{dt}\overline{\langle x(t) \rangle} = \left[1-\frac{2 \overline{\langle x(t) \rangle}}{n}\right]Dn \,.
\end{align}
Solving this equation and using 
Eq.~\eqref{eq:imbalance-x-relation}, we obtain 
for the dynamics of average imbalance
\begin{align}
    \overline{\mathcal{I}(t)} = 1-\frac{2 \overline{\langle x(t) \rangle}}{n}=e^{-2Dt} \,.
    \label{eq:QREM-imbalance-decay}
\end{align}

We discuss now the behavior of the diffusion constant $D(W,n)$ when the system approaches the transition from the ergodic side.
With increasing disorder, $D$ decreases until the transition from ergodicity to localization takes place. This transition takes place when the correlation volume $N_\xi$, which grows with increasing disorder $W$, reaches the Fock-space volume $N=2^n$, see Refs.~\cite{Herre2023,Scoquart2024} and Appendices \ref{app:RRG} and \ref{app:QREM}.
The corresponding disorder strength is denoted by $W_c^{\rm QREM}(n)$. With increasing $n$, this disorder approaches from below the thermodynamic-limit transition point in the corresponding RRG model with coordination number $n$, i.e., $W_c^{\rm QREM}(n) / W_c^{\rm RRG}(n,\infty) \to 1$ as $n \to \infty$.

The scaling of the width $\Gamma$ and the diffusion constant $D$ is determined, up to subleading factors, by the behavior of the correlation volume $N_\xi$:
\begin{equation}
N_\xi \sim \frac{W n^{1/2}}{\Gamma} \,, \qquad D \sim \frac{\Gamma}{n} \,.
\label{eq:relations-Nxi-Gamma-D}
\end{equation}

When the system is large enough, $n > n_* \approx 22$, the finite-size transition takes place in the critical regime of the thermodynamic-limit transition, which means that $W_c^{\rm QREM}(n) > (1/2) W_c^{\rm RRG}(n,\infty)$. In this case, there is a range of disorder $W$ for which the system is still ergodic and, at the same time, the disorder satisfies $W > (1/2) W_c^{\rm RRG}(n,\infty)$. 
In this range of disorder, the correlation volume is given by Eq.~\eqref{eq:RRG-N-xi-critical} (with $m \mapsto n$) and
\begin{align}
        D \sim N_\xi^{-1} \sim \exp{-\pi \ln n \left[\frac{3}{2}\left(1-\frac{W}{W_c}\right)\right]^{-1/2}},
        \label{eq:QREM_D_critical}
    \end{align}
    where we keep only the leading exponential factor.
(A more accurate formula for $D$ in this regime includes an additional subleading factor, which is a power of $\ln N_\xi$.) 

For QREM systems that are accessible to exact numerical simulations, we have $n < n_*$ and the behavior \eqref{eq:QREM_D_critical} cannot be observed. In this case, the range of disorder on the ergodic side of the finite-size transition point is in the precritical regime (from the point of view of the thermodynamic-limit transition), $n^{1/2} < W < (1/2) W_c^{\rm RRG}(n,\infty)$. The correlation volume in this regime is given by \eqref{eq:QREM-precrit-N-xi}, which implies, in combination with Eq.~\eqref{eq:relations-Nxi-Gamma-D}, the following behavior of the diffusion constant:
   \begin{align}
        D \sim \frac{1}{n^{1/2}W}
        \exp\left[-\frac{W}{(6n)^{1/2}} \right].
        \label{eq:QREM_D_precritical}
    \end{align}
The exponential decay \eqref{eq:QREM-imbalance-decay} proceeds until the imbalance saturates at its limiting value $\overline{\mathcal{I}}_\infty$, which we are going to discuss now.  According to Eq.~\eqref{eq:I_inf-1}, $\overline{\mathcal{I}}_\infty$ can be written as 
\begin{eqnarray}
\overline{\mathcal{I}}_\infty &=& \sum_\beta
N F( r(\alpha,\beta))
\left(1-\frac{2 r(\alpha,\beta)}{n}\right)
\nonumber \\
& = & \sum_{x} N F(x) C(x) (1-2x/n)
\,,
\label{eq:QREM-I-infty-averaged}
\end{eqnarray}
where 
\begin{equation}
F(x) = \overline{|\braket{I}{\alpha}|^2 |\bra{I}\ket{\beta}|^2 \: \delta_{r(\alpha,\beta),x}} - \frac{1}{N^2} \,.
\label{eq:QREM-I-infty}
\end{equation}
and $C(x)$ is the number of vertices $\beta$ of the graph at a distance $x$ from a given vertex $\alpha$.
The correlation function $F(x)$ was studied for eigenstates $\ket{I}$  in the ergodic phase of the RRG model in Ref.~\cite{tikhonov19statistics}. It was found there that
\begin{equation}
NF(x)C(x) \sim \frac{N_\xi}{N} \, x^{-3/2} \,,
\label{eq:RRG-F}
\end{equation}
 Expecting the same behavior for QREM and substituting Eq.~\eqref{eq:RRG-F} into Eq.~\eqref{eq:QREM-I-infty}, we see that the sum over $x$ in Eq.~\eqref{eq:QREM-I-infty} is dominated by $x \sim 1$, yielding
 \begin{equation}
\overline{\mathcal{I}}_\infty \sim \frac{N_\xi}{N} \,.
\label{QREM-I-infty-result}
 \end{equation}
This result suggests that the limiting value $\mathcal{I}_\infty$ of the generalized imbalance exhibits the same behavior as the inverse participation ratio (IPR) $P_2$ or the infinite-time return probability to the initial point on the graph. 

Using Eq.~\eqref{QREM-I-infty-result}, we can estimate the time $t_{\rm sat}$ of saturation of the exponential decay   \eqref{eq:QREM-imbalance-decay} of the imbalance,
\begin{equation}
t_{\rm sat} \simeq \frac{1}{2D} (n \ln 2 - \ln N_\xi).
\label{eq:QREM-t-sat}
\end{equation}
When the system is sufficiently deep in the ergodic phase (i.e., $W$ well below $W_c(n)$), this time is much smaller than the Heisenberg time $t_H$, Eq.~\eqref{eq:tH}.

The following comment is in order at this point. The above derivation of the estimate 
\eqref{QREM-I-infty-result} uses Eq.~\eqref{eq:RRG-F} for the correlation function \eqref{eq:QREM-I-infty}, which holds for averaging over eigenstates $\ket{I}$ near a given energy in the spectrum (in our case, this is the band center, $E=0$, where density of states is sharply peaked). On the other hand, our Eq.~\eqref{eq:I_inf-1} contains summation over all eigenstates $\ket{I}$ in the spectrum. For $W$ well below $W_c(n)$, nearly all states in the band are ergodic but an exponentially small fraction of states in the band tails is expected to be localized. Such a localized state $\ket{I}$ would give a contribution to the sum in Eq.~\eqref{eq:I_inf-1} if the initial basis states $\ket{\alpha}$ happens to have a substantial overlap with $\ket{I}$. This exponentially rare event would give an additional small contribution to the average Eq.~\eqref{QREM-I-infty-result}. To minimize the effect of the localized states in the tails, we choose as the initial state $\alpha$ in our numerical analysis a basis state with an energy close to zero, see Sec.~\ref{sec:imbalance-numerics}. This ensures that eigenstates $\ket{I}$ that give a sizeable contribution to Eq.~\eqref{eq:I_inf-1} are located close to the band center (in the energy window $\Gamma$, within which the states are efficiently mixed), and the contribution of eigenstates with other energies (in particular, localized tails) is negligible. This comments applies also to other models that we study.

 We consider now fluctuations of the asymptotic ($t\to \infty$) value of the imbalance $\mathcal{I}_\infty$ in the ergodic phase. In the  large-$t$ limit,  the distribution of $\ket{\psi(t)}$ over the hypercube Fock-space graph is 
 approximately uniformly on volumic scales from the correlation volume $N_\xi$ to the volume $N$ of the whole graph.
 The distribution of Hamming distances $x=r(\alpha,\beta)$ with respect to a given state $\ket{\alpha}$ on this graph is given by 
\begin{align}
p(x) = \frac{1}{2^n}\frac{n!}{(n-x)!x!} \simeq \sqrt{\frac{2}{n\pi}}\exp\left[-\frac{2(x-\frac{n}{2})^2}{n}\right].
\label{P-Hamming}
\end{align}
This yields 
\begin{equation}
\langle x \rangle = \frac{n}{2}\,, \qquad \langle x^2 \rangle-\langle x \rangle^2 = \frac{n}{4} \,,
\end{equation}
and thus, for the quantum variance \eqref{eq:imbalance-vq} of the imbalance, 
\begin{align}
v_q(t\to\infty) = \frac{1}{n} \,.
\label{vq-QREM-ergodic}
\end{align}

At the same time, the mesoscopic fluctuations of $\langle x\rangle$ are expected to be exponentially small in this regime, because $\langle x\rangle$ results from averaging over a state $\ket{\psi(t)}$ with exponentially many ($ \sim N / N_\xi$) independent components of comparable amplitude:
\begin{align}
\overline{\langle x \rangle^2}-\overline{\langle x \rangle}^2 \sim \frac{N_\xi}{N} \, n = \frac{N_\xi}{2^n} \, n \,,
\label{eq:meso_flucts_ergodic_QREM}
\end{align}
so that 
\begin{equation}
v_m(t\to\infty) \sim \frac{N_\xi}{2^n \, n} \,, \qquad N_\xi \sim \frac{W n^{1/2}}{\Gamma}\,.
\label{vm-QREM-ergodic}
\end{equation}
Here $\Gamma$ is the energy width within which states are efficiently mixed; ergodicity implies that $\Gamma$ is much larger than the many-body level spacing, which corresponds to the condition $N_\xi \ll N$.

\subsubsection{Localized phase}
\label{sec:analytics-QREM-loc}

Now, we consider the localized phase, i.e., the disorder $W$ well above $W_c^{\rm QREM}(n)$. The mechanism of the localization transition on RRG and QREM is peculiar in comparison with a conventional $d$-dimensional Anderson model (e.g., in $d=3$ dimensions). In the conventional Anderson transition, when the disorder changes towards the transition point $W_c$ from the localized side, the system effectively spreads fractally over the localization volume of a radius of order of localization length $\xi_{\rm loc}$ that diverges at $W \to W_c$. One manifestation of this is the fractal growth of the IPR $P_2$ \cite{evers08}.
In this sense, the delocalization happens gradually within the increasing length scale $\xi_{\rm loc}$ until it reaches the system size at the transition point. On the other hand, on tree-like graphs, like RRG, the delocalization is driven by system-wide resonances that are exponentially rare in the localized phase but proliferate when $W$ reaches the transition point. Eigenstates of the system remain close to basis states (i.e. infinite-disorder eigenstates) in the whole localized phase, up to the transition point, with only a parametrically small admixture of other basis states. In particular, the IPR $P_2$ remains close to unity. This behavior is also expected for QREM at large $n$.  

Thus, the wave function $\ket{\psi(t)}$  in the localized phase of QREM will remain close to the initial state $\ket{\psi(0)}=\ket{\alpha_0}$. The dominant correction will come from admixture of basis states at a Hamming distance $x=1$. 
The level spacing of these states is $\sim W / n^{1/2}$, and they are connected to the initial state by a matrix element $T \sim 1$.
Thus, with a probability $\sim n^{1/2}/W$, one of these states will form a strong resonance with $\ket{\alpha}$. In view of the scaling \eqref{eq:RRG-maintext-Wc-rescaled-equation} of the QREM critical disorder, this probability is small in the localized phase. To the leading order, one can thus neglect the probability of two or more such resonances. Similarly, the probability of, say, second-order resonance (with a state at Hamming distance $x=2$) will have an additional smallness and can be discarded. 

We thus focus on a dominant admixture to $\ket{\alpha}$ of a state at Hamming distance unity. It will give, in the long-time limit, the value of $\langle x \rangle$ in the range $[0, 1/2]$.  (More precisely, there will be Rabi oscillations around this value; we discard them by assuming time averaging over a sufficient interval.)  The value $\langle x \rangle$ will be random, i.e., dependent on the disorder realization. Typically, it will be small but the variances that we want to calculate will be determined by those realizations for which  $\langle x \rangle \sim 1/2$.  The probability of such events is $\sim n^{1/2}/W$, as discussed above. Therefore, to the leading order, we have at $t\to \infty$
\begin{eqnarray}
\overline{\langle x \rangle} &=& \overline{\langle x^2 \rangle} = c_1 \, \frac{n^{1/2}}{W} \,, 
\label{eq:QREM-loc-x-av-1}
\\
\overline{\langle x \rangle^2} &=&  c_2 \, \frac{n^{1/2}}{W} \,,
\label{eq:QREM-loc-x-a-2}
\end{eqnarray}
where $c_1, c_2 \sim 1$ are numerical coefficients.
In Eq.~\eqref{eq:QREM-loc-x-av-1}, we have used that the quantum operator $\hat{x}$ takes only values $x=0$ and $x=1$ in this case, so that $x^2 = x$. Thus, we obtain for the average imbalance $\overline{\mathcal{I}}_\infty = 1 - 2\overline{\langle x \rangle}/n$ and for its
quantum and mesoscopic variances,
Eqs.~\eqref{eq:imbalance-vq} and \eqref{eq:imbalance-vm}, the following results
\begin{eqnarray}
1 - \overline{\mathcal{I}}_\infty & = & \frac{2c_1}{n^{1/2} W} \,,
\label{eq:QREM-loc-average-imb} \\
v_q (t\to \infty) & = &  \frac{4 (c_1 - c_2)}{n^{3/2} W} \,, 
\label{eq:QREM-loc-vq} \\
v_m (t\to \infty) &=&  \frac{4 c_2}{n^{3/2} W} \,.
\label{eq:QREM-loc-vm}
\end{eqnarray}
In Eq.~\eqref{eq:QREM-loc-vm}, we have neglected $\overline{\langle x \rangle}^2$, which contains an additional small factor $\sim n^{1/2} /W$ in comparison to  $\overline{\langle x \rangle^2}$.

To calculate the numerical coefficients $c_1$ and $c_2$, we use the distributions of the diagonal and hopping matrix elements in the QREM, which are given by
Eq.~\eqref{eq:distrib_energies_hoppings_1D}. 
The result is 
\begin{equation}
c_1 =   \frac{\pi\sqrt{3}}{2} \,, \qquad c_2 = \frac{\pi\sqrt{3}}{8} \,.
\label{eq:1D-loc-c1-c2}
\end{equation}
The calculation of these coefficients turns out to be essentially the same as  that in the localized phase of the 1D model, Sec.~\ref{sec:analytics-1D-loc}. The above results can be obtained from 1D results by including a rescaling factor $2\sqrt{6/\pi n}$, which takes into account the ratio of the densities of states on sites directly coupled to a site with zero energy in the QREM and 1D models, $\gamma^{\rm QREM}(0)=(6/\pi n)^{1/2}W^{-1}$ and $\gamma^{\rm 1D}(0)= 1/2W$. An important difference is that, in the QREM, 
Eqs.~\eqref{eq:QREM-loc-average-imb}, 
\eqref{eq:QREM-loc-vq}, and \eqref{eq:QREM-loc-vm} correspond to the effect of a single strong resonance, while their counterparts in the localized phase of the 1D model, Eqs. (\ref{1D-Iinfty}), (\ref{eq94}), and (\ref{eq95}), hold for a finite density of strong resonances.

Analyzing corrections to Eqs.~\eqref{eq:QREM-loc-average-imb}, 
\eqref{eq:QREM-loc-vq}, and \eqref{eq:QREM-loc-vm} coming from higher-order resonances, we find that the parameter controlling the smallness of these corrections 
is $W_c^{\rm RRG}(n, \infty) / W \ll 1$.  

\subsubsection{Transition region}
\label{sec:analytics-QREM-trans}

We turn now to the transition region, $W\approx W_c^{\rm QREM}(n)$. At the transition,
 the correlation volume $N_\xi$ is equal to the system size: $N_\xi \simeq N = 2^n$. 
 To estimate the long-time asymptotics $\overline{\mathcal{I}}_\infty$  of the average imbalance, as well as quantum and mesoscopic fluctuations of the imbalance, we develop a simplified picture of the state 
 $\ket{\psi(t)}$ (at $t \to \infty$) in this regime, which however captures the key properties. Specifically, as discussed above, we will typically have one strong system-wide resonance, i.e., 
 \begin{align}
\ket{\psi(t)} \simeq a\ket{\alpha} + b\ket{\beta}, \qquad |a|^2 + |b|^2 = 1\,,
\label{eq:QREM-transition-state}
\end{align}
with $|a|\sim |b|$ and $r(\alpha,\beta) \approx n/2$.
To estimate $v_q$ at the transition, consider a maximally strong resonance , i.e., a state \eqref{eq:QREM-transition-state} with $|a|=|b|=1/\sqrt{2}$. This state is characterized by  
\begin{align}
&\langle x \rangle =\frac{n}{4}\,, \qquad
\langle x ^2\rangle -\langle x \rangle^2 = \frac{n^2}{8}- \frac{n^2}{16} = \frac{n^2}{16} 
\end{align} 
and thus
\begin{equation}
\mathcal{I}_\infty = \frac{1}{2}\,, \qquad 
v_q = \frac{1}{4} \,.
\end{equation}
Taking into account fluctuations of the resonance strength (i.e., amplitudes $a$ and $b$), as well as a finite probability of $\ket{\psi(t)}$ involving $2,3,\ldots$ strong resonances
 will not change the dominant $n^2$ scaling of $\overline{\langle x ^2\rangle -\langle x \rangle^2}$ but  will reduce the numerical prefactor. Thus, we will have at the transition, $W\approx W_c^{\rm QREM}(n)$,
 \begin{eqnarray}
 \overline{\mathcal{I}}_\infty &\simeq & 1/2 \,, \\
 \overline{\langle x ^2\rangle -\langle x \rangle^2} &\simeq & C_q n^2\,, \qquad
 v_q(t \to \infty) \simeq 4C_q \,,
  \label{eq:QREM-transition-vq}
\end{eqnarray}
with a numerical coefficient $C_q$ satisfying $C_q < 1/16$. 

 To estimate the mesoscopic fluctuations,  let us again consider the states of the form \eqref{eq:QREM-transition-state} and take into account that the strength of the resonance $|b/a|$ fluctuates from one realization of disorder to the other. Consider a simplistic model of this, with a half of the states having a maximally strong resonance, $|b/a|=1$, and another half of the states having no strong resonance at all, which can be approximated as $|b/a|=0$. This would give for the mesoscopic variance of $\langle x \rangle$
 \begin{equation}
\overline{\langle x \rangle^2} - \overline{\langle x \rangle}^2 = \frac{n^2}{64} \,.
 \end{equation}
 Clearly, this is only a caricature of the true physical situation of mesoscopic fluctuations at the transition $W\approx W_c^{\rm QREM}(n)$. Indeed, as discussed above, there should be, first, a continuous distribution of $|b/a|$ and, second, a finite probability of formation of more than a single resonance. This can, however, only modify a numerical coefficient (making it larger or smaller) but can not change the scaling with $n$. Thus, we have
 \begin{align}
\overline{\langle x \rangle^2} - \overline{\langle x \rangle}^2  \simeq C_m n^2, \qquad v_m(t \to \infty) \simeq 4C_m \,,
 \label{eq:QREM-transition-vm}
\end{align}
with a rough estimate $C_m \sim 1/64$ for the coefficient $C_m$.

Comparing the results for the long-time value of the imbalance and its fluctuations  at the transition, $W\approx W_c^{\rm QREM}(n)$, with those in the ergodic 
(Sec.~\ref{sec:analytics-QREM-erg})
and localized (Sec.~\ref{sec:analytics-QREM-loc}) phases, we make the following qualitative observations:
\begin{itemize}
\item[(i)] The average imbalance $\overline{\mathcal{I}}_\infty$ is small in the ergodic phase, close to unity in the localized phase, and crosses the line $\overline{\mathcal{I}}_\infty=1/2$ in the transition region;
\item[(ii)] Both quantum and mesoscopic  variances of the imbalance, $v_q$ and $v_m$, are of order unity in the transition region (up to a numerical smallness of coefficients $4C_q < 1/4$ and $4C_m \sim 1/16$), while they are parametrically small (for large $n$) both in the ergodic and localized phases. Therefore,  $v_q$ and $v_m$ should have a maximum at the transition (at least, when $n$ is sufficiently large). 
\end{itemize}
Thus, the three conditions, namely $\overline{\mathcal{I}}_\infty=1/2$, maximum of $v_q$, and maximum of $v_m$, can serve as indicators of the transition point.

Of course, for a relatively small $n$, there are additional sources of finite-size corrections that are not taken into account in the above analysis. Thus, it is important to complement these analytical considerations with a numerical study, which is done in Sec.~\ref{sec:numerics} below.

\subsection{1D model}
\label{sec:analytics-1D}

As discussed in Sec.~\ref{sec:intro}, most analytical works predict
\begin{equation}
W_c^{\rm 1D}(n) = {\rm const} \,, \qquad n \to \infty \,,
\end{equation}
for the MBL transition in 1D models. In analogy with the structure of Sec.~\ref{sec:analytics-QREM}, we discuss here analytical predictions for the imbalance and its fluctuations in our 1D model, starting with the ergodic phase, and then continuing to the localized phase and the vicinity of the transition point. 

\subsubsection{Ergodic phase}
\label{sec:1D-analytics-ergodic}

On the ergodic side of the transition, $W<W_c^\text{1D}(n)$, the dynamics is governed by the Griffiths phenomenon due to the presence of rare regions with strong disorder. For models with extensive conserved quantities (energy in our model), the
real-space relaxation of the corresponding density is  subdiffusive. 
The subdiffusion can be described in terms of an effective, momentum-dependent diffusion coefficient $D(q)$ that obeys
\begin{align}
    D(q)\propto q^\beta, \quad\quad \beta>0.
\end{align}
This yields the power-law behavior for the characteristic spreading in real space, characterized by exponent $0<\gamma_X< 1$:
\begin{align}
    \overline{ (\Delta r)^2  (t)} \propto t^{\gamma_X}, \qquad \gamma_X = \frac{2}{2+\beta}.
\end{align}
In this regime, the averaged imbalance is expected to decay as a power law of time:
\begin{align}
    \overline{\mathcal{I}(t)} \propto t^{-\gamma_I},\quad\quad 0<\gamma_I \leq \frac{1}{2}.
    \label{I-gammaI}
\end{align}
The subdiffusive transport and the power-law behavior of the imbalance on the ergodic side of the MBL transition have been found numerically in various one-dimensional models  (see, e.g., Ref.~\cite{Sierant2024} for a recent review). 
The exponents $\beta$ and $\gamma_I$ were found to change continuously as a function of $W$, with $\beta$ diverging and $\gamma_I$ approaching zero when $W$ approaches the transition point. 

Anayltically, the power-law decay 
\eqref{I-gammaI} of the imbalance, with
$\gamma_I = 1/(1+\beta)$ for a system with extensive conserved quantities, was derived in Ref.~\cite{Gopalakrishnan2016a}. The corresponding mechanism is directly 
 provided by the Griffiths effects due to the presence of  exponentially rare regions (``traps''), in which the relaxation times are exponentially large.  
An alternative mechanism leading to Eq.~\eqref{I-gammaI} was explored in Ref.~\cite{Popperl2022}. Within this mechanism, the memory effects associated with the return probability in subdiffusive dynamics are responsible for producing the power-law tail of imbalance, yielding a somewhat slower decay, $\gamma_I = 1/(2+\beta)$. When the system approaches the transition point, $\beta$ becomes large and the two contributions to imbalance decay with almost the same exponents. 

Beyond the Heisenberg time $t_H$, Eq.~\eqref{eq:tH}, the imbalance saturates at a small value $\overline{\mathcal{I}}_\infty\ll 1$ determined by the system size. 
If one assumes that the power-law decay \eqref{I-gammaI} of the imbalance holds in the whole time range up to a time of order of $t_H$, one gets an estimate \begin{equation}
\overline{\mathcal{I}}_\infty\sim t_H^{-\gamma_I} \,.
\label{eq:1D-I-infty-ergodic}
\end{equation}

The quantum and mesoscopic fluctuations of the residual imbalance $\mathcal{I}_\infty$ in the ergodic phase of the 1D model are obtained in the same way as for the QREM, see
the analysis in the end of Sec.~\ref{sec:analytics-QREM-erg}. Since the two models are defined on the same hypercube Fock-space graph, their distribution functions of the Hamming distances (\ref{P-Hamming}) are equivalent. The subsequent analysis, leading to Eq.~\eqref{vq-QREM-ergodic} for $v_q(t\to \infty)$ and Eq.~\eqref{vm-QREM-ergodic} for $v_m(t\to \infty)$, applies also for the 1D model, with the following word of caution concerning Eq.~\eqref{vm-QREM-ergodic}. It was assumed in derivation of Eq.~\eqref{vm-QREM-ergodic} that $N/N_\xi$ components contributing to the state $\ket{\psi(t)}$ can be viewed as essentially independent, which leads to an exponential suppression $\propto 2^{-n}$ of $v_m(t\to\infty)$. The status of this assumption (similar to that made in justification of the eigenstate thermalization hypothesis) is not fully clear for the 1D model (and also QD model, see below), because the number of independent parameters of the Hamiltonian is of order $n$. In any case, the mesoscopic  variance $v_m(t\to\infty)$ should be strongly suppressed compared to the quantum variance $v_q(t\to\infty)$ in the ergodic phase. Our numerical results below fully confirm this and, moreover, show that this suppression in the 1D and QD models is similar to that in the QREM, supporting the validity of Eq.~\eqref{vm-QREM-ergodic} also for the 1D and QD models. 

\subsubsection{Localized phase}
\label{sec:analytics-1D-loc}

We now turn to the localized phase. 
Our analysis here has close connections to that in Refs.~\cite{DeTomasi2021, roy2021fock} where a different 1D model was considered.

For strong disorder, ${W\gg W_c^\text{1D} (n) \sim 1}$, various properties of the system are governed by rare strong resonances. While this bears some similarity to the physics of the localized phase of the QREM, 
Sec.~\ref{sec:analytics-QREM-loc},
there is an essential difference. For QREM deeply in the localized phase, we had at most one (if at all) strong resonance. On the other hand, in the localized phase of the 1D model, there will be a finite {\it concentration}  $\sim 1/W$ of resonant spins. 
This is because, in the 1D model,
spins that are not nearest neighbors do not interact to each other. 
As a consequence, the average imbalance in the 1D model is predicted to saturate to a finite ($n$-independent) value  
$0<\overline{\mathcal{I}}_\infty(W)<1$ in the thermodynamic limit $n\to \infty$. 
Indeed, each of these resonant spins gives a contribution $\sim -1/n$ to the imbalance, and the total number of resonant spins is $\sim n/W$ for $W \gg 1$, yielding an $n$-independent result

\begin{align}
    1-\overline{\mathcal{I}}_\infty \sim \frac{n}{W}\,\frac{1}{n} \sim \frac{1}{W}.
    \label{eq:1D-loc-imbalance-estimate}
\end{align}
It is worth emphasizing a direct relation 
of this result to a fractal scaling of  the IPR in the MBL phase, which is governed by the same resonances,
\begin{align}
    P_2\sim N^{-\tau_2(W)},
\end{align}
with the multifractal exponent $\tau_2(W)\sim 1/W$. 

In this way, we can also estimate the scaling of imbalance fluctuations. Each of $\sim n/W$ strong resonances gives a  contribution $\sim {\cal O}(1)$ to the variance $\langle x ^2\rangle-\langle x \rangle^2$ at long times. The total variance is thus $\sim n/W$, which yields for the quantum variance of the imbalance $v_q \sim 1/nW$. To estimate mesoscopic fluctuations, we notice that the number of strong resonances $\sim n/W$ in Eq.~\eqref{eq:1D-loc-imbalance-estimate} should exhibit fluctuations according to Poisson statistics, with variance equal to the mean value. Thus, the mesoscopic variance of the imbalance scales as $v_m \sim 1/nW$, i.e., in the same way as $v_q$.

In Appendix \ref{app:1D-loc}  we calculate numerical coefficients in these formulas for our 1D model. 
The result for the average imbalance reads
\begin{align}
  1-\overline{\mathcal{I}}_\infty =  \frac{2}{n}\overline{\langle x \rangle}\simeq  \frac{\pi^{3/2}}{2\sqrt{2}\, W} \,,
  \label{1D-Iinfty}
\end{align}
and the long-time imbalance fluctuations are given by
\begin{align}
&v_q (t\to \infty)= \left(\frac{2}{n}\right)^2\left[\overline{\langle x ^2\rangle-\langle x \rangle^2}\right] \simeq \frac{3\pi^{3/2}}{4\sqrt{2}\, n W},
\label{eq94} 
\\
&v_m (t\to \infty)= \left(\frac{2}{n}\right)^2\left[\overline{\langle x\rangle^2}-\overline{\langle x \rangle}^2\right] \simeq \frac{\pi^{3/2}}{4\sqrt{2}\, n W}.
\label{eq95}
\end{align}
The multifractal exponent characterizing the scaling of the (typical) IPR $P_2$ is, to the $1/W$ order,
\begin{align}
\tau_2(W) &=
-\frac{1}{\ln N}\, \overline{\ln P_2}
= \frac{\pi^{3/2}(\sqrt{2}-1)}{2\ln \!2 \, W}.
\label{eq:1D-loc-main-tau-2-final}
\end{align}

\subsubsection{Transition region}

Development of a fully controllable theory of the MBL transition in 1D systems, including the corresponding critical behavior of observables, remains a challenging goal. Several phenomenological theories of the transition have been put forward
\cite{roeck17,Thiery2017a,goremykina2019analytically,morningstar2019renormalization,morningstar2020a}. While differing in various aspects, they have a very important common feature: the critical point of the transition is essentially a continuation of the MBL phase. From this point of view, the MBL transition in 1D systems is similar to the localization transition in the  RRG model (and its relatives, including QREM). 
To discuss properties of the 1D model in the transition region, we will thus proceed as for the QREM (Sec.~\ref{sec:analytics-QREM-trans}) and approach the transition region from the localized side.

In the localized phase, the localized eigenstates in a 1D chain are formed via $p(W) n$ strong resonances, with $p(W) \sim W^{-1}$, as discussed in Sec.~\ref{sec:analytics-1D-loc}.
While strengths of the resonances are statistically distributed, we assume here, for the sake of an estimate, that there are $p(W)n$ maximally strong resonances, while all other resonances can be neglected. 
It is understood that the function $p(W)$ has a limit $p_c$ satisfying $0 < p_c < 1$ at the (thermodynamic-limit) transition point $W_c^{\rm 1D}$. This corresponds to the value 
$\overline{\mathcal{I}}_\infty = 1- p_c$ of the imbalance in the localized phase just near the transition. Thus, every eigenstate is effectively spread over $\sim 2^{p_cn}$ basis states.

As $W$ decreases to the transition region, $W\simeq W_c^\text{1D}(n)$, system-wide resonances  proliferate, representing instability of the localized phase. It is commonly understood, and supported by numerical simulations, that system-wide resonances drive the MBL transition \cite{villalonga2020, Ghosh2022resonance, Morningstar2022avalanches, 
Herre-thesis, Ha2023, long2023phenomenology, colbois2024statistics}.
To obtain estimates for the average imbalance $\overline{\mathcal{I}}_\infty(W)$, as well as for quantum and mesoscopic fluctuations, we will assume that at $W\simeq W_c^\text{1D}(n)$, every localized eigenstate (residing on $2^{p_cn}$ basis states formed by $p_c$ resonances) finds resonance with another such state, located maximally far apart in the Fock space ($x \simeq  n/2$). 

Within this ``caricature'' of eigenstates in the transition region, the distribution $P^{(\alpha)}(x,t)$ of Hamming distances, Eq.~\eqref{eq:radial_distrib},  acquires at long times a double-peak structure. Specifically, one peak corresponds to  resonances characterstic for the localized phase, with a maximum at $x=p_cn/2$ and the width $\sim \sqrt{p_cn}/2$, while the other peak corresponds to system-wide resonances and is centered at $x=n/2$ with the width $\sim \sqrt{n}/2$. 
Under this simplifying assumption, we get, for $W \simeq W_c^{\rm 1D}$,
\begin{align}
\overline{\langle x \rangle} \simeq \frac{1}{2}\left( \frac{p_cn}{2} + \frac{n}{2} \right) = \frac{1+p_c}{4}n,
\end{align}
and thus
\begin{align}
\overline{\mathcal{I}}_\infty(W_c^{\rm 1D})\simeq \frac{1 - p_c}{2}.
\label{eq:analytics-1D-imbalance-crit}
\end{align}
For the quantum fluctuations of saturated ($t \to \infty$) imbalance, we estimate
\begin{align}
\overline{\langle x ^2\rangle -\langle x \rangle^2}&\simeq \left(\frac{1-p_c}{4}\right)^2 n^2  + \left(\frac{n}{4}+\frac{n p_c}{4}\right)\frac{1}{2}.
\label{1D-q-fluct-x}
\end{align}
In the large-$n$ limit, the first term is dominant, and we get
\begin{align}
v_q (t \to \infty) \simeq \left(\frac{1-p_c}{2}\right)^2, \qquad W\simeq W^\text{1D}_c.
 \label{eq:1D-transition-vq}
\end{align}
However, for numerically accessible (in the sense of exact computations) system sizes $n\lesssim 20$,
the correction term in Eq.~(\ref{1D-q-fluct-x}) can be comparable to the leading term.

Finally, we estimate the mesoscopic fluctuations. For the part of the distribution over $x$ that is related to system-wide resonances, the consideration is similar to the ergodic regime [see the reasoning above Eq.~\eqref{eq:meso_flucts_ergodic_QREM}]. There is an exponentially large number ($\sim 2^{p_c n}$) of independent states here, thus the corresponding contribution to the mesoscopic fluctuations should be exponentially small. This leaves us with the same result as in the localized phase discussed above (with an additional factor $1/2$ and $W\to W_c^\text{1D}(n)$): the critical region is similar to the localized phase in this sense. We thus expect that, for a finite $n$, when the transition has a finite width, the maximum of $v_m$ will be on the larger-$W$ side of the transition region, where system-wide resonances are not yet important, with 
\begin{align}
\overline{\langle x \rangle^2} - \overline{\langle x \rangle}^2 \simeq \frac{n}{4}p_c 
\end{align}
and thus
\begin{align}
 v_m (t \to \infty) \simeq \frac{p_c}{n}, \qquad W\simeq W^\text{1D}_c. 
 \label{eq:1D-transition-vm}
\end{align}

\subsection{QD model}
\label{sec:analytics-QD}

Similarly to QREM, the critical disorder $W_c^{\rm QD}(n)$ in the QD model grows as a power-law function of $n$ (likely with a logarithmic correction). While the exact exponent of the power-law scaling  of $W_c^{\rm QD}(n)$ is not known with certainty, there are upper and lower bounds leaving a relatively narrow window \cite{Scoquart2024}. The upper bound, $W_c^{\rm QD}(n) < W_c^{\rm uQD}(n)$ is provided by the uncorrelated quantum-dot (uQD) model, which is obtained from the QD model by removing correlations of Fock-space hopping matrix elements. This makes the RRG-like approximation applicable, leading to the equation
$W = \sqrt{\pi}\: n\ln (4W/\sqrt{\pi})$ for the critical disorder in the uQD model, which yields $W_c^{\rm uQD}(n)  \simeq \sqrt{\pi}\, n\ln n$ in the large-$n$ limit \cite{Scoquart2024}.  
The lower bound for the QD model obtained in Ref.~\cite{Scoquart2024} was
$W_c^{\rm QD}(n) \gtrsim n^{3/4} (\ln n)^{-1/4}$. It turns out that a better bound can be derived, with the same power of $n$ but a different power of the logarithm: 1/2 instead of -1/4, see Appendix \ref{app:QD-Wc-lower-bound}. We thus have the following window for the scaling of the transition point in the QD model (omitting numerical coefficients):
\begin{equation}
n^{3/4} (\ln n)^{1/2} \lesssim W_c^{\rm QD}(n) \lesssim n \ln n \,.
\label{sec:analytics-QD-Wc-bounds}
\end{equation}

\subsubsection{Ergodic phase}
\label{sec:analytics-QD-ergodic}

Both the upper and lower bounds for $W_c^{\rm QD}$ are obtained from considerations akin to the physics of models on tree-like graphs, such as the RRG model and QREM. This leads to an expectation that the time decay of the imbalance $\overline{\mathcal{I}(t)}$  in the ergodic phase of the QD model is analogous to that in QREM,
Eq.~\eqref{eq:QREM-imbalance-decay}. 
The Griffiths physics related to rare strong-disorder spatial spots, which is responsible for slow decay of the imbalance in 1D model, is not relevant to the QD model because of all-to-all couplings of spins.

Concerning  the variances of the saturated imbalance, $v_q(t\to \infty)$ and $v_m(t\to \infty)$, the arguments leading to Eqs.~\eqref{vq-QREM-ergodic} and 
\eqref{vm-QREM-ergodic} are generic for the ergodic phase of the class of spin models that we consider. Thus, these formulas apply also to the QD model, with a word of caution concering 
Eq.~\eqref{vm-QREM-ergodic} in the end of Sec.~\ref{sec:1D-analytics-ergodic}.

\subsubsection{Localized phase}

The results for the average imbalance as well as quantum and mesoscopic fluctuations that we have derived for the 1D model, Eqs.~ \eqref{1D-Iinfty}, \eqref{eq94}, and \eqref{eq95}, apply to the localized phase of the QD model as well. Indeed, the distribution of Fock-space hopping matrix elements and the statistics of energies of basis states coupled directly to a given basis state, which determine the statistics of resonances, is the same in both models. 
The difference is in the range of applicability of these formulas. For the QD model they definitely apply for $W \gg W_c^\text{uQD}(n) \simeq \sqrt{\pi} n \ln n$, as, under this condition, the dominant contribution is provided by a single strong first-order resonance. We do not attempt in this work to analyze the disorder region $W_c^\text{QD}(n) \ll W \lesssim W_c^\text{uQD}(n)$
[i.e., the part of the localized phase closer to the transition point $W_c^{\rm QD}(n)$]. While this region is formally parametrically broad at large $n$, it is relatively narrow in practice for moderately large $n$.

\subsubsection{Transition region}

Our discussion of the MBL transition region, $W \simeq W_c^{ \rm QD}(n)$ is necessarily rather speculative. Indeed, even the scaling of the transition point $W_c^{\rm QD}(n)$ with $n$ is not fully understood, although the corresponding window between the bounds  is relatively narrow. We argue, however, that the character of the transition bears similarity to that in the QREM and in the 1D model. Specifically, in the localized phase close to the transition, there is a certain number $n_{\rm res}^{(c)}$ of strong single-spin resonances, which implies that $\sim 2^{n_{\rm res}^{(c)}}$ eigenstates substantially contribute to the basis state $\alpha$. At the transition point, system-wide resonances proliferate, so that typically each of these eigenstates forms such a resonance. For QREM, we had $n_{\rm res}^{(c)}=0$, while for the 1D model we had 
$n_{\rm res}^{(c)}= p_c n$, with an $n$-independent $p_c$. In the QD model, the situation is intermediate, although in a sense ``closer to QREM''. Specifically, if we assume that the lower bound in Eq.~\eqref{sec:analytics-QD-Wc-bounds} yields in fact the true scaling of $W_c^\text{QD}(n)$, we get 
\begin{equation}
n_{\rm res}^{(c)} \sim n^{1/4} (\ln n)^{-1/2} \,.
\end{equation}

\begin{figure*}[ht!]
    \centering
    \includegraphics[width=0.45\textwidth]{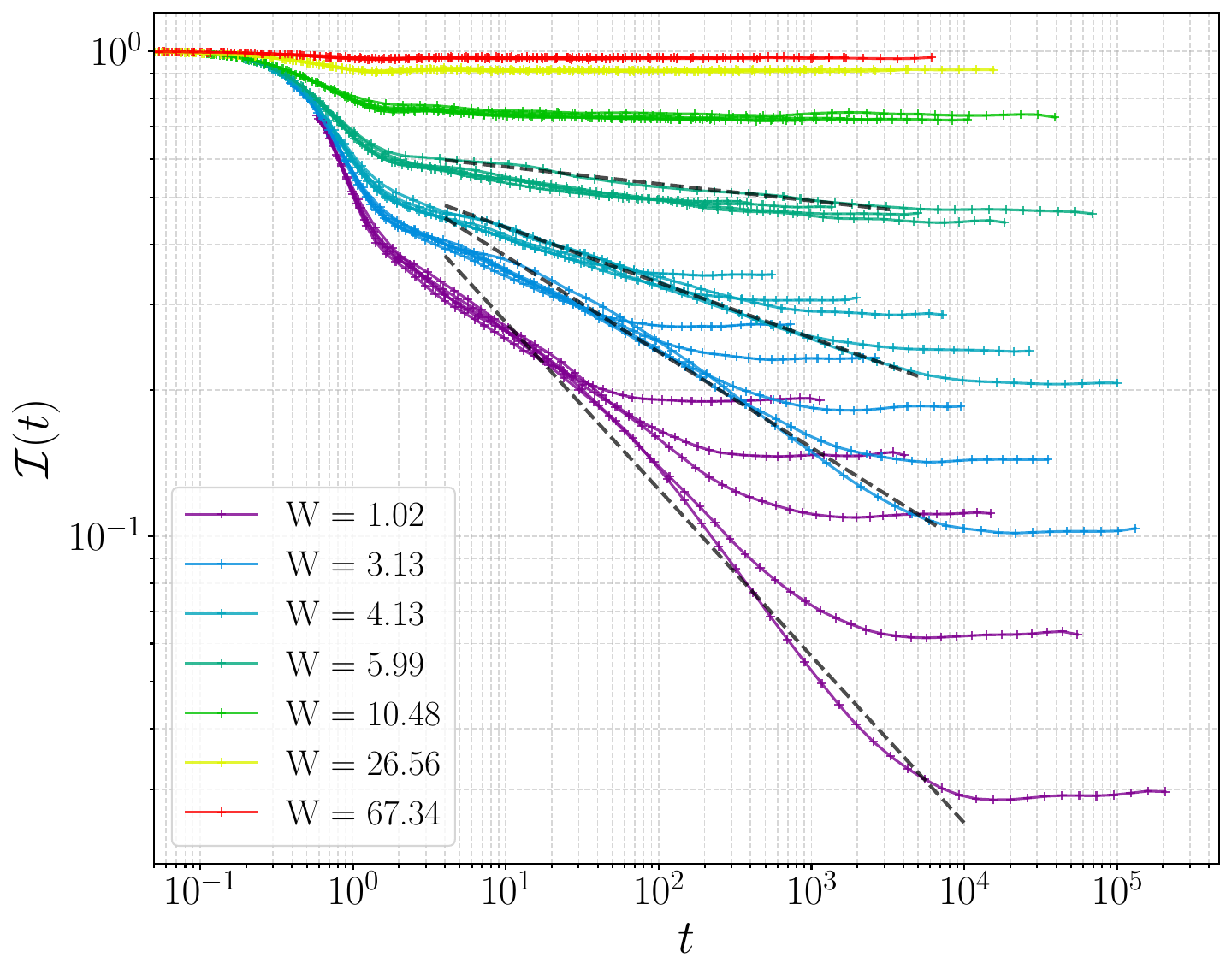}\hspace{0.5cm}
    \includegraphics[width=0.45\textwidth]{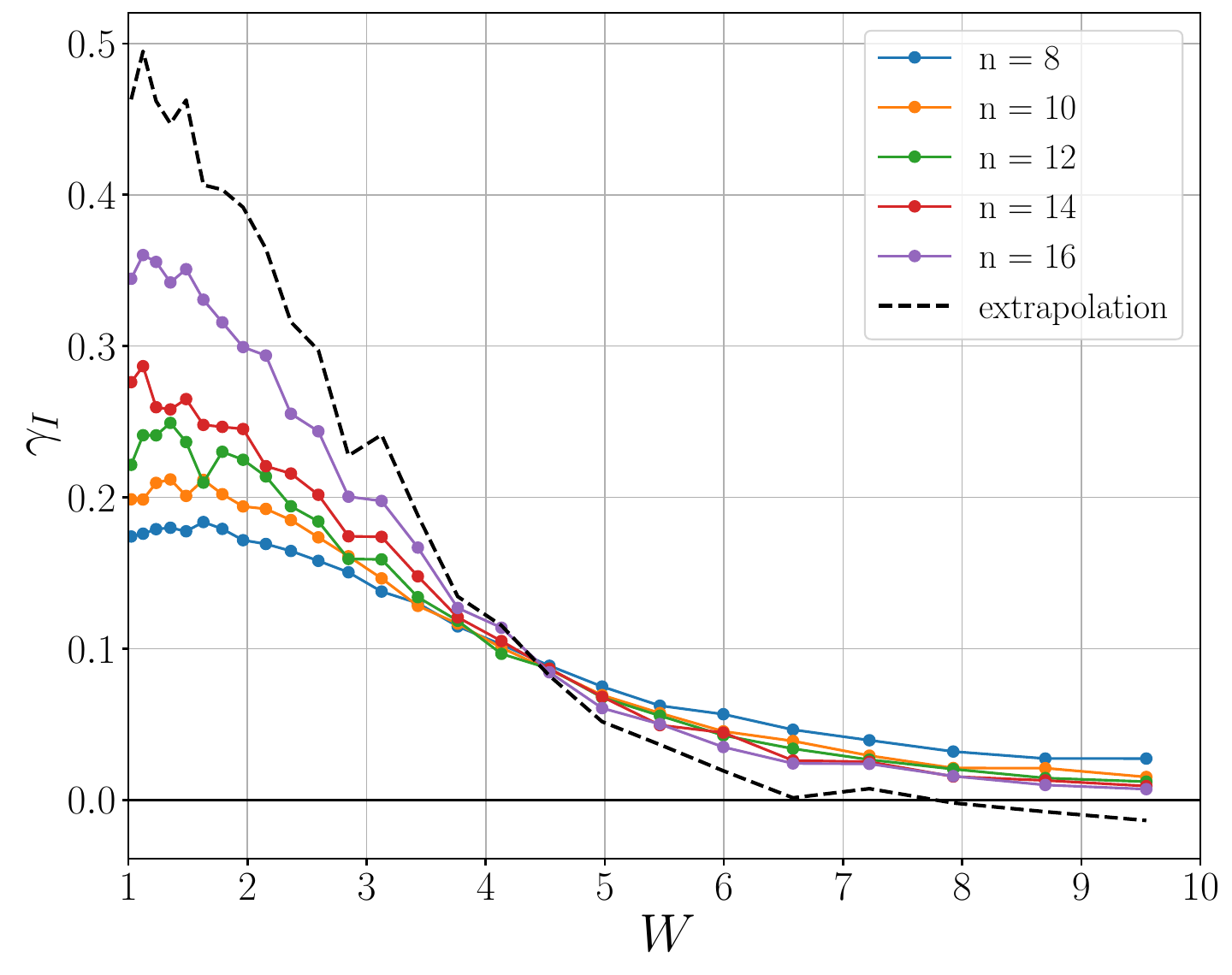}
    
    \caption{ Left: Dynamics of average imbalance in the 1D model for several values of disorder $W$ (see legend) and for system sizes $n=8$, 10, 12, 14, and 16.  For $n=16$, power-law fits to Eq.~\eqref{I-gammaI} are shown, which are used to obtain the exponent $\gamma_I$ shown in the right panel.  
    Right: Power-law exponent $\gamma_I$ for the 1D model as a function of the disorder strength $W$, as obtained by a fit to Eq.~\eqref{I-gammaI} in the time range from $t = 4$ to  $t = 0.1 \, t_H $. The result of extrapolation to $n\to \infty$ according to $\gamma_I(W,n) = \gamma_I(W,\infty) + c(W)/n $ is shown by a dashed line.
    }
    \label{fig:fits_gamma_1D}
\end{figure*}

\begin{figure*}[ht!]
    \centering
    \includegraphics[width=0.32\textwidth]{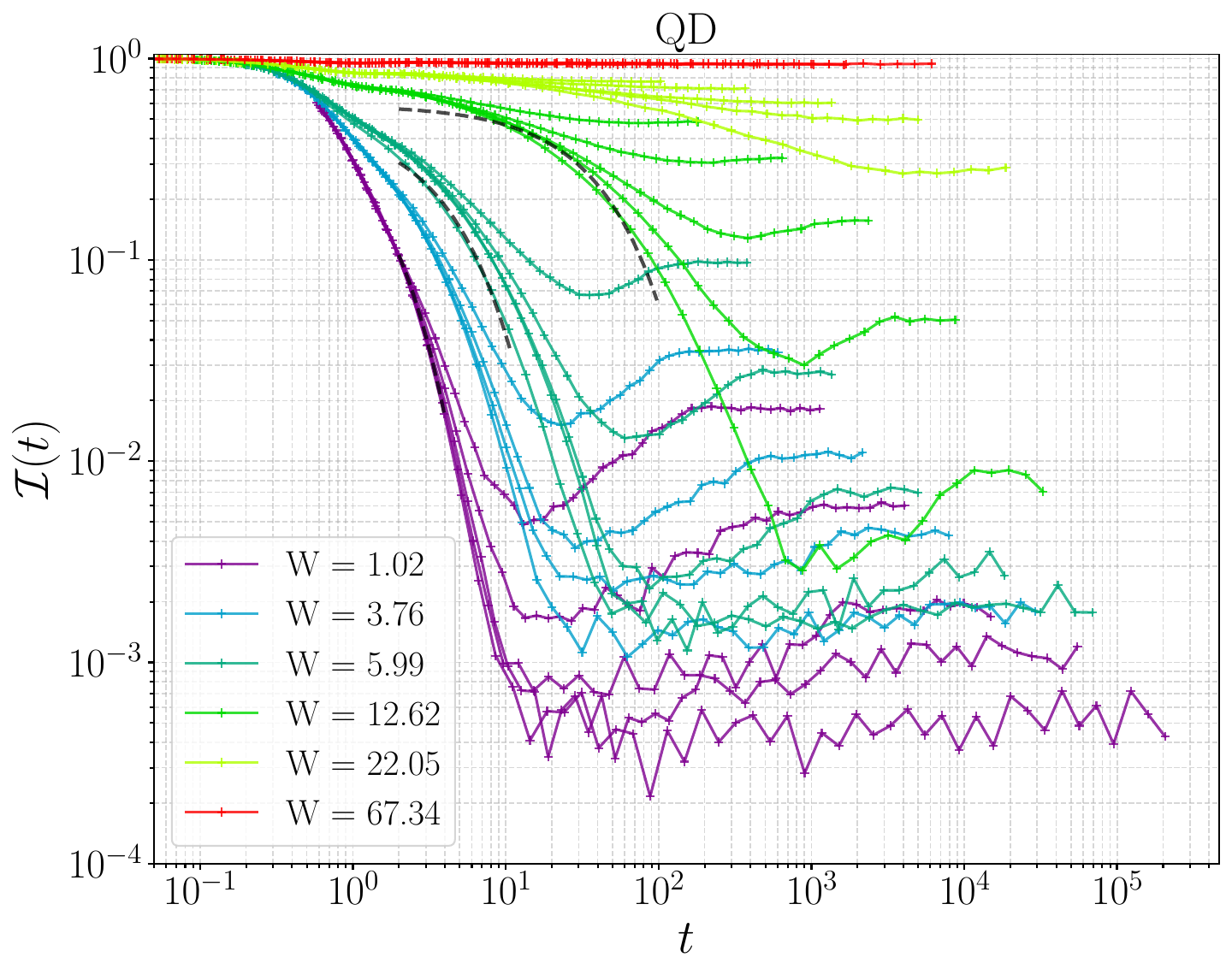}\hfill
    \includegraphics[width=0.32\textwidth]{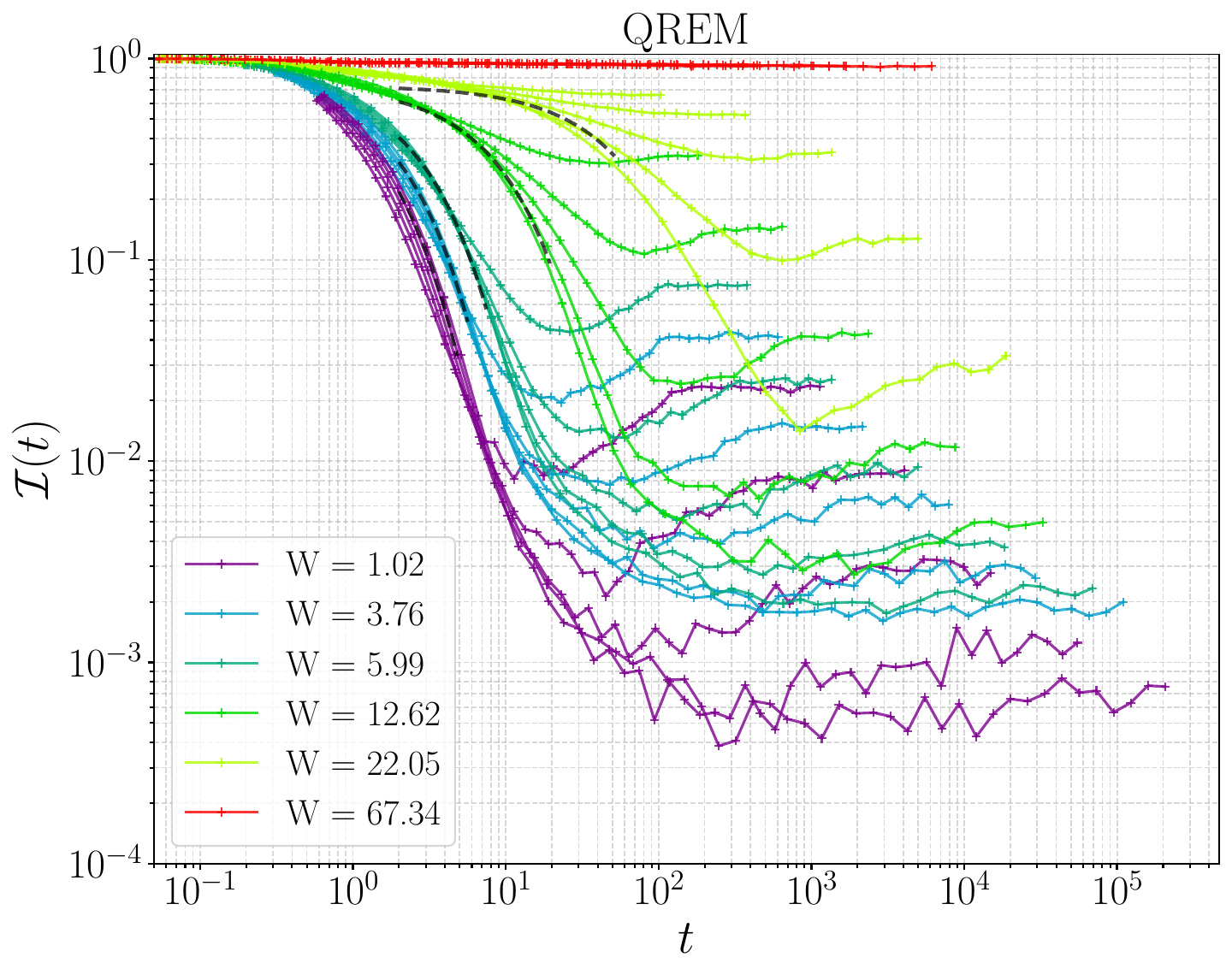}\hfill
    \includegraphics[width=0.324\textwidth]{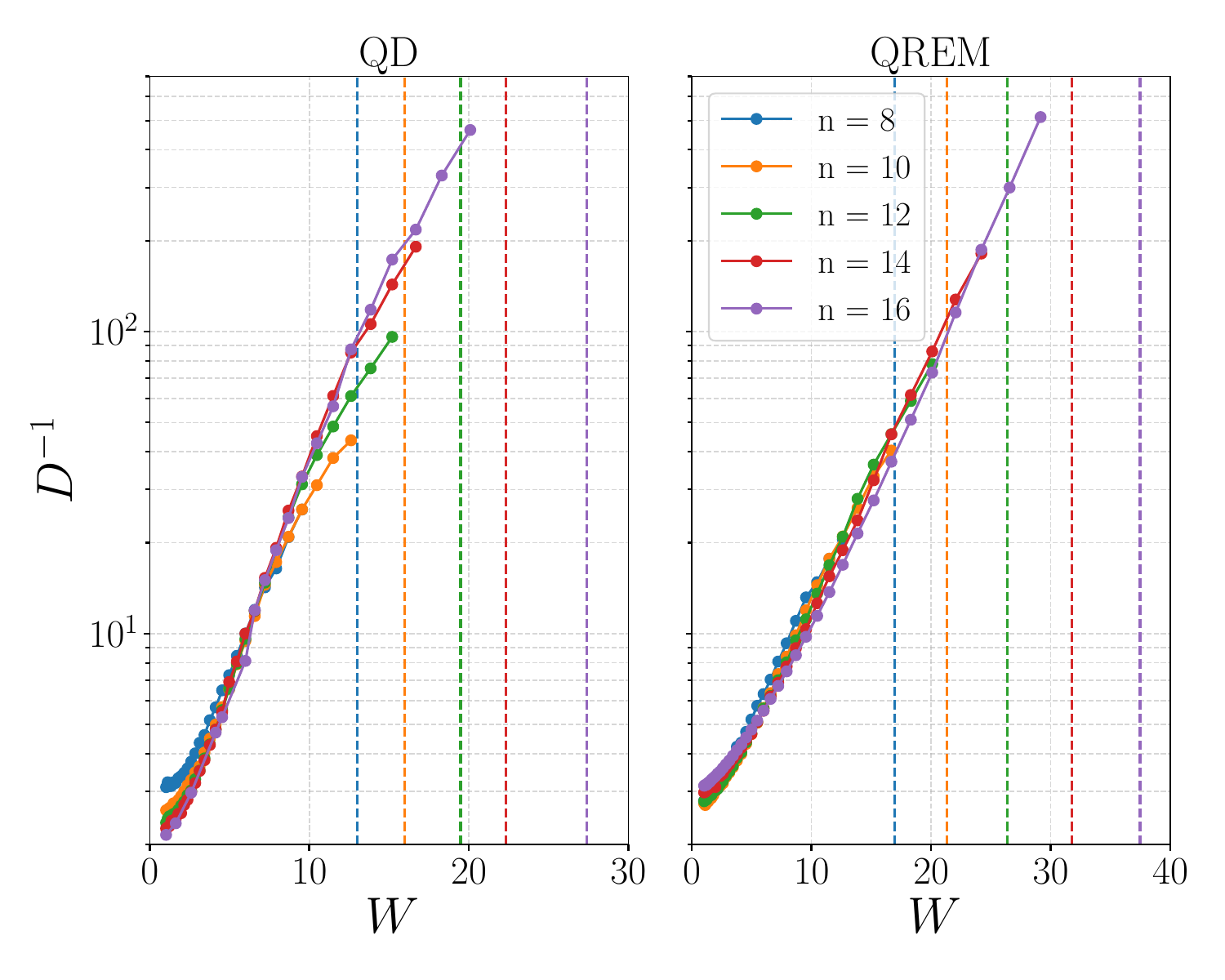}
    \caption{ Left two panels: Dynamics of average imbalance in the QD model and QREM for several values of disorder (see legend) and for system sizes $n=8$, $10$, $12$, $14$, and $16$.  For $n=16$, exponential fits to Eq.~\eqref{eq:QREM-imbalance-decay} are shown, which are used to obtain the diffusion coefficient $D$.  
    Right: Inverse of the diffusion coefficient $D$ for the QD model and QREM as functions of the disorder strength $W$, as obtained by a fit to Eq.~\eqref{eq:QREM-imbalance-decay} in the time range from $t=2$ to the time at which  
$\overline{\mathcal{I}(t)} = 4 \overline{\mathcal{I}}_\infty$.
    Vertical lines indicate the estimates of the critical disorder
    $W_c^{\rm QD}(n)$ and $W_c^{\rm QREM}(n)$  extracted from $\overline{\mathcal{I}}_\infty$.   }
    \label{fig:fits_gamma_QREM}
\end{figure*}

Extending the analysis of QREM and 1D model, we obtain the estimates 
\begin{equation}
\overline{\mathcal{I}}_\infty \simeq \frac{1 - n_{\rm res}^{(c)}/n}{2} \simeq \frac{1}{2} 
\label{eq:QD-transition-I_infty}
\end{equation}
for the average imbalance, and 
\begin{eqnarray}
v_q(t \to \infty) & \simeq & \left( \frac{1 - n_{\rm res}^{(c)}/n}{2}\right)^2 \simeq \frac{1}{4} \,, 
\label{eq:QD-transition-vq} \\
v_m(t \to \infty) & \simeq &  \frac{1}{4} \: \frac{2^{ - n_{\rm res}^{(c)}} }{n_{\rm res}^{(c)}} + \frac{n_{\rm res}^{(c)}}{n^2} 
\label{eq:QD-transition-vm}
\end{eqnarray}
for the fluctuations. 
Here,
we have taken into account that the mesoscopic fluctuations at the transition in QREM and in the 1D model are governed by different mechanisms. Specifically, for the QREM, these are fluctuations coming from the system-wide resonances, while for the 1D model, this is the same mechanism as in the localized phase.  In Eq.~\eqref{eq:QD-transition-vm}, we have summed contributions of both types for the QD model. In the asymptotic limit of very large $n$, the second term will win, since the first one gets exponentially small. At the same time, since $n_{\rm res}^{(c)}$ grows very slowly with $n$, the first term is important, and is in fact larger, for realistic values of $n$. 


\section{Numerical results}
\label{sec:numerics}

 \begin{figure*}[ht!]
    \centering    \includegraphics[width=0.315\textwidth]{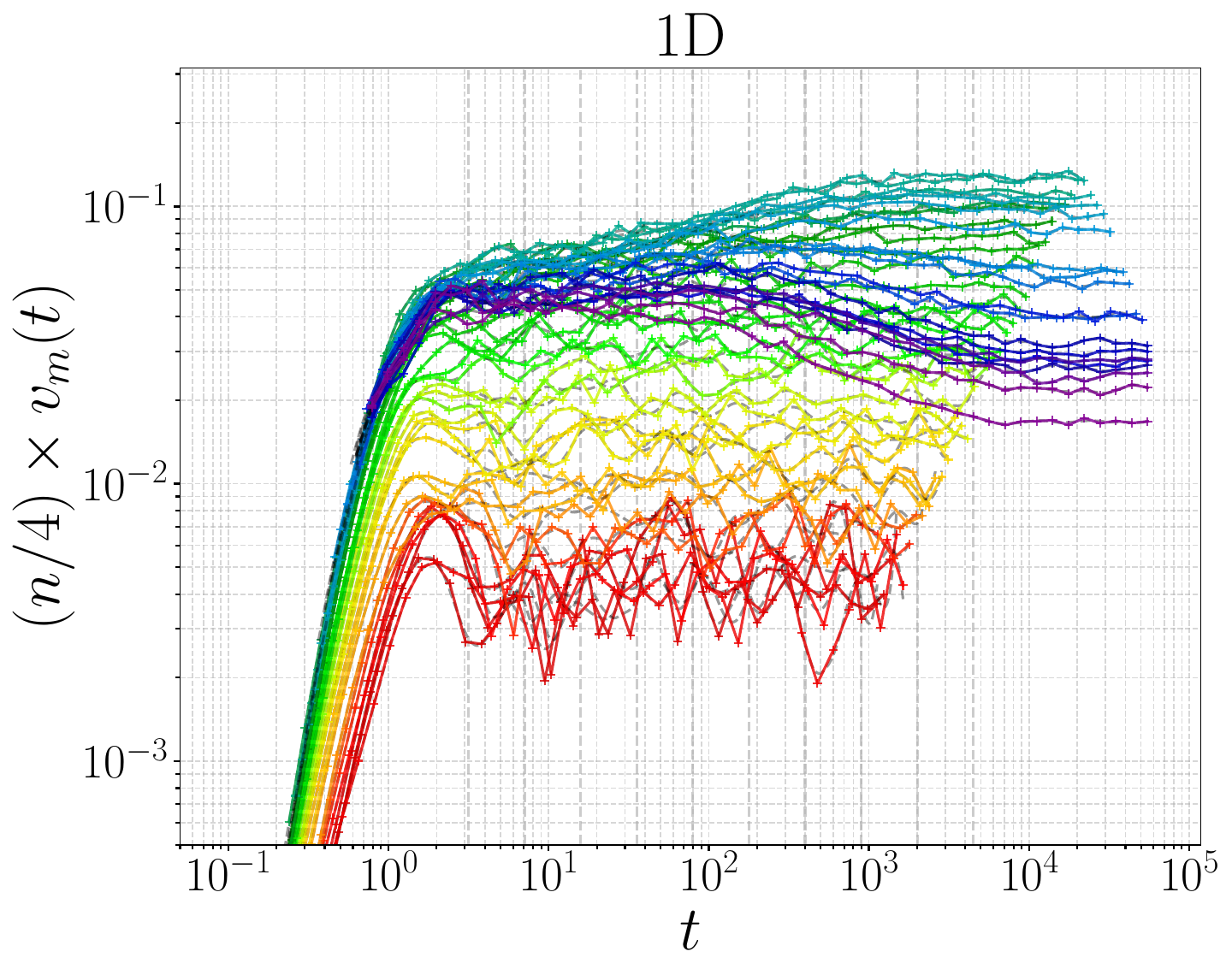}\hfill
    \includegraphics[width=0.315\textwidth]{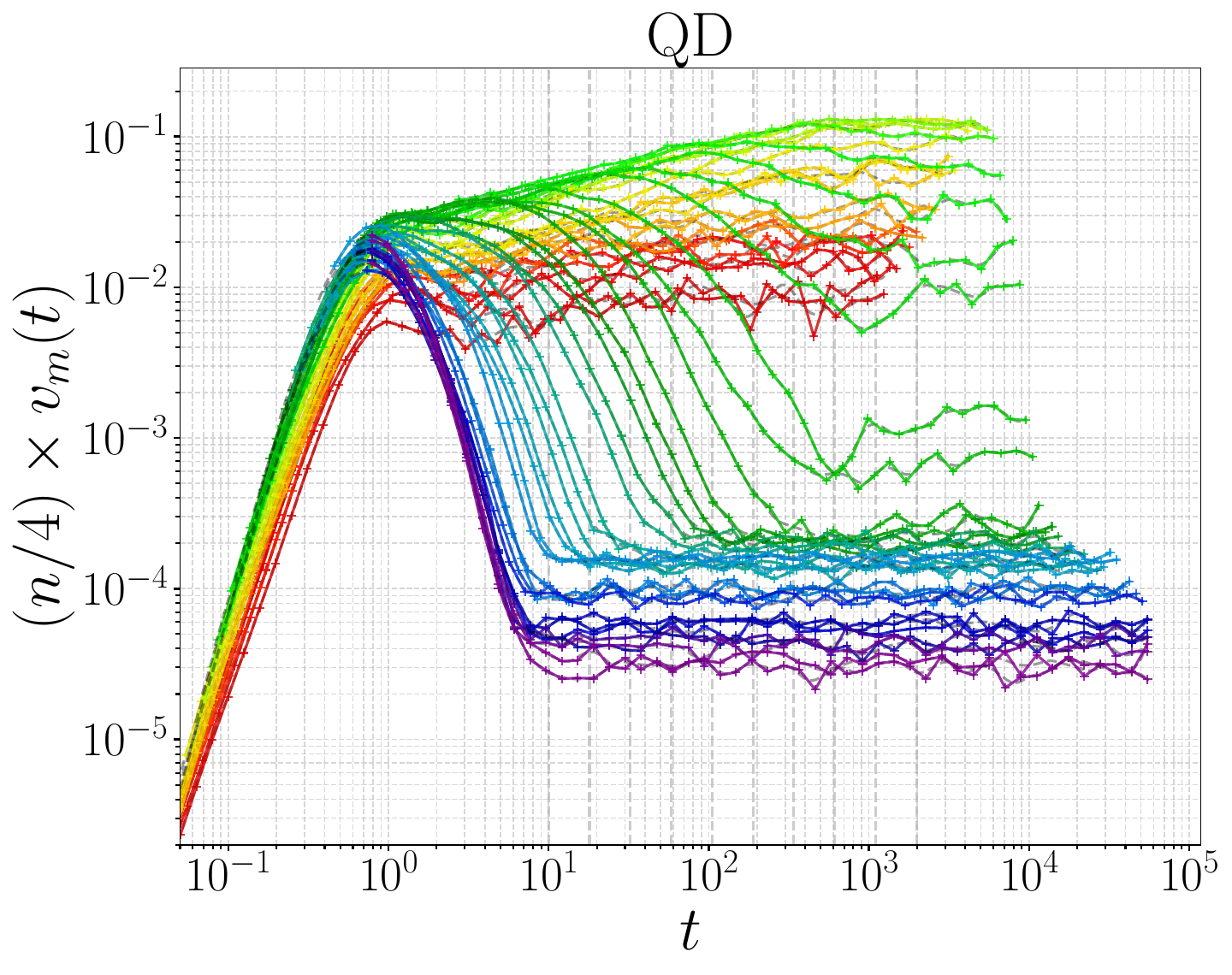}\hfill
    \includegraphics[width=0.35\textwidth]{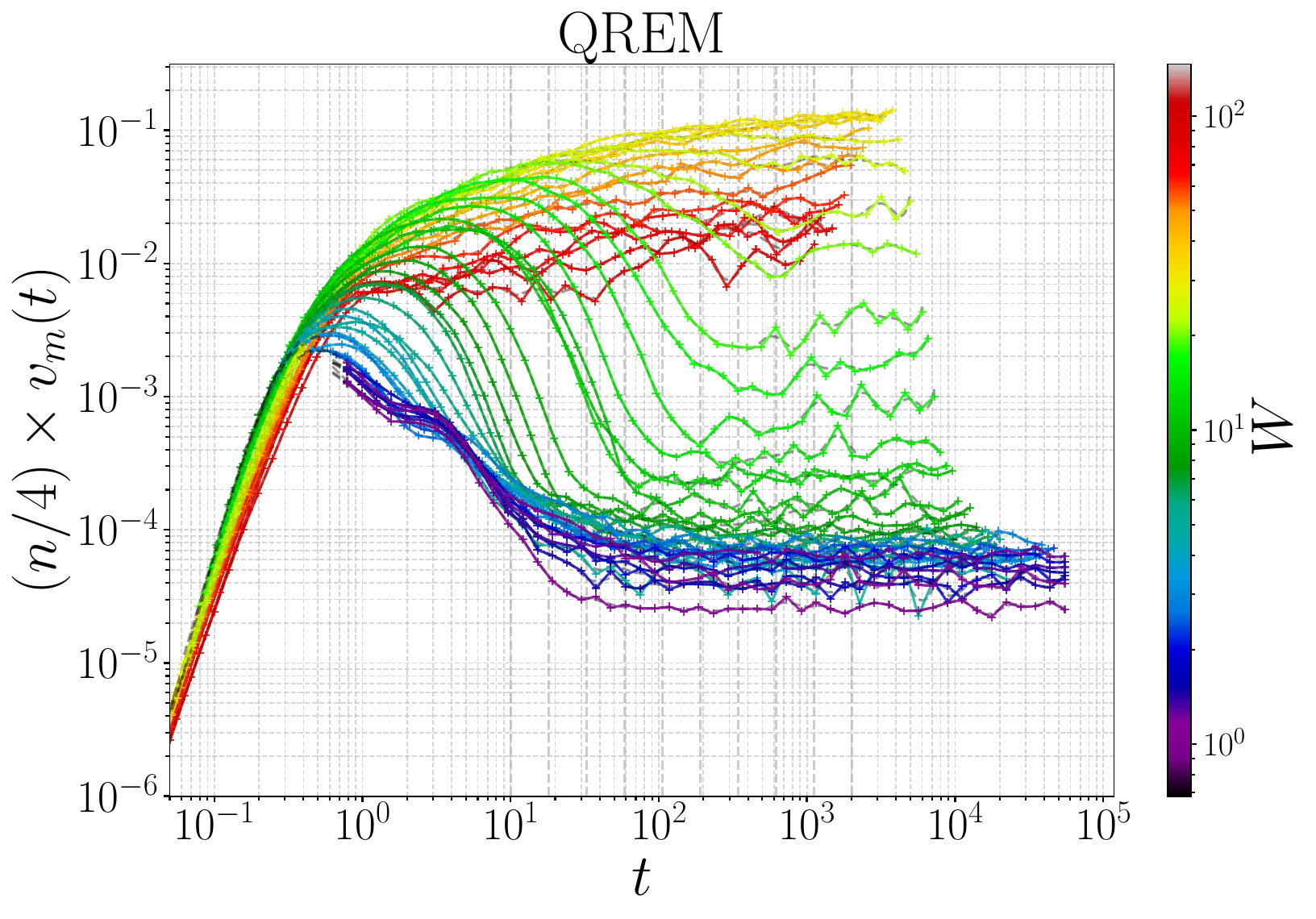}
    \caption{ Time evolution of the mesoscopic variance $v_m(t)$ for the 1D model (left), QD model (center) and QREM (right) with $n=14$ spins. The disorder ranges from $W\simeq 1$ to $W\simeq 100$ for all models.      The maximum times exceed the estimated Heisenberg time $t_H$, Eq.~\eqref{eq:tH}, for all values of $W$.  Vertical dashed lines mark times $t$, data for which are shown in Fig.~\ref{fig:cuts_meso_variance_1D_QD_QREM}.
    }
    \label{fig:time-evol_meso_variance_1D_QD_QREM}
\end{figure*}

 \begin{figure*}[ht!]
    \centering
    \includegraphics[width=0.33\textwidth]{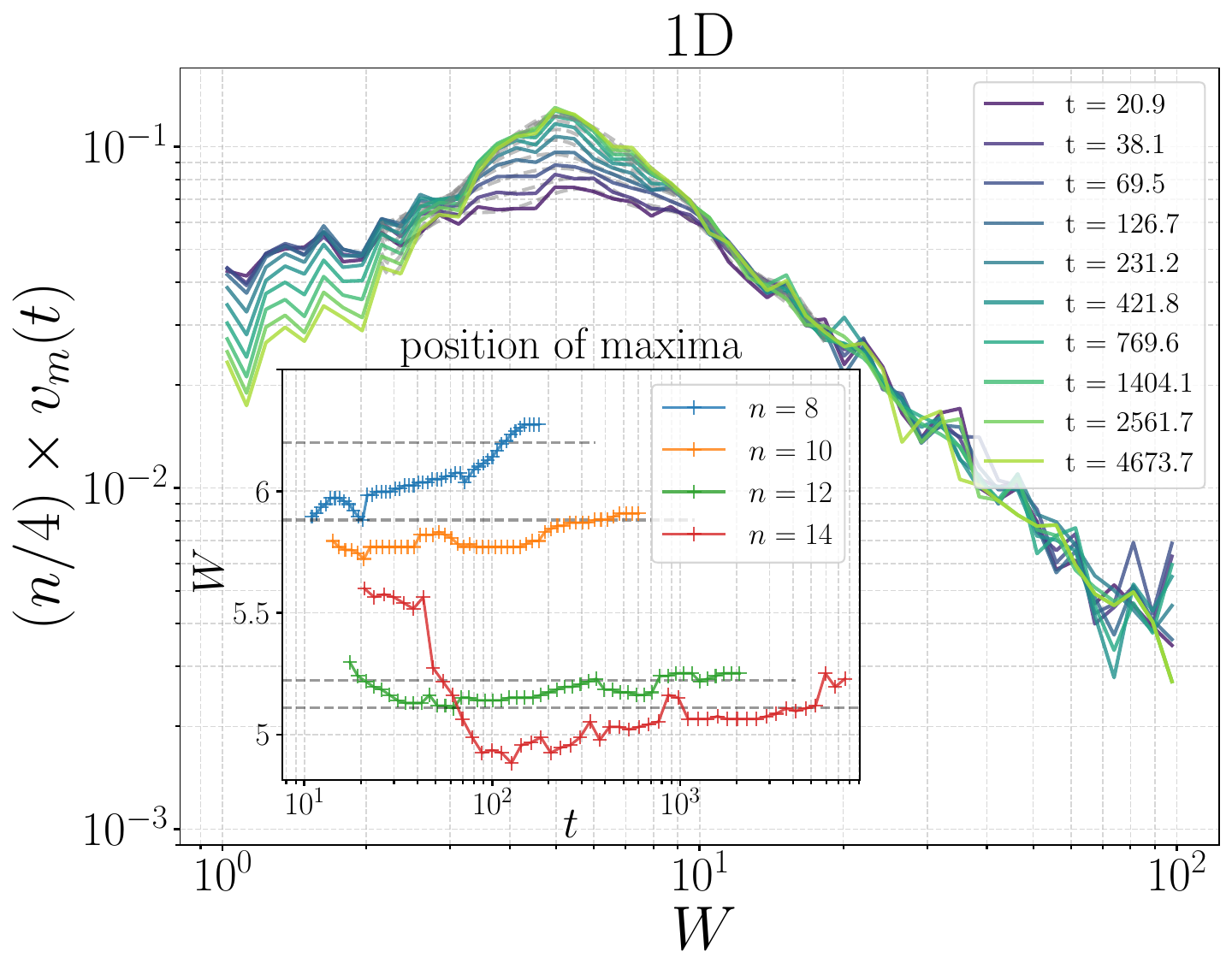}\hfill
    \includegraphics[width=0.33\textwidth]{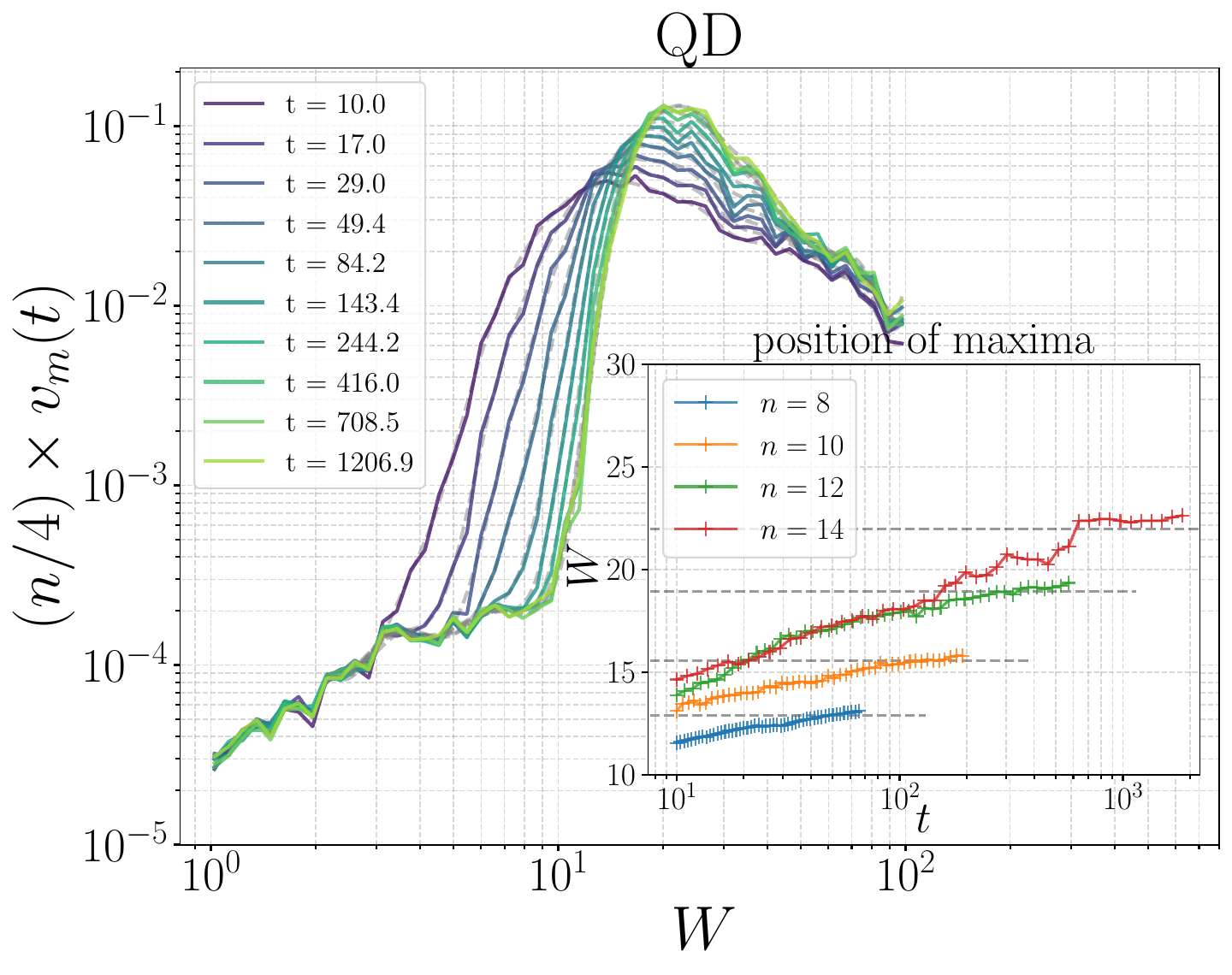}\hfill
    \includegraphics[width=0.33\textwidth]{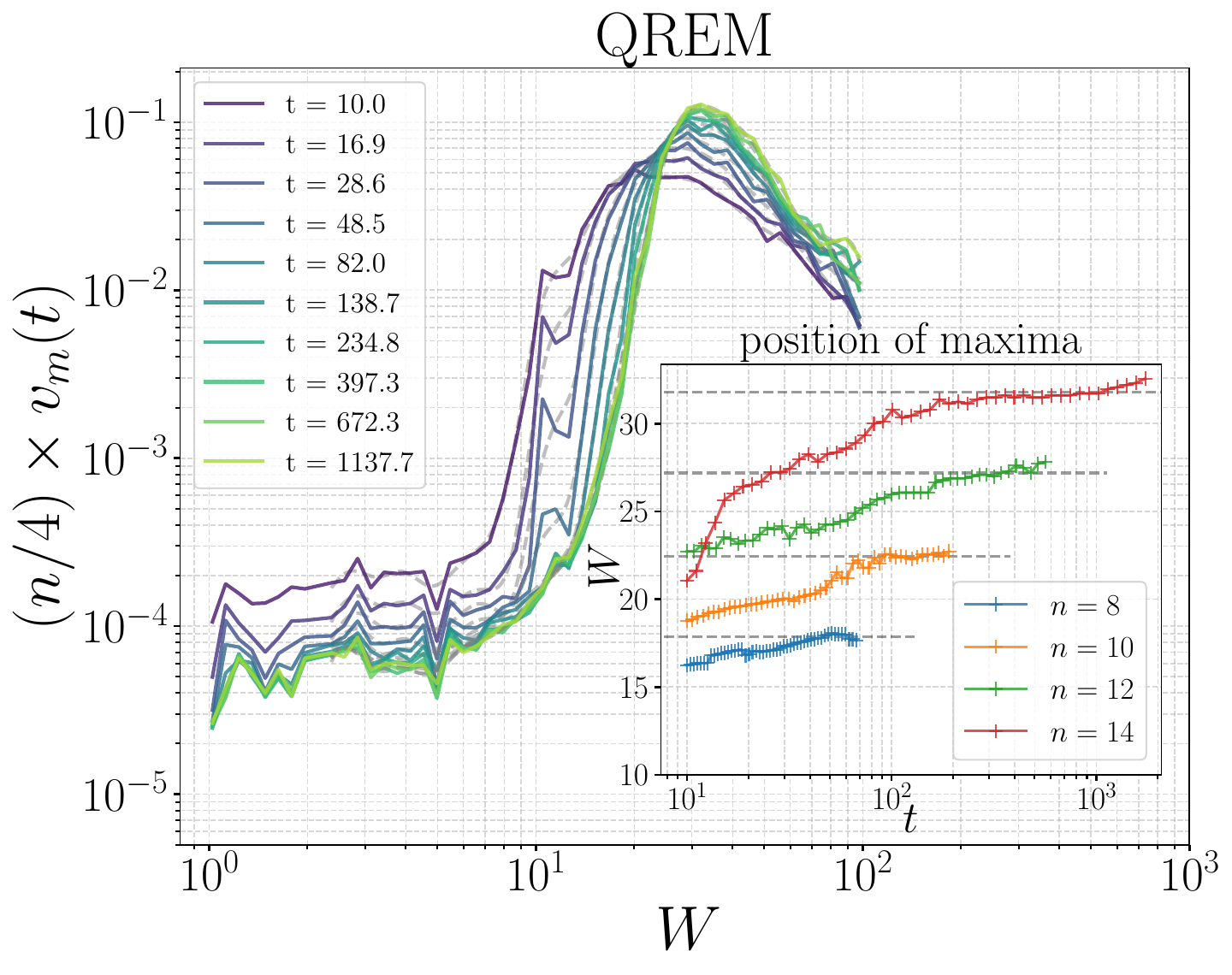}
    \caption{Disorder dependence of the mesoscopic variance $v_m(t)$ for the 1D model (left), QD model (center) and QREM (right) with $n=14$ spins for selected values of time $t$.  
    This is an alternative representation of the data  shown in Fig.~\ref{fig:time-evol_meso_variance_1D_QD_QREM}. 
    Insets show how the disorder $W$ providing a maximum to $v_m(t)$ develops with time for different system sizes $n$.
     }
\label{fig:cuts_meso_variance_1D_QD_QREM}
\end{figure*}

In Sec.~\ref{sec:analytics}, we have presented analytical results for the imbalance dynamics $\overline{\mathcal{I}(t)}$ and for the long-time limiting values of the average imbalance,
$\overline{\mathcal{I}}_\infty$, and its quantum and mesoscopic variances,
$v_q(t \to \infty)$ and $v_m(t \to \infty)$, for QREM, 1D model, and QD model. These results show that, for all these models, the imbalance and its fluctuations provide indicators of the MBL transitions. Specifically, the asymptotic imbalance at the transition   is $\overline{\mathcal{I}}_\infty \simeq 1/2$ for the QREM and QD model, and $\overline{\mathcal{I}}_\infty \simeq (1-p_c)/2$ for the 1D model, while the quantum and mesoscopic variances $v_q(t \to \infty)$ and $v_m(t \to \infty)$ are predicted to exhibit maxima in the transition region for all the models.

In this Section, we present detailed computational study of the imbalance and its fluctuations 
(see Appendix \ref{app:numerics_details} for details of methodology)
and confront the results with analytical predictions. In Sec.~\ref{sec:transition}, we will use numerical results to determine $W_c(n)$ for MBL transitions in the models under consideration. 

\subsection{Dynamics of the imbalance}
\label{subsec:dynamics_imbalance_numerics}

\subsubsection{Average imbalance}
\label{subsubsec:time-evol_average_numerics}

We have already showed the time dependence of average imbalance for all three models with $n=14$ in Fig.~\ref{fig:time-evol1D_QD_QREM} and 
discussed it in Sec.~\ref{sec:imbalance-numerics} above.
We inspect now the dynamics of average imbalance $\overline{\mathcal{I}(t)}$ in more detail.

In the left panel of 
 Fig.~\ref{fig:fits_gamma_1D}, we show 
 $\overline{\mathcal{I}(t)}$ in the 1D model, for different system sizes, and for several values of disorder. For relatively weak disorder, a power-law decay of the imbalance is clearly observed, which proceeds up to a time of order of Heisenberg time $t_H$, so that the saturation value drops exponentially 
 with $n$. These findings are in full agreement with analytical arguments in Sec.~\ref{sec:1D-analytics-ergodic}, see
Eq.~\eqref{I-gammaI} and
Eq.~\eqref{eq:1D-I-infty-ergodic}. 
At the same time, for strong disorder, the behavior of imbalance $\overline{\mathcal{I}(t)}$ is independent of $n$, with a saturation value close to unity, in agreement with predictions for the MBL phase, Sec.~\ref{sec:analytics-1D-loc}.

\begin{figure*}[ht!]
    \centering
    \includegraphics[width=0.315\textwidth]{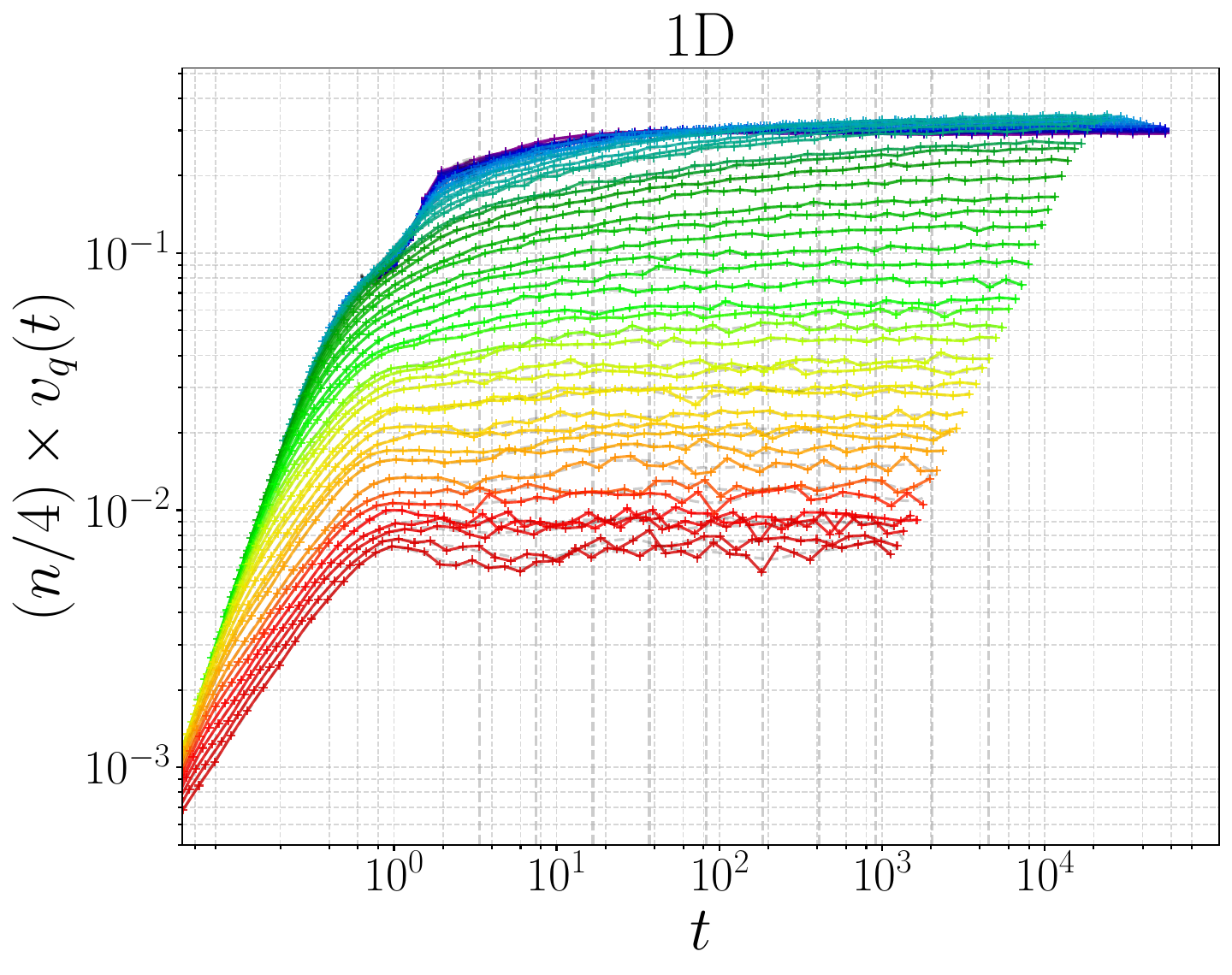}\hfill
    \includegraphics[width=0.315\textwidth]{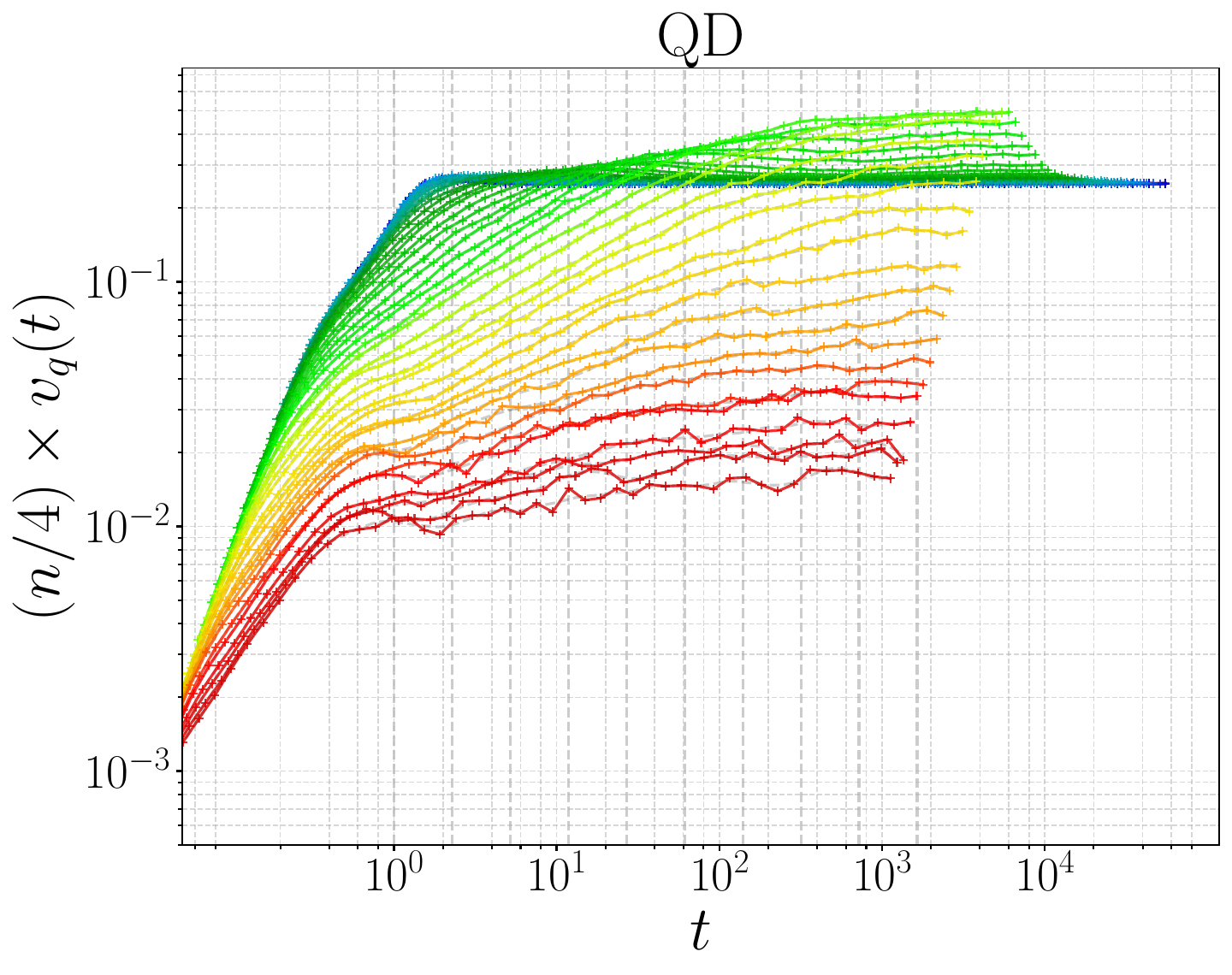}\hfill
    \includegraphics[width=0.35\textwidth]{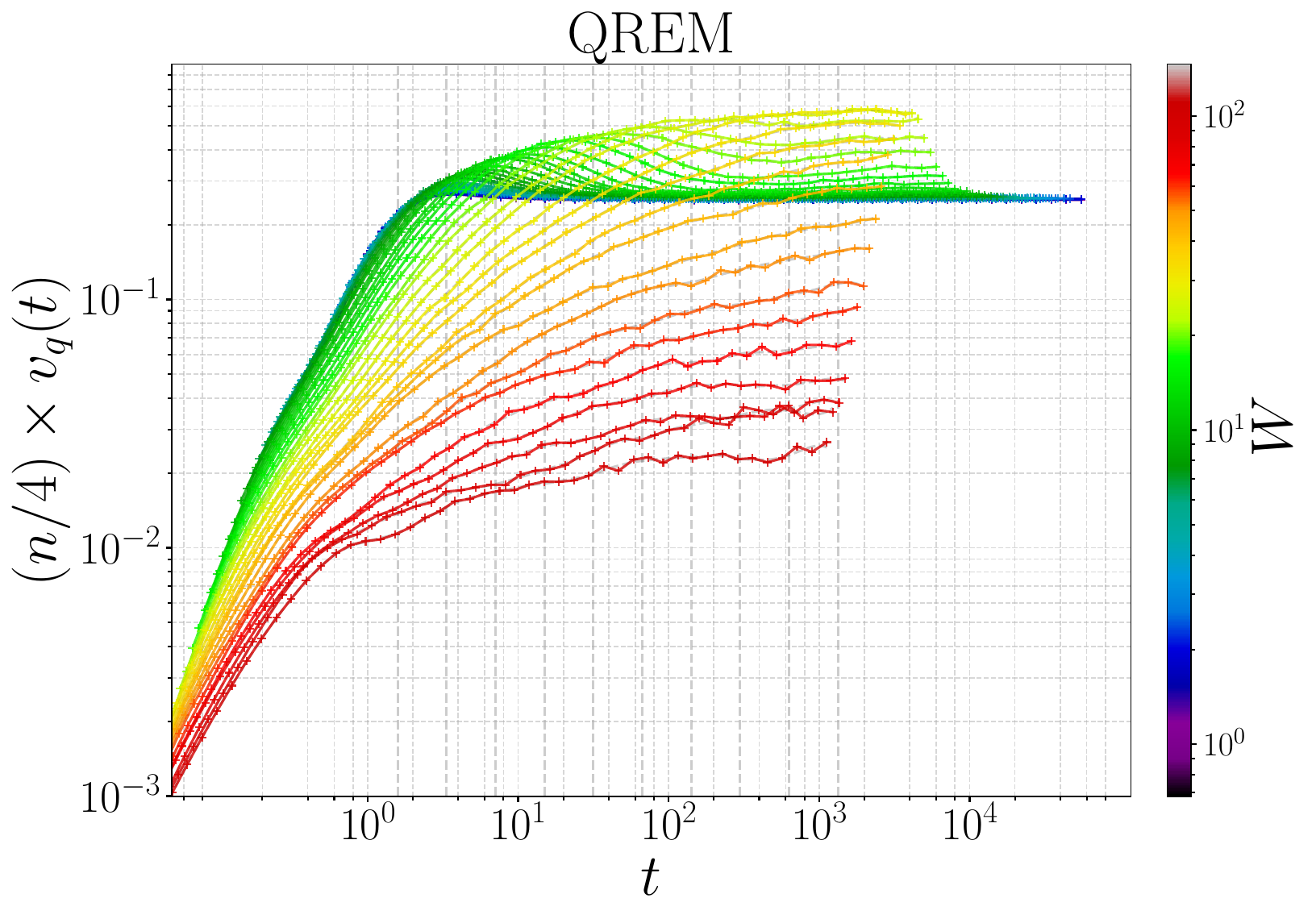}
    \caption{
    Time evolution of the quantum variance $v_q(t)$ for the 1D model (left), QD model (center) and QREM (right) with $n=14$ spins. The disorder ranges from $W\simeq 1$ to $W\simeq 100$ for all models.      The maximum times exceed the estimated Heisenberg time $t_H$, Eq.~\eqref{eq:tH}, for all values of $W$.  Vertical dashed lines mark times $t$, data for which are shown in Fig.~\ref{fig:cuts_quantum_variance_1D_QD_QREM}.
    }
    \label{fig:time-evol_quantum_variance_1D_QD_QREM}
\end{figure*}
\begin{figure*}[ht!]
    \centering
    \includegraphics[width=0.33\textwidth]{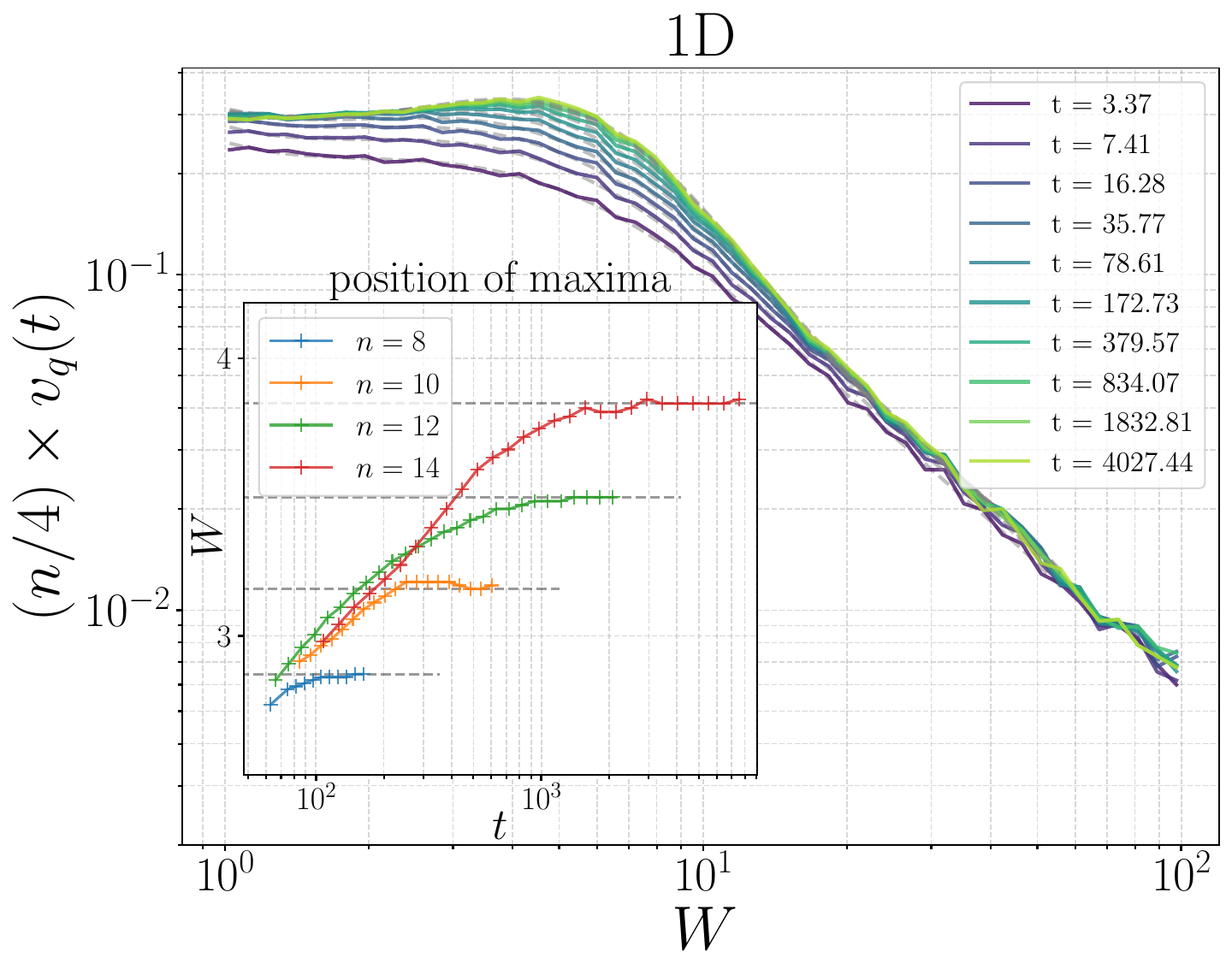}\hfill
    \includegraphics[width=0.33\textwidth]{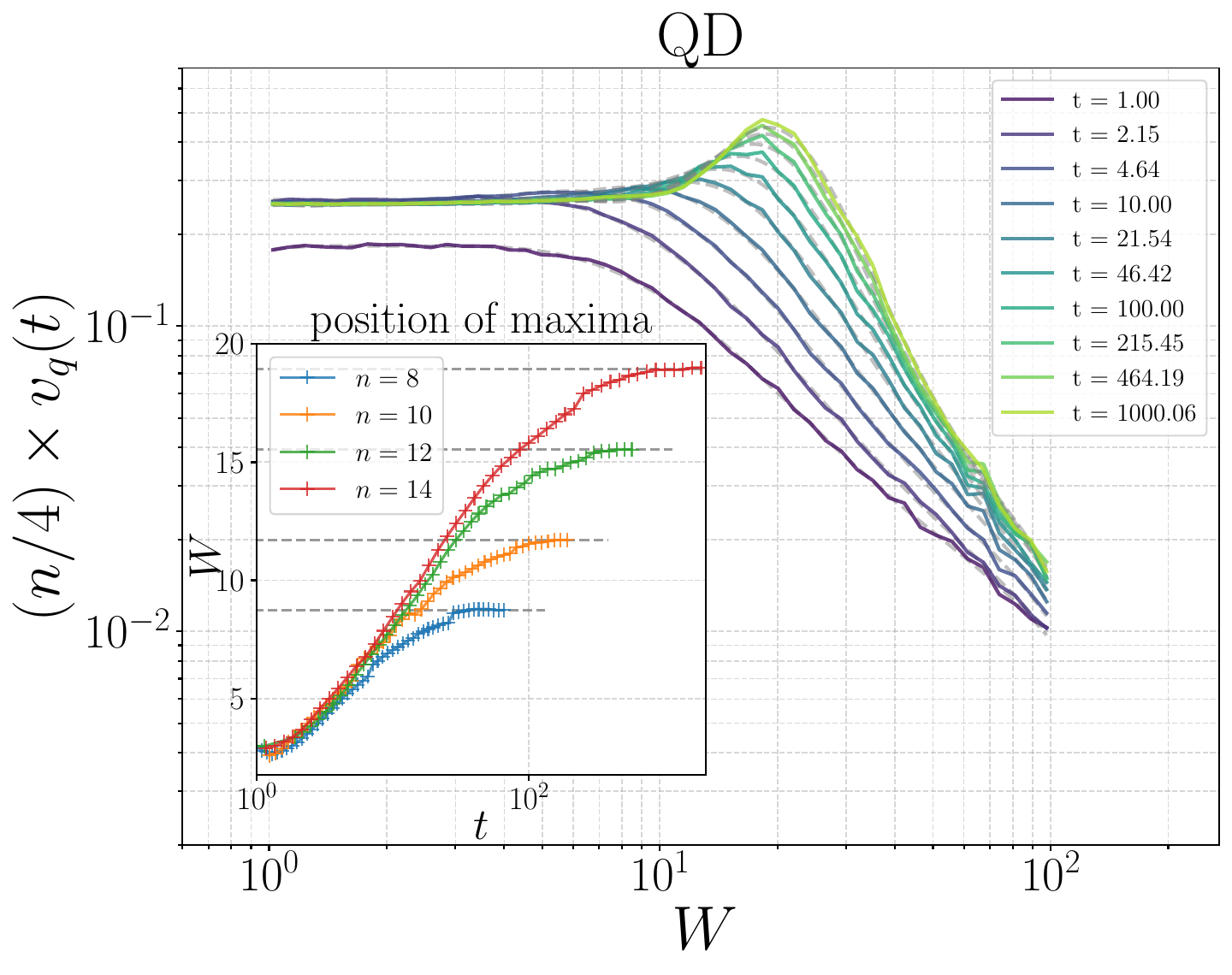}\hfill
    \includegraphics[width=0.33\textwidth]{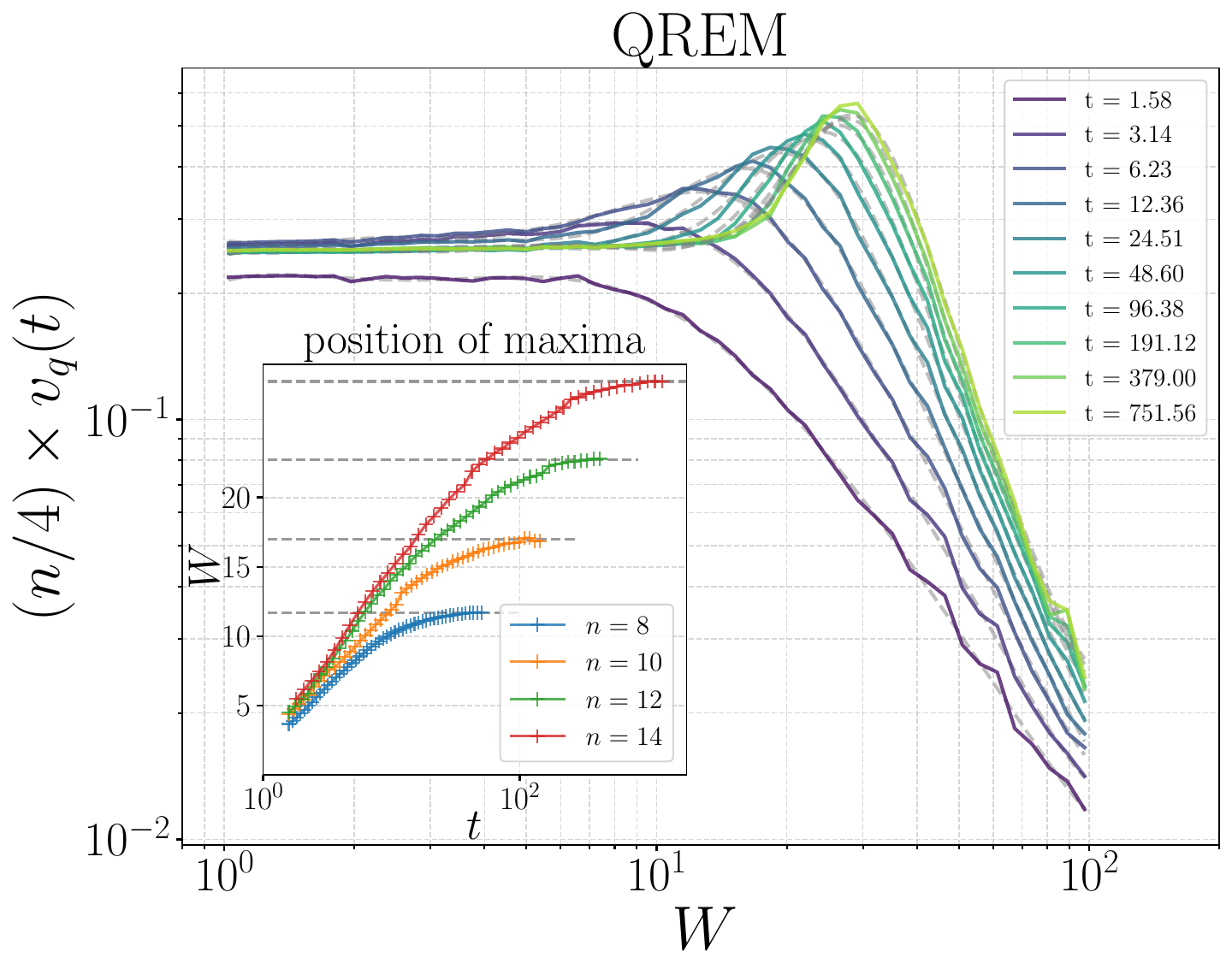}
    \caption{
    Disorder dependence of the quantum variance $v_q(t)$ for the 1D model (left), QD model (center) and QREM (right) with $n=14$ spins.  
    This is an alternative representation of the data  shown in Fig.~\ref{fig:time-evol_meso_variance_1D_QD_QREM}.  
    Insets show how the disorder $W$ providing a maximum to $v_q(t)$ develops with time for different system sizes $n$.}  \label{fig:cuts_quantum_variance_1D_QD_QREM}
\end{figure*}

In the right panel of 
 Fig.~\ref{fig:fits_gamma_1D}, we present the exponent $\gamma_I$ obtained from a power-law fit of $\overline{\mathcal{I}(t)}$ in the range of time from $t = 4$ to  $t = 0.1 \, t_H $. For $W < 4$, the fitted exponent has a tendency to increase with increasing $n$. At the same time, for $W > 6$ the exponent decreases with increasing $n$ as $\gamma_I\propto 1/n$.  This implies that $\gamma_I = 0$ in the $n \to \infty$ limit in this range of disorder. The vanishing of $\gamma_I$ reflects the long-time saturation of the imbalance at an $n$-independent value for these values of $W$, which is a manifestation of the MBL phase. Apparent non-zero values $\gamma_I\propto 1/n$ result from some dynamics of the imbalance at relatively short times, which are included in the fit window. The dependence $\gamma_I(W)$ in the range from $W\approx 2$ to $W \approx 6$ becomes steeper with increasing $n$.
Extrapolating to $n \to \infty$ via 
$\gamma_I(W,n) = \gamma_I(W,\infty) + c(W)/n $ yields $\gamma_I(W,\infty) = 0$ for $W \ge 6.5$, thus providing an estimate of the critical disorder $W_c^{\rm 1D}(n \to \infty) \approx 6.5 $ in the thermodynamic limit. We will see below that various indicators point to approximately the same value of the transition point at large $n$.

 In the left two panels of 
 Fig.~\ref{fig:fits_gamma_QREM}, the dynamics of average imbalance  
 $\overline{\mathcal{I}(t)}$ for QD model and QREM is shown for different $n$ and several disorder values.  We have already seen in Fig.~\ref{fig:time-evol1D_QD_QREM} a fast decay of $\overline{\mathcal{I}(t)}$ in the ergodic phases of these models. It is in full consistency with 
 analytical arguments in Sec.~\ref{sec:analytics-QREM-erg} and \ref{sec:analytics-QD-ergodic} that predict
 the exponential decay law \eqref{eq:QREM-imbalance-decay}, with the decay rate determined by the diffusion constant $D$. 
 In the right two panels of 
 Fig.~\ref{fig:fits_gamma_QREM}, we show the disorder dependence of the diffusion constant $D$ for both models and for various system sizes. The results are in a good agreement with analytically predicted exponential dependence, $\ln D^{-1} \propto W$, see Eq.~\eqref{eq:QREM_D_precritical}, in the range of system sizes $n$ accessible to exact numerical simulations.  

\subsubsection{Imbalance fluctuations}
\label{subsubsec:time-evol_variance_numerics}

In Fig.~\ref{fig:time-evol_meso_variance_1D_QD_QREM}, the mesoscopic variance $v_m(t)$ as a function of time $t$ is shown for 1D, QD, and QREM models with $n=14$ spins, for various disorder strength.  We use in this figure the same color code for disorder as in Fig.~\ref{fig:time-evol1D_QD_QREM} for the dynamics of the average imbalance. It is seen that, after an initial increase at short times, the variance $v_m(t)$ develops a strong dependence on disorder $W$, with a maximum at an intermediate disorder. This is in full consistency with analytical results in Sec.~\ref{sec:analytics} that predict that, in the long-time limit,  $v_m(t)$ and $v_q(t)$ have maxima at the MBL transition, $W \simeq W_c(n)$.  

\begin{figure*}[ht!]
    \centering
    \includegraphics[width=0.33\textwidth]{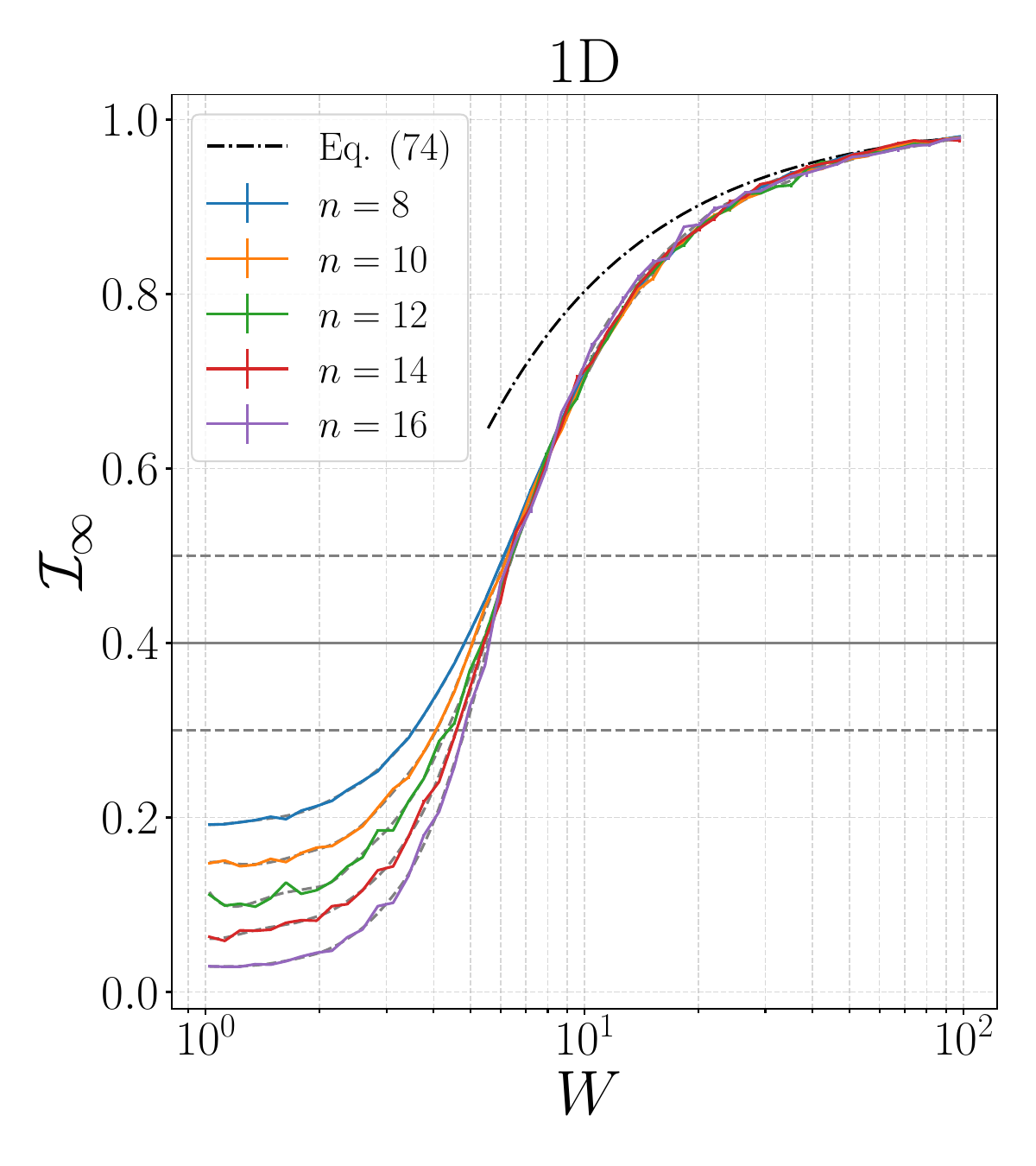}\hfill
    \includegraphics[width=0.33\textwidth]{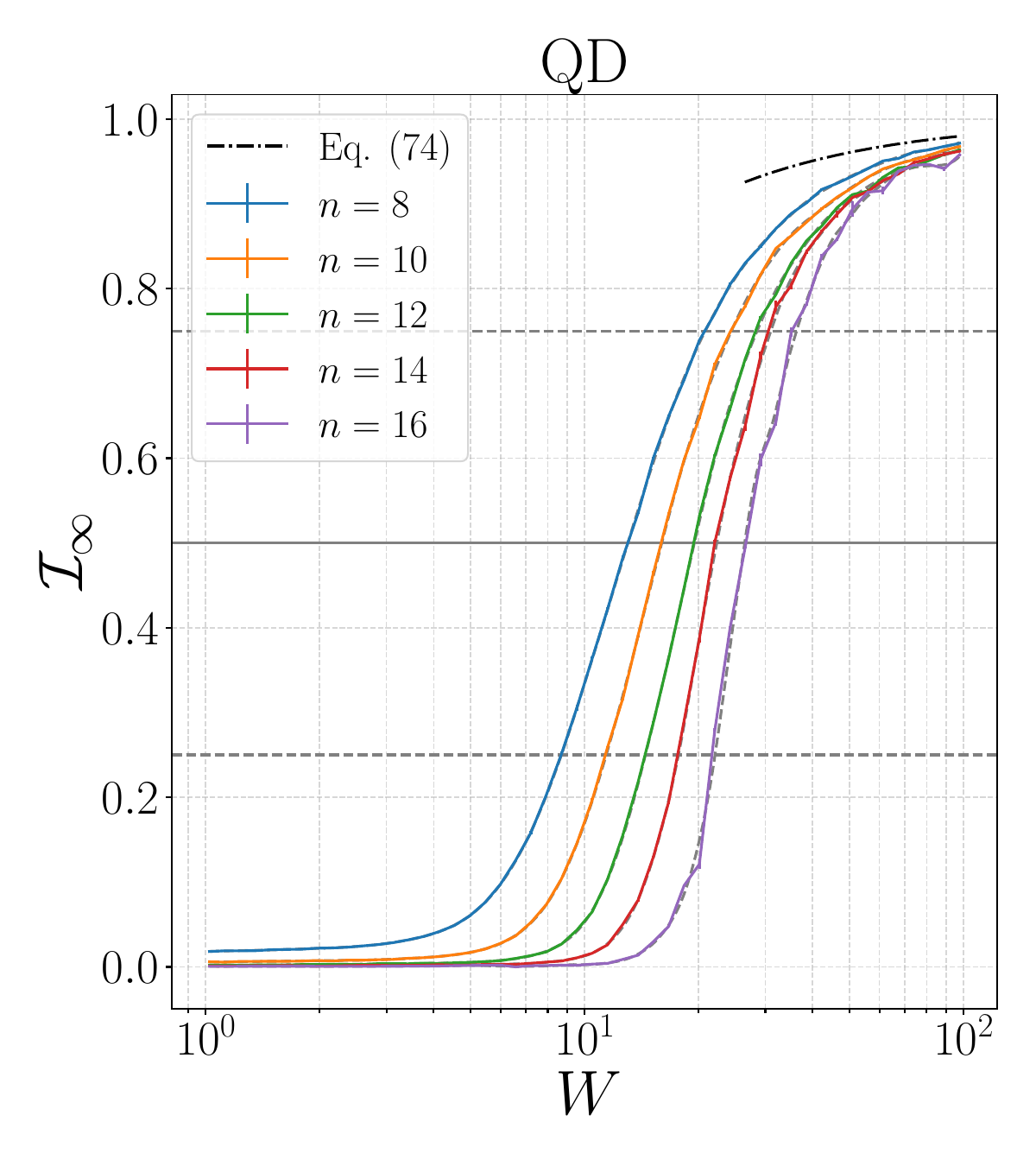}
    \hfill
    \includegraphics[width=0.33\textwidth]{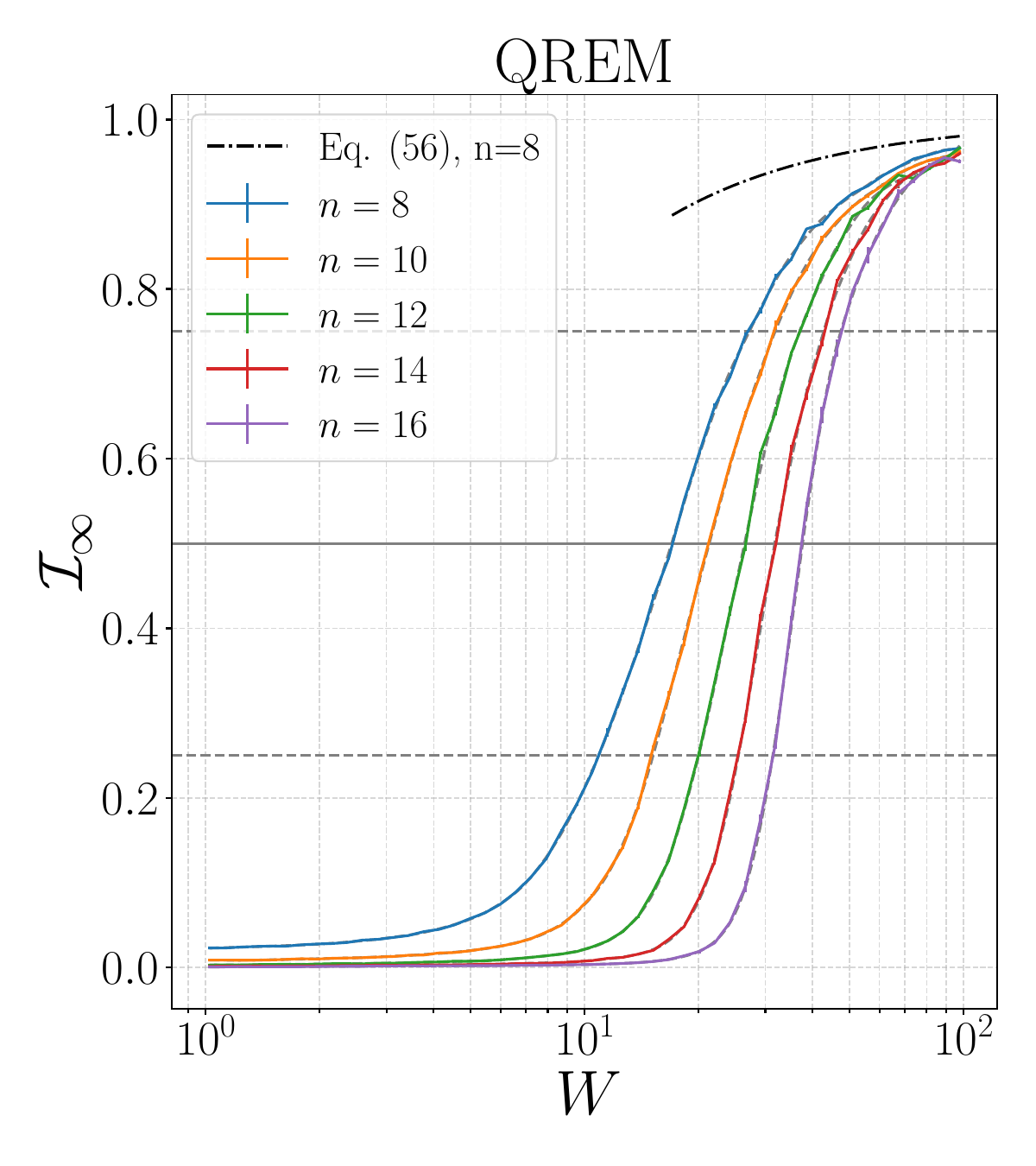}
    \caption{ Average imbalance at infinite time, $\overline{\mathcal{I}}_\infty$
    for system sizes from $n=8$ to $n=16$, extracted from the time evolution of the 1D model (left), QD model (center), and QREM (right).  The large-$W$ asymptotics of $\overline{\mathcal{I}}_\infty$, Eqs.~\eqref{1D-Iinfty} for 1D and QD models and 
Eq.~\eqref{eq:QREM-loc-average-imb} for QREM are shown by dashed line.  Horizontal full and dashed lines serve to extract the positions of the MBL transition, $W_c(n)$ and the transition widths $\Delta W / W_c$, see text. 
}   \label{fig:Infinite_time_imbalance_1D_QD_QREM}
\end{figure*}

\begin{figure*}[ht!]
    \centering
    \includegraphics[width=0.33\textwidth]{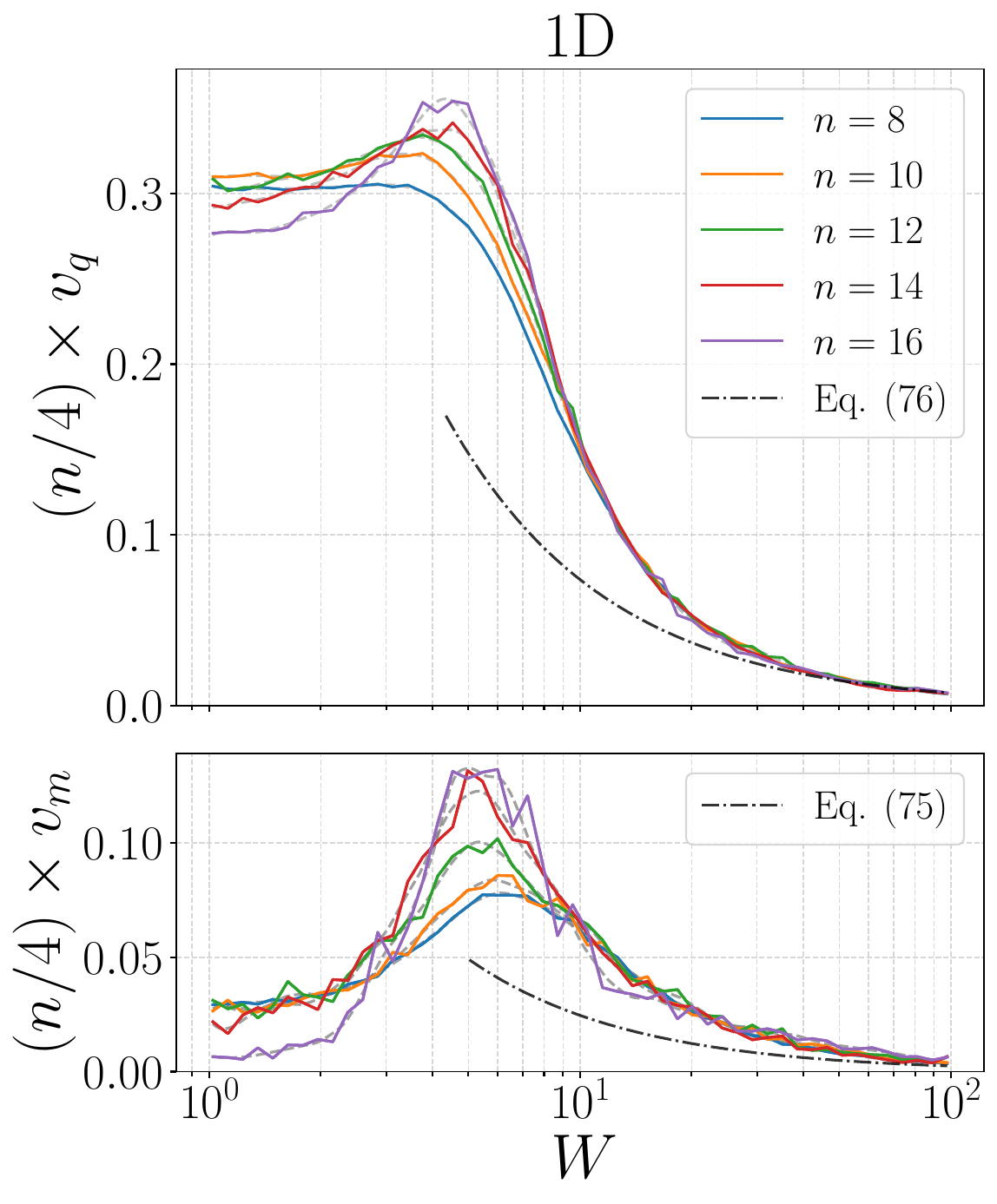}\hfill
    \includegraphics[width=0.33\textwidth]{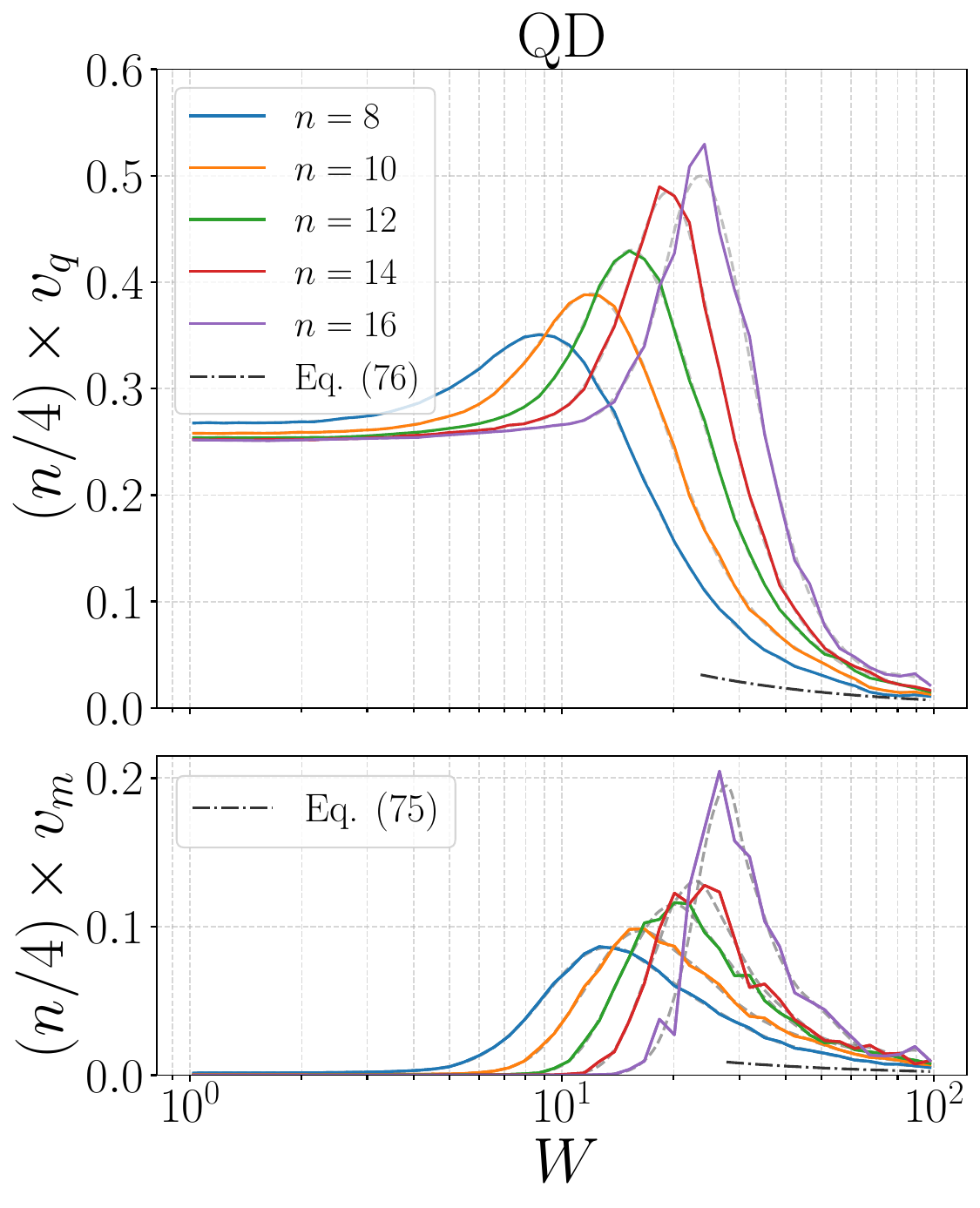}\hfill
    \includegraphics[width=0.33\textwidth]{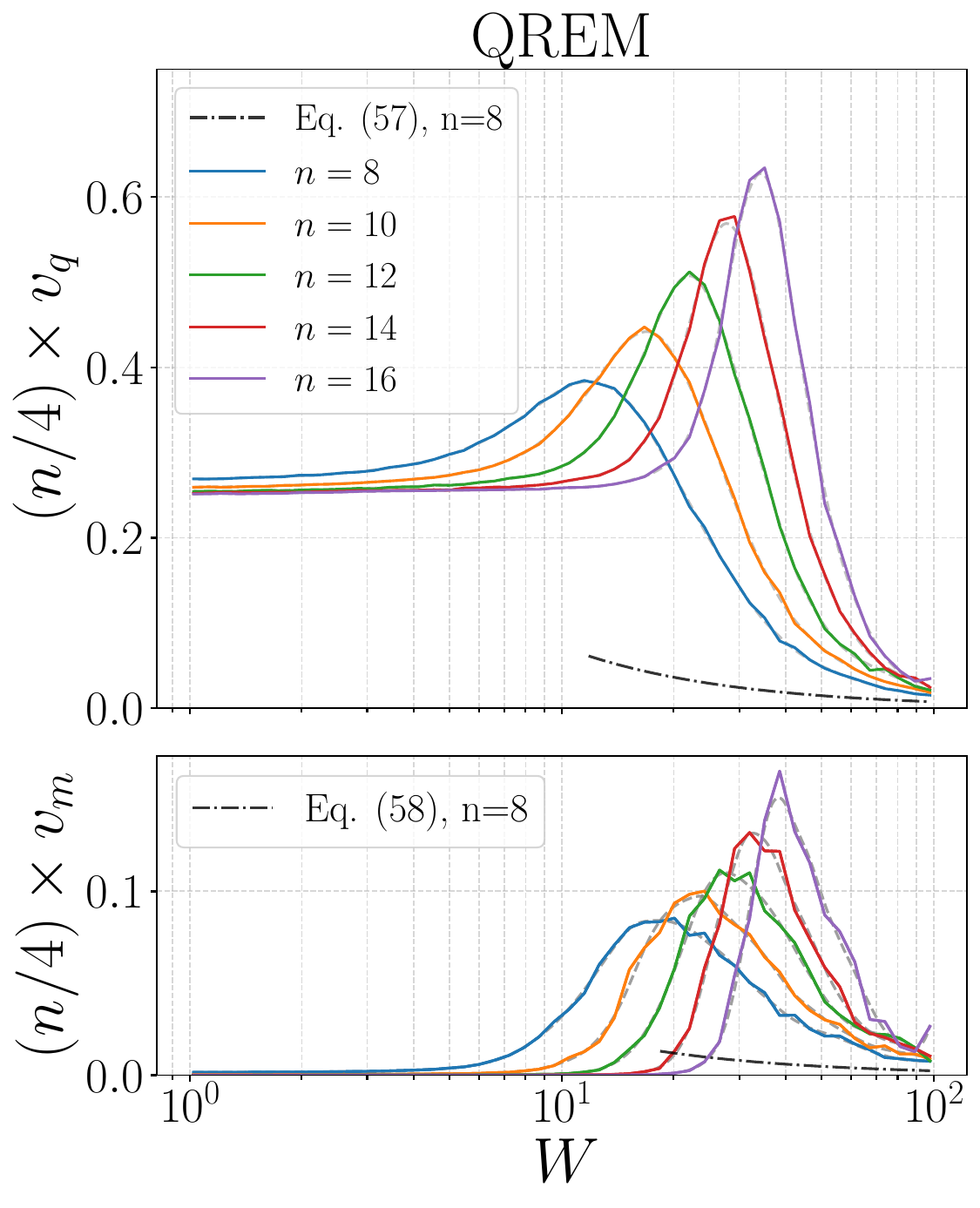}
    \caption{Quantum (top) and mesoscopic (bottom) imbalance variances $v_q(t\to \infty)$ and $v_m(t\to \infty)$, multiplied by $n/4$,
    for system sizes from $n=8$ to $n=16$, extracted from the time evolution of the 1D model (left), QD model (center) and QREM (right).  The large-$W$ asymptotics 
    Eqs.~\eqref{eq94} and \eqref{eq95}  for the 1D and QD models and
Eqs.~\eqref{eq:QREM-loc-vq}, \eqref{eq:QREM-loc-vm}, and \eqref{eq:1D-loc-c1-c2} for QREM are shown by dashed lines. 
    }
    \label{fig:Infinite_time_v_m_1D_QD_QREM}
\end{figure*}

In Fig.~\ref{fig:cuts_meso_variance_1D_QD_QREM}, the same data are plotted as functions of $W$ for selected times $t$. It is seen that, for the 1D model, the position of the maximum only weakly depends on time already for short times,
$t \sim 10$, becoming nearly constant (within statistical fluctuations) at $t \simeq 100$, roughly two orders of magnitude smaller than the Heisenberg time for the corresponding disorder ($W\simeq 5$). 
This is also illustrated in the inset.
On the other hand, for the QD and QREM models, the position of the maximum changes much more prominently, exhibiting an approximately logarithmic increase with time up to $t \sim 10^3$.
This increase can be explained if one recalls the decay law of the imbalance in the ergodic phases of these models,
 Eq.~\eqref{eq:QREM-imbalance-decay}. For a not-too-long time, this decay leads to a substantial suppression of the imbalance only for $D \gtrsim t^{-1}$. Using the exponential dependence of the diffusion constant $D$ on disorder in the pre-critical regime,  Eq.~\eqref{eq:QREM_D_precritical},
 we come to a characteristic disorder strength $W$ that grows logarithmically with time $t$, explaining the numerically observed drift of the position of the maximum. 
 
For larger systems, $n > n_* \approx 22$, we predict that at long times, the system will probe  the critical regime \eqref{eq:QREM_D_critical}; this will imply a saturation of the logarithmic drift of the position of the maximum for the QREM and QD models
at the corresponding crossover time that grows as a power law of $n$ (i.e., is much shorter than the exponentially large Heisenberg time).

\begin{figure*}
    \centering
    \includegraphics[width=0.33\textwidth]{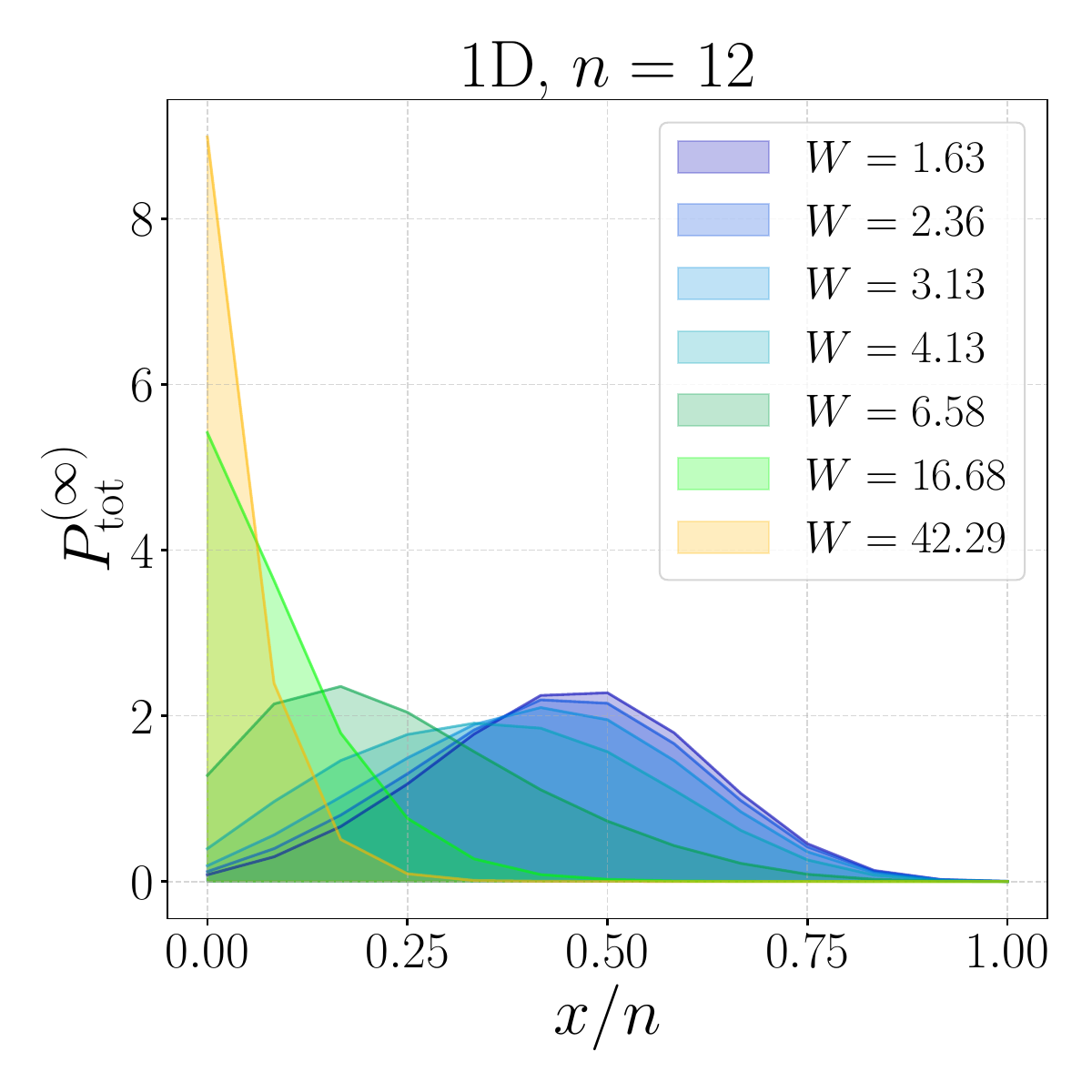}\hfill
    \includegraphics[width=0.33\textwidth]{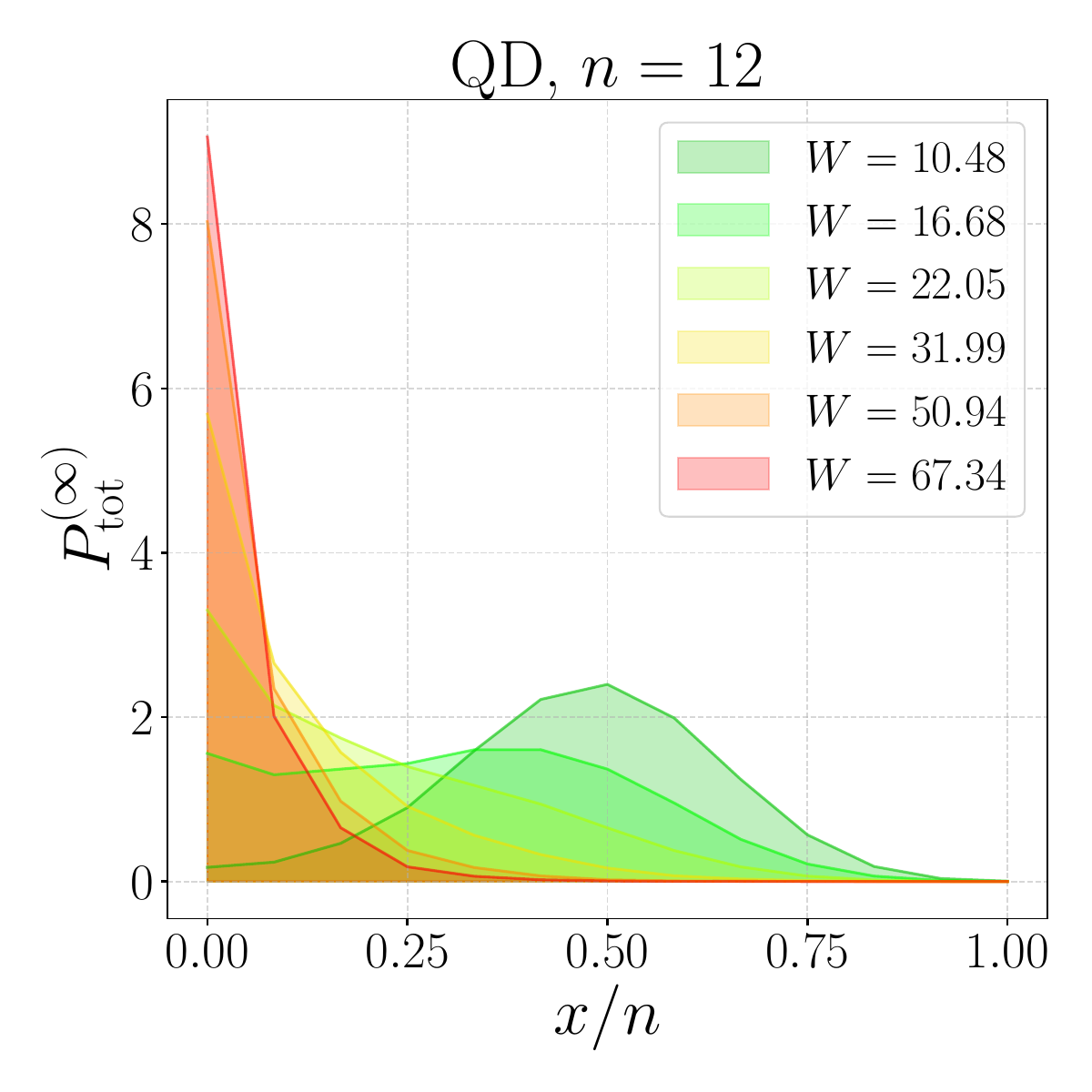}\hfill
    \includegraphics[width=0.33\textwidth]{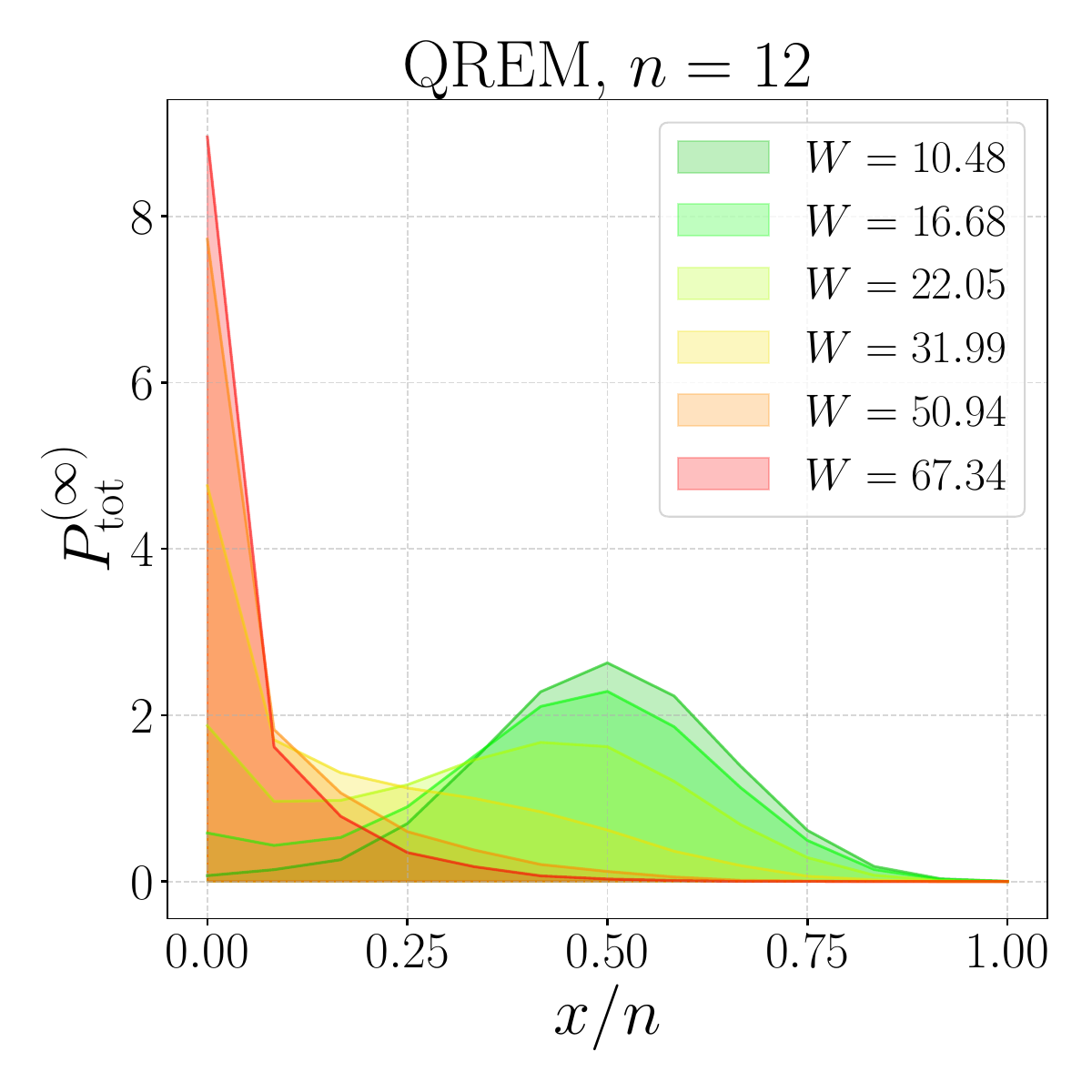}\\
    \includegraphics[width=0.33\textwidth]{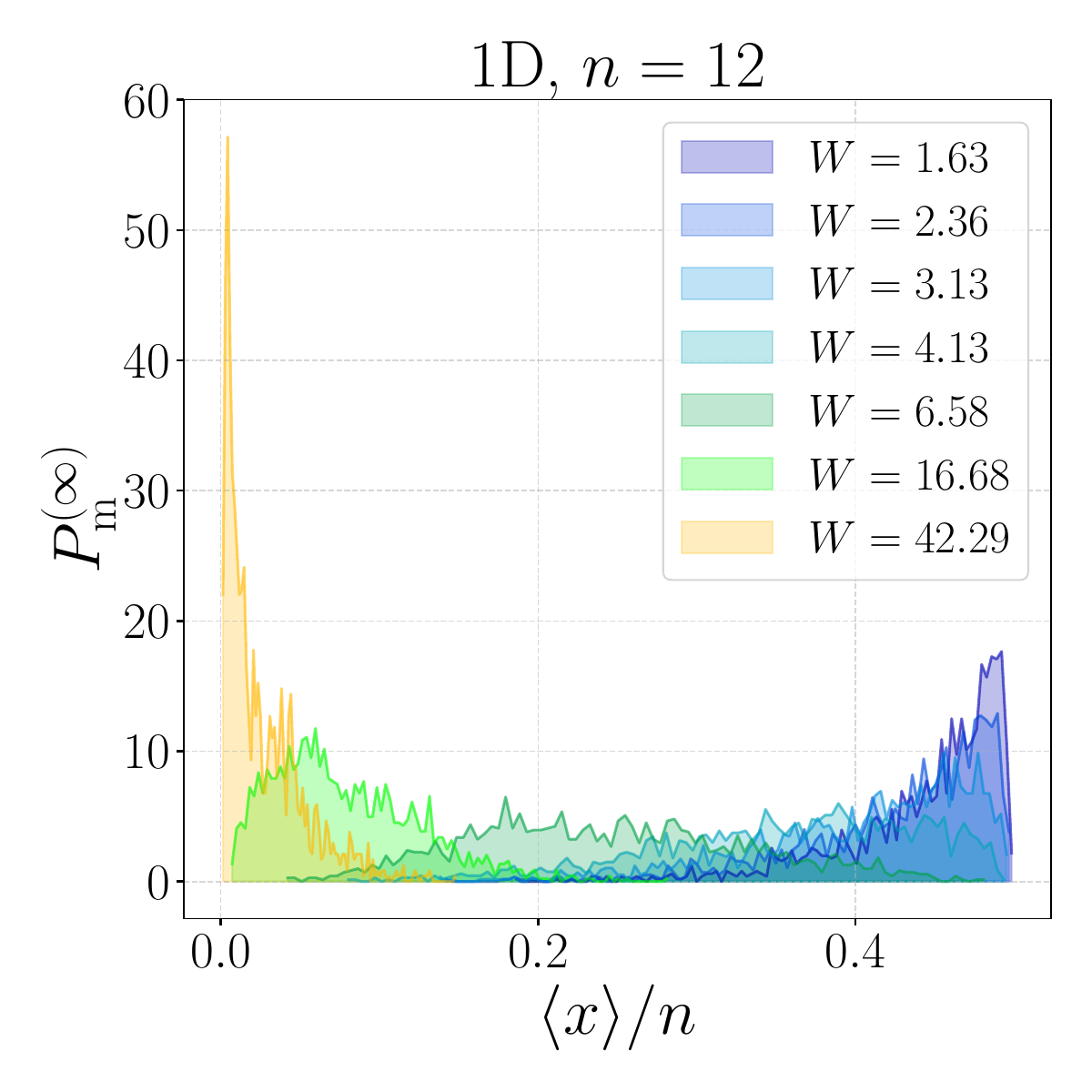}\hfill
    \includegraphics[width=0.33\textwidth]{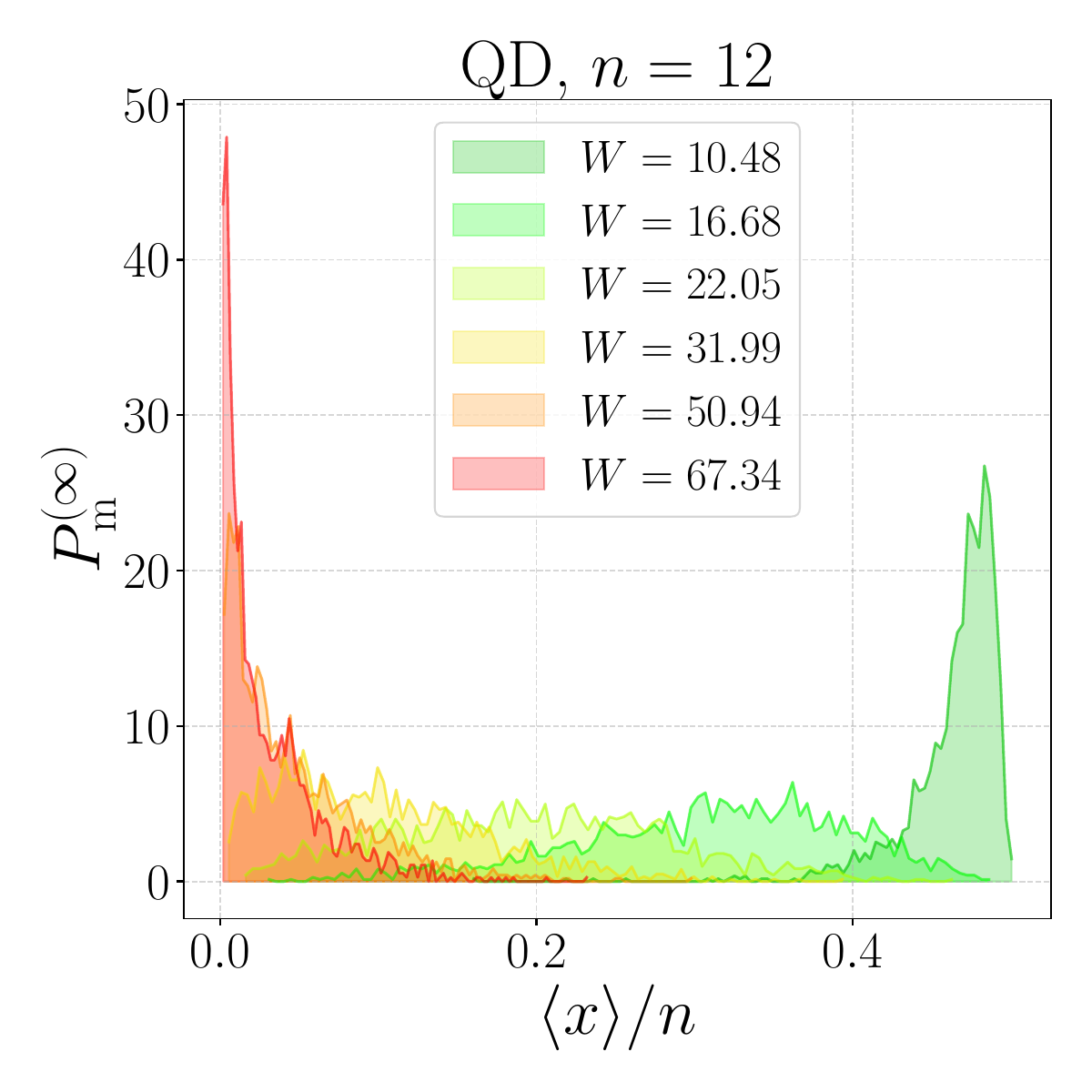}\hfill
    \includegraphics[width=0.33\textwidth]{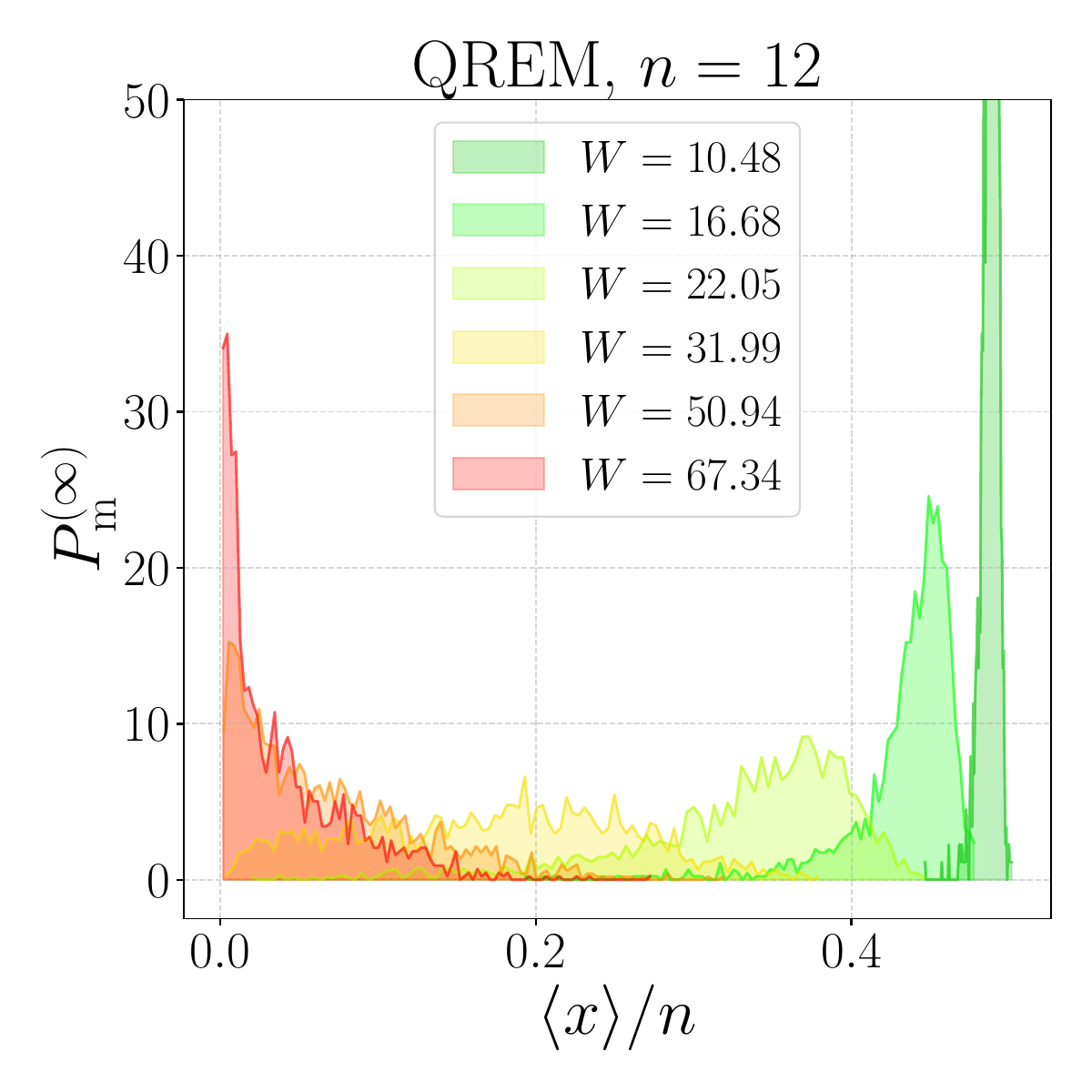}
    \caption{Total (top) and mesoscopic (bottom) distributions of Hamming distance $x$ at infinite time, $P_\text{tot}^{(\infty)}(x/n)$ and $P_\text{m}^{(\infty)}(\langle x \rangle/n)$,  for 1D model (left), QD model (middle), and QREM (right) with $n=12$. The data is averaged over 1600 realizations of disorder. 
    }
    \label{fig:imbalance_statistics_1D}
\end{figure*}

In Fig.~\ref{fig:time-evol_quantum_variance_1D_QD_QREM} and
Fig.~\ref{fig:cuts_quantum_variance_1D_QD_QREM}, the results for the quantum variance $v_q(t)$ are presented. In agreement with analytical predictions, we find $nv_q(t) \simeq 1$ for sufficiently long times in the ergodic phase of all models. Further, as analytically expected, a maximum is formed, providing a further indicator of the MBL transition. For QD and QREM models, the position of the maximum shows the same logarithmic drift at intermediate times as for $v_m(t)$. For the 1D model, the maximum in $v_q(t)$ is less pronounced than for mesoscopic fluctuations, because $v_q$ is much larger than $v_m$ in the ergodic phase. However, the maximum becomes more pronounced with increasing system size $n$, since $nv_q$ at the maximum grows with increasing $n$, in consistency with analytical predictions.


\subsection{Long-time asymptotic values of imbalance and its variances}
\label{subsec:infinite_time_numerics}

We focus now on long-time asymptotic values of the imbalance, $\overline{\mathcal{I}}_\infty$, and its mesoscopic and quantum fluctuations, $v_m(t\to \infty)$ and $v_q(t\to \infty)$.
In Fig.~\ref{fig:Infinite_time_imbalance_1D_QD_QREM},  the  infinite-time imbalance is shown as a function of disorder for various system sizes. For all the models, we observe a kink form of $\overline{\mathcal{I}}_\infty(W)$, which is a manifestation of the MBL transition. It is seen that the transition becomes sharper with increasing $n$. Further, a clear difference between the 1D model, on one side, and the QD and QREM models, on the other side, is observed. For the latter two models, the kink moves to the right with increasing $n$, in full consistency with the analytical prediction of the power-law growth of $W_c(n)$. In contrast, for the 1D model, the lines $\overline{\mathcal{I}}_\infty(W)$ for different $n$ are virtually indistinguishable at $W\ge 7$, which is a manifestation of $W_c^{\rm 1D}(n)$ being $n$-independent in the large-$n$ limit.
In the MBL phase (i.e., for strong disorder), the results are in agreement with the predicted behavior
$1-\overline{\mathcal{I}}_\infty \propto 1/W$, see Eq.~\eqref{1D-Iinfty} for the 1D and QD models and 
Eq.~\eqref{eq:QREM-loc-average-imb} for QREM. 
In particular, for the 1D model, the numerical data are essentially indistinguishable from the asymptotic formula 
\eqref{1D-Iinfty} (without any fitting of the numerical coefficient) for $W > 30$. For the QD and QREM models, for which $W_c(n)$ is substantially higher,
the data nicely approaches the analytical large-$W$ asymptotics 
at our strongest disorder $W\simeq 100$. 

Figure \ref{fig:Infinite_time_v_m_1D_QD_QREM} 
displays the quantum and mesoscopic variances of the imbalance at $t \to \infty$ as  functions of disorder. Comparing this figure with Fig.~\ref{fig:Infinite_time_imbalance_1D_QD_QREM}, we see that the maxima of $v_q(t\to\infty)$ and $v_m(t\to\infty)$ exhibit the same behavior as the positions of the kinks in the average imbalance. (A more accurate comparison will be performed 
in Sec.~\ref{sec:transition} below.)  In the MBL phase, the behavior at strong disorder is in good agreement with the analytical $1/W$ asymptotics,
Eqs.~\eqref{eq94} and \eqref{eq95}  for the 1D and QD models and
Eqs.~\eqref{eq:QREM-loc-vq}, \eqref{eq:QREM-loc-vm}, and \eqref{eq:1D-loc-c1-c2} for QREM.  Deeply  in the ergodic phase, the results are in agreement with 
Eq.~\eqref{vq-QREM-ergodic} for $v_q(t\to \infty)$ and with the exponential suppression of $v_m(t\to \infty)$,
Eq.~\eqref{vm-QREM-ergodic}. 
The values and $n$-dependence in the transition region (i.e., at maxima) are also 
consistent with analytical estimates in Sec.~\ref{sec:analytics}, see
Eqs.~\eqref{eq:QREM-transition-vq} and \eqref{eq:QREM-transition-vm} for QREM,
Eqs.~\eqref{eq:1D-transition-vq} and
\eqref{eq:1D-transition-vm} for 1D model, and
Eqs.~\eqref{eq:QD-transition-vq} and
\eqref{eq:QD-transition-vm} for QD model.

In Sec.~\ref{sec:dynamics-gener-imbalance}, we have discussed representation of the imbalance in terms of exact eigenstates of the Hamiltonian. The formulas strongly simplify in the $t \to \infty$ limit, see Eq.~\eqref{eq:I_inf-1} for the quantum average of the imbalance. We have performed exact diagonalization for system sizes $n =8$, 10, 12, and 14 to verify the numerical results for the imbalance and its fluctuations obtained as a large-time limit of quantum dynamics. 
The $t\to \infty$ results obtained by the two approaches are identical (within statistical uncertainty). 

In Fig.~\ref{fig:imbalance_statistics_1D}, distributions of $t\to\infty$ imbalance  obtained by exact diagonalization are shown. More specifically, we plot distribution functions of the Hamming distance $x$ divided by $n$, which is related to the imbalance via 
$\hat{x} / n = (1 - \hat{\mathcal{I}}^{(\alpha)})/2$, see Eq.~\eqref{eq:x-operator-def}. The top panels present total  distributions 
$P_{\rm tot}^{(\infty)}(x/n)$, which are obtained by  disorder averaging of the distributions $P^{(\alpha)}(x,t)$, Eq.~\eqref{eq:radial_distrib}, at $t\to \infty$. The term ``total'' means that these distributions involve both quantum and mesoscopic fluctuations. 
The bottom panels present mesoscopic distributions $P_{\rm m}^{(\infty)}(\langle x \rangle/n)$ of the quantum average $\langle x \rangle / n$. The figure nicely illustrates the evolution of distributions from the ergodic to the MBL phase with increasing disorder for all three models. 

\begin{figure*}[ht!]
    \centering
    \includegraphics[width=0.50\textwidth]{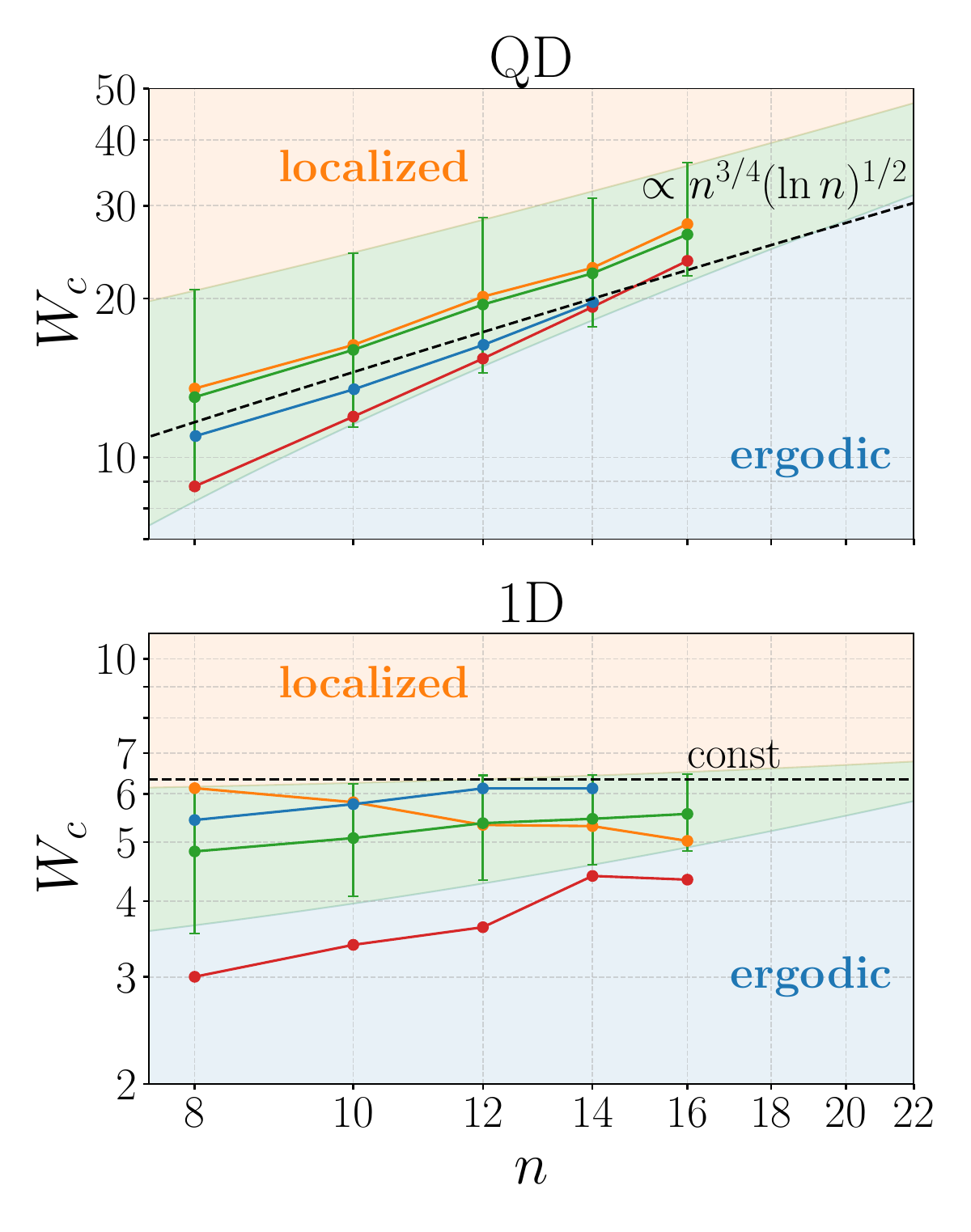}\hfill
 \includegraphics[width=0.50\textwidth]{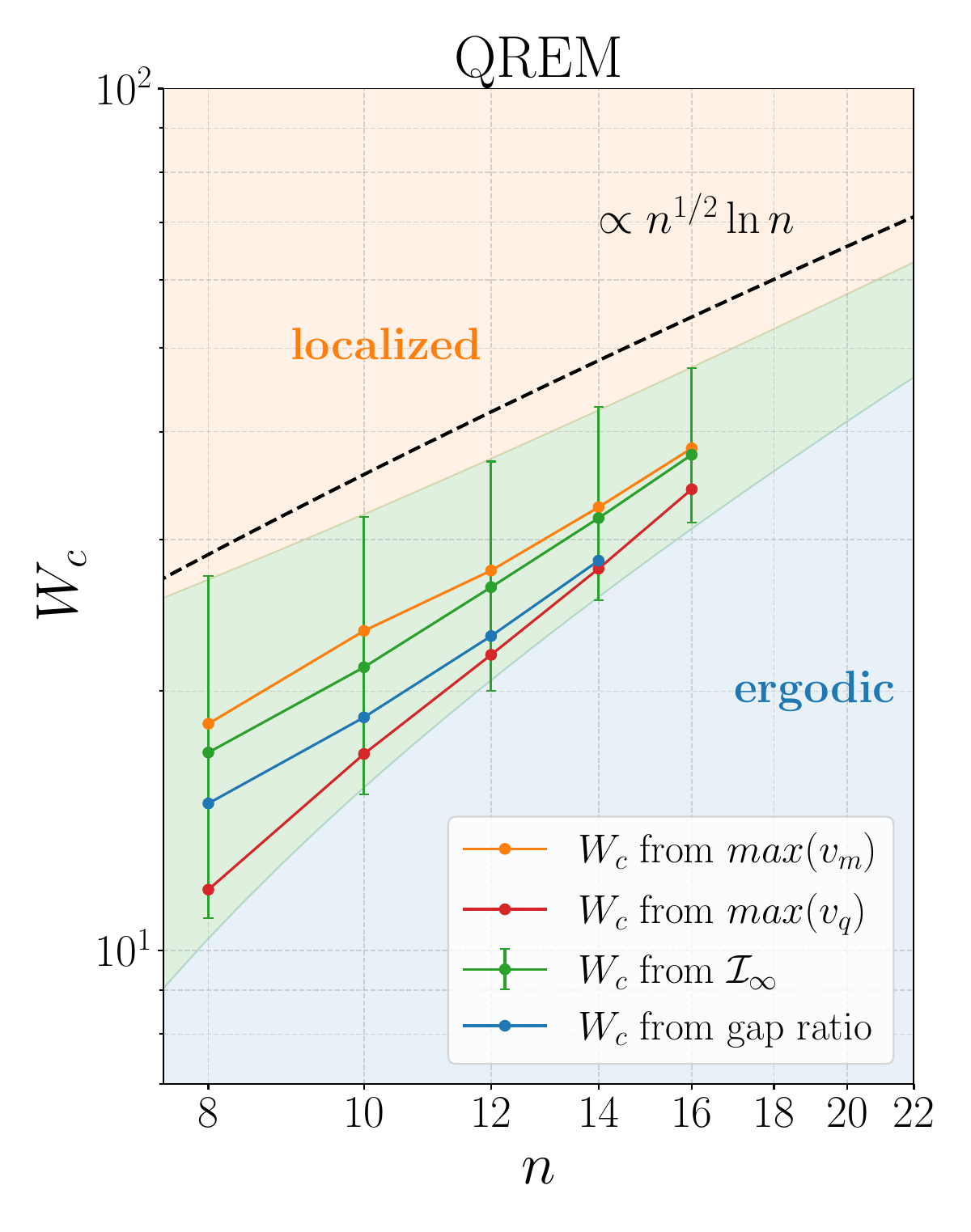}
    \caption{System-size-dependent values of the critical disorder $W_c(n)$ for the 1D model and QD model (left) and the QREM (right), extracted from several indicators: $t \to \infty$ average imbalance and maxima of the quantum and mesoscopic variances obtained from the time evolution out of an initial basis state $\ket{\alpha}$ in this paper, as well as the adjacent gap ratio obtained via exact diagonalization in Ref.~\cite{Scoquart2024}).  System sizes range from $n=8$ to $n=16$. 
    }
    \label{fig:Wc_vs_n_1D_QD_QREM}
\end{figure*}


\begin{figure*}[ht!]
    \centering
    \includegraphics[width=0.5\textwidth]{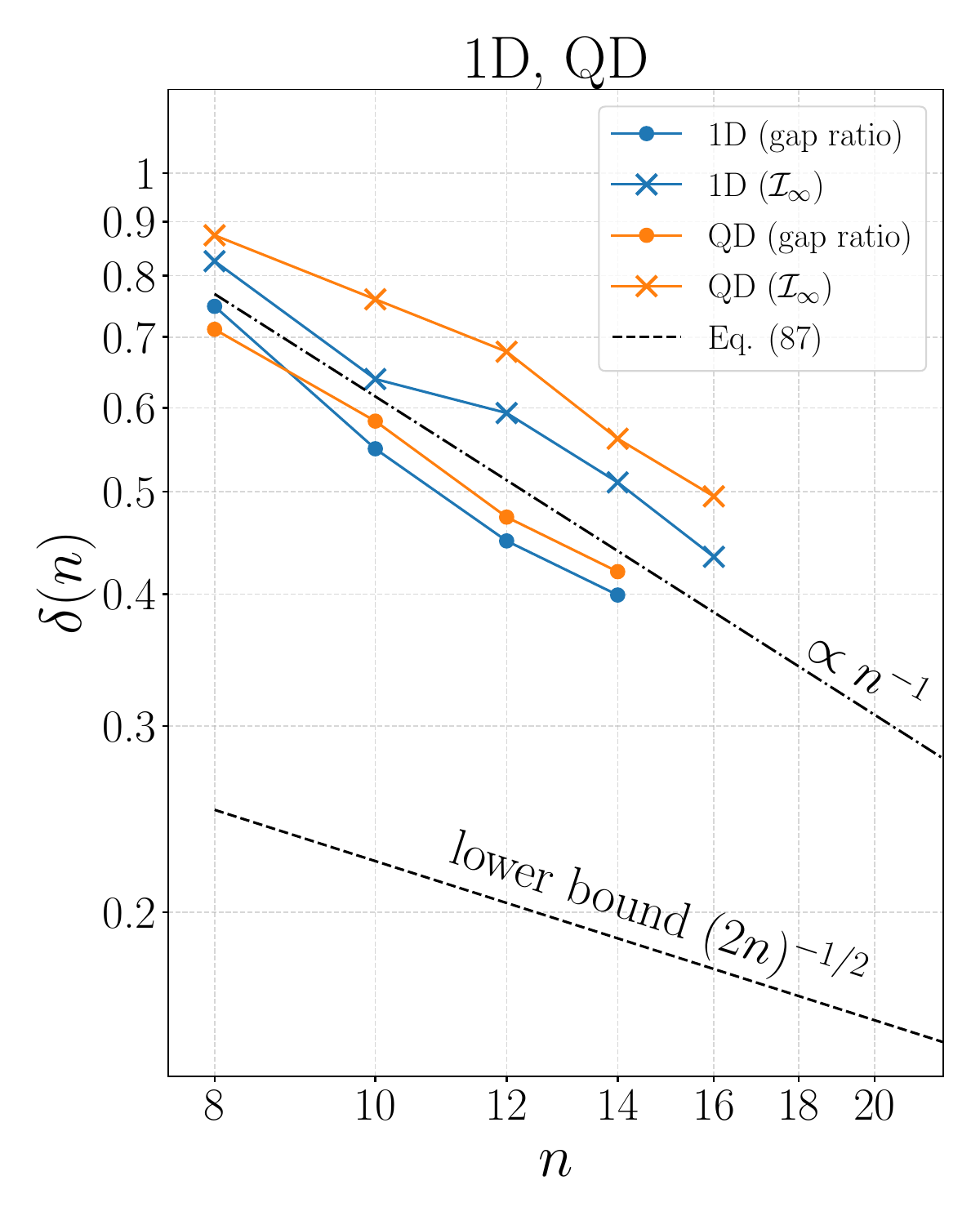}\hfill
    \includegraphics[width=0.5\textwidth]{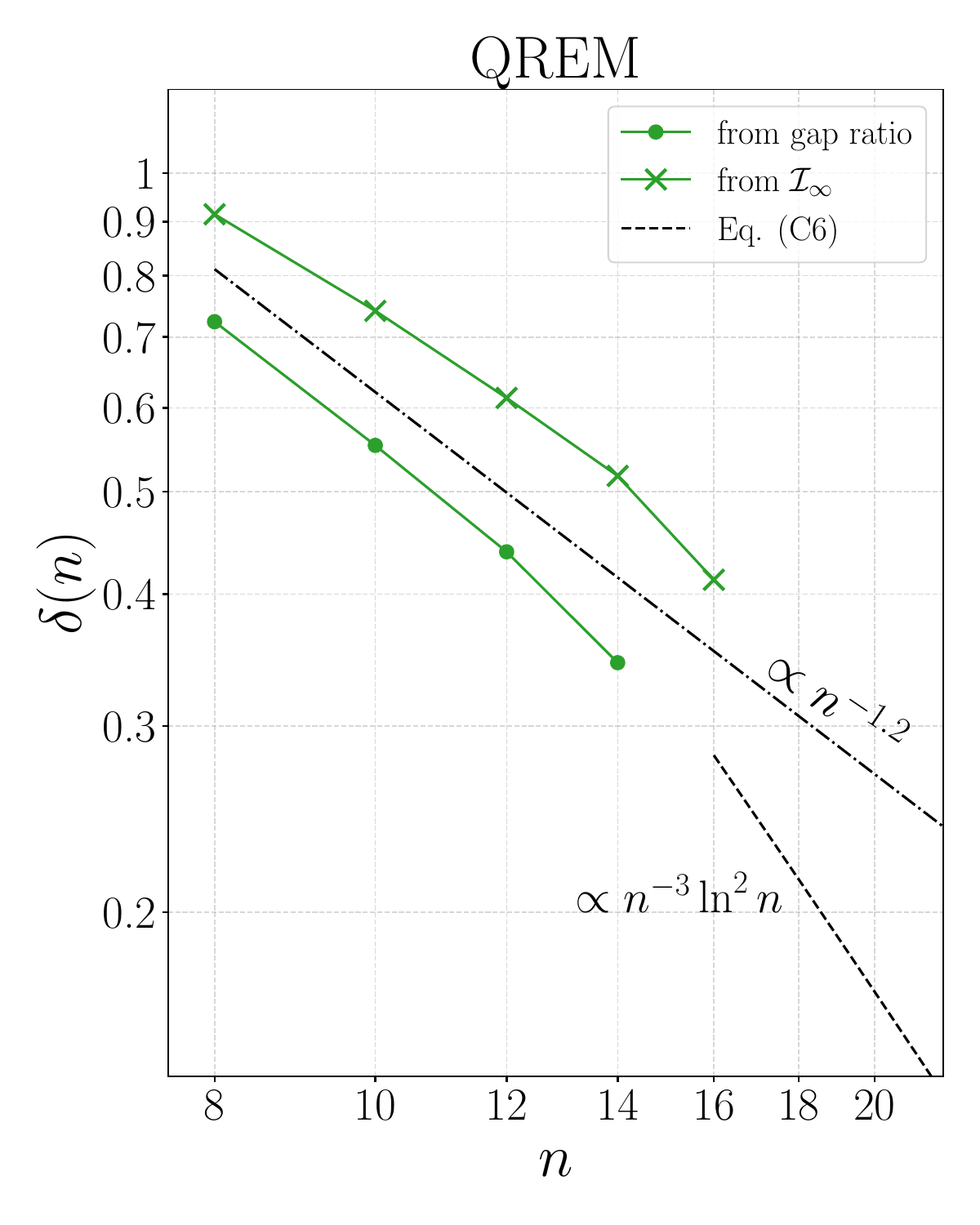}
    \caption{Transition width $\delta(n)$, Eq.~\eqref{eq:transition-width-delta}, as a function of $n$, for 1D and QD models (left) and for QREM (right), as obtained from $t\to\infty$ values of the averaged imbalance (this work) and from results for the gap ratio (Ref.~\cite{Scoquart2024}). 
The lines $\sim n^{-\mu}$ with $\mu=1$ in the left panel and $\mu = 1.2$ in the right panel are guides to the eye. For 1D and QD models (left panel), the lower bound \eqref{eq:transition-width-bound} is also shown; it crosses the line $\sim 1/n$
   at $n_b \approx 80$. For QREM (right panel), the large-$n$ asymptotics 
   \eqref{eq:QREM-maintext-Wc-finite-size-width} is also shown.
   }
  \label{fig:Relative_width_1D_QD_QREM}
\end{figure*}

\section{Phase diagrams of the MBL transitions}
\label{sec:transition}

The numerical results presented in Sec.~\ref{sec:numerics}, in combination with analytical results of Sec.~\ref{sec:analytics}, allow us to determine the phase diagrams of MBL transitions in the considered models by using several indicators of the transition point $W_c(n)$. The first of them is the average imbalance $\overline{\mathcal{I}}_\infty(W)$. In the QREM and QD model, $\overline{\mathcal{I}}_\infty(W)$ exhibits, in the $n\to \infty$ limit, a jump from 0 to 1 at the transition, which is smeared at finite $n$. We thus use the condition $\overline{\mathcal{I}}_\infty(W)= 1/2$ to get the position of the transition point in these two models. For the 1D model, the jump is from zero to a value smaller than unity, see Eq.~\eqref{1D-Iinfty} and
Fig.~\ref{fig:Infinite_time_imbalance_1D_QD_QREM}. We find numerically that the limiting value of the imbalance on the MBL side is $\approx 0.6$. For the transition point, we use the condition $\overline{\mathcal{I}}_\infty(W)= 0.4$.
Two further indicators of $W_c(n)$ are provided by the positions of the maxima of the quantum and mesoscopic fluctuations of the imbalance, $v_q(t\to \infty)$ and $v_m(t\to\infty)$. Finally, in addition to the results of this work for the imbalance, we also include the level-statistics indicator as obtained in Ref.~\cite{Scoquart2024}. Specifically, it is determined by the point where the mean adjacent gap ratio crosses the line that is half-way between the Poisson and Wigner-Dyson values. 

The obtained phase diagrams in the $n - W$ plane are shown in Fig.~\ref{fig:Wc_vs_n_1D_QD_QREM}.
We see that, for each of the models, the values $W_c(n)$ mutually agree, indicating a direct transition between the ergodic and MBL phases. In particular, the agreement between the imbalance and the level-statistics indicators excludes an intermediate phase that would be non-ergodic but, at the same time, would allow propagation up to the largest Hamming distance in Fock space.

For the QD model and QREM, we observe a fast, power-law increase of $W_c(n)$, in full agreement with analytical predictions (shown in the figure): $W_c^{\rm QREM}$ grows as $n^{1/2 } \ln n$, see  Sec.~\ref{sec:analytics-QREM}
and Appendix \ref{app:QREM}, and 
$W_c^{\rm QD}$ grows (at least) as $n^{3/4 } \ln^{1/2} n$, see
Sec.~\ref{sec:analytics-QD} and Appendix \ref{app:QD-Wc-lower-bound}. 
At the same time, the results for the 1D model look very different: they yield $W_c^{\rm 1D}$ that is essentially constant as a function of disorder. Estimating the critical disorder $W_c^{\rm 1D}(n \to \infty)$ using the data for the averaged imbalance (which have considerably smaller statistical uncertainty than those for maxima of the variances) and using an extrapolation $W_c^{\rm 1D}(n) = W_c^{\rm 1D}(\infty) - a/n$, we obtain 
an estimate $W_c^{\rm 1D}(n \to \infty) \approx 6.33$, 
in good agreement with the estimate 
$W_c^{\rm 1D}(n \to \infty) \approx 6.5$
obtained above by extrapolating the exponent $\gamma_I$ in Fig.~\ref{fig:fits_gamma_1D}. 

 We analyze now the transition width. For this purpose, we consider a window of values of $\overline{\mathcal{I}}_\infty(W)$ chosen symmetrically around the value that we use to determine the transition point. Specifically, we choose the window to be $[0.25;0.75]$ for QD and QREM models, and $[0.3;0.5]$ for the 1D model. This determines the disorder window $[W_-(n);W_+(n)]$ associated with the transition, yielding the transition width $\ln (W_+(n)/W_-(n)) \simeq \Delta W(n) / W_c(n)$. 
In Fig.~\ref{fig:Relative_width_1D_QD_QREM}, we plot the transition width as a function of $n$ for all three models that we study. In addition to the width obtained from our results for the imbalance, we also include in this figure the results of Ref.~\cite{Scoquart2024}  obtained in an analogous way from data on levels statistics. The above definition of the width contains a freedom of choosing a window of the variable. For a reasonable choice of the window, this freedom does not affect the scaling of the transition width with $n$ but introduces a constant factor. To compensate for this freedom, we introduce in the definition of the transition width a factor $R = (\Delta X)_{\rm t} / 2 (\Delta \overline{X})_{\rm w}$, where  $(\Delta X)_{\rm t}$ is the total jump of a variable $X$  at the transition (in the thermodynamic limit) and $(\Delta \overline{X})_{\rm w}$ is the chosen window width, see
Appendix \ref{app:Harris-criterion} for detail.
We thus characterize the transition width by 
\begin{equation}
\delta (n) = R \ln (W_+(n)/W_-(n))
\simeq R \: \Delta W(n) / W_c(n)\,.
\label{eq:transition-width-delta}
\end{equation}

It is seen in Fig.~\ref{fig:Relative_width_1D_QD_QREM} that the transition widths obtained from two observables (imbalance and level statistics) are essentially the same. Further, for all the models, we observe a power-law sharpening of the transition with increasing system size, $\delta(n) \sim n^{-\mu}$. Interestingly, the exponent $\mu$ for all the models is close to unity in the considered range of $n$. It should be emphasized, however, that, for all these models, this is {\it not} the asymptotic (large-$n$) behavior. For the QREM, the actual
asymptotics is predicted to be 
$\delta  \sim n^{-3} \ln^2 n$, see
Eq.~\eqref{eq:QREM-maintext-Wc-finite-size-width}. However, it holds only for larger system sizes, $n > n_* \approx 22$.  For smaller values of $n$ (pre-critical regime), the effective exponent $\mu$ is reduced compared to its asymptotic value and $\mu \approx 1$ is in agreement with analytical expectations \cite{Scoquart2024}.

For 1D and QD models, the asymptotic behavior of the width should be very different. Indeed, in these models, the disorder of strength $W$ is implemented via only $n$ independent energy variables $\epsilon_i$ (in contrast to QREM where all $2^n$ variables $E_\alpha$ are independent). This leads to an important bound (extension of Harris criterion) on the sharpness of the transition \cite{chandran2015finite,Chayes1989correlation}. Specifically, we use the upper bound on a derivative of the mean value of an observable serving as a transition indicator given by Eq. (2) of Ref.~\cite{chandran2015finite} (see also
Lemma 2.3 in Ref.~\cite{Chayes1989correlation}). 
In our case, the observables are the mean imbalance and the mean adjacent gap ratio. This leads, for the 1D and QD models, to the following lower bound on the transition width defined above~\footnote{As pointed out in Ref.~\cite{chandran2015finite},  the numerical coefficient 
in front of $n^{-1/2}$ in the bound formally diverges for a box distribution of $\epsilon_i$. We used a Gaussian distribution of $\epsilon_i$, with the same variance, to obtain the coefficient in the bound \eqref{eq:transition-width-bound}.}:
\begin{equation}
\delta^{\rm 1D}(n)\,, \ \delta^{\rm QD}(n) \ge \frac{1}{\sqrt{2n}} \,.
\label{eq:transition-width-bound}
\end{equation}
Details of derivation of Eq.~\eqref{eq:transition-width-bound} are presented in Appendix \ref{app:Harris-criterion}, where we also discuss implications of this bound for the form of scaling at the transition at sufficiently large $n$.
The bound \eqref{eq:transition-width-bound}, which is also shown in Fig.~\ref{fig:Relative_width_1D_QD_QREM}, implies that the numerically observed behavior $\delta(n) \sim n^{-\mu}$ with $\mu \approx 1$ holds only  for not too large $n$, since it would violate the bound for $n > n_b$. Extrapolating line $\sim 1/n$ approximating the numerical results for  $\delta(n)$ until a crossing with the bound, we obtain an estimate $n_b \sim 80$. (An analogous estimate was performed in Ref.~\cite{chandran2015finite} by using data for different 1D models and different observables \cite{kjall2014many,Luitz2015}; it yielded substantially larger values, $n_b \sim 500 - 5000$.) Thus, we expect that the actual critical behavior in 1D and QD models can be observed for system sizes $n \gtrsim 80$. Clearly, this argument provides only the upper bound on the crossover value of $n$; it is possible that the crossover to the actual asymptotic behavior in fact takes place already for smaller system sizes. 

\section{Summary and outlook}
\label{sec:summary}

In this paper, we have studied MBL transitions in a family of interacting  spin-$\frac12$ (or, equivalently, qubit) models with random interactions and random Zeeman energies. The models are of single-spin-flip type, so that the associated Fock-space graphs are formed by vertices and edges of the $n$-dimensional hypercube, where $n$ is the number of spins.
Two of the models that we considered involve only interactions of pairs of spins, thus belonging to a class of models that can be naturally implemented in quantum simulators. 
One of these models is a 1D chain with nearest-neighbor interactions, the other one is the QD model with all-to-all pair interactions. 

The third model is (a version of) the QREM. While a direct experimental implementation of QREM for a relatively large number of spins does not appear to be realistic, its investigation in combination with more realistic, pair-spin interaction models (1D and QD) is very instructive. First, the MBL transition in QREM can be studied analytically by using its connection to the RRG model, thus providing a benchmark for numerical studies of various observables. Second, the MBL transition in QREM and in more realistic models have many common features. In particular, the physics of the QD model turns out to be in many respects very similar to that of QREM. Third, despite common features, the analytics predicts a very different behavior for the scaling of the MBL transition in the 1D model as compared to QREM, which was one of the central points of our study. 

The observable on which we focus in the present paper is the generalized imbalance characterizing propagation in the Fock space out of an initial basis state $\ket{\alpha}$; the corresponding operator is given by 
Eq.~\eqref{eq:imbalance-3}. More specifically, we consider the averaged imbalance, as well as its quantum and mesoscopic fluctuations. Importantly, while the imbalance has a clear physical meaning in the Fock-space representation of the problem, it can also be efficiently determined by real-space measurements. 

We show that, for all considered models, the average imbalance and its quantum and mesoscopic fluctuations provide excellent indicators for the position of the MBL transition $W_c(n)$,
see analytical results in Sec.~\ref{sec:analytics}
and numerical results in Figs.~\ref{fig:Infinite_time_imbalance_1D_QD_QREM} and
\ref{fig:Infinite_time_v_m_1D_QD_QREM}.
Combining these findings with earlier results on levels statistics, we determine phase diagrams of the MBL transitions in the $n - W$ plane, Fig.~\ref{fig:Wc_vs_n_1D_QD_QREM}. Our results provide evidence of a single transition between the ergodic and MBL phases for each of the models, without any intermediate phase that would be non-ergodic but, at the same time, exhibiting propagation up to the maximal Hamming distance in Fock space. For QREM and QD model, $W_c(n)$ grows as a power law of $n$ (with logarithmic corrections), in agreement with analytical predictions $W_c^{\rm QREM}(n) \sim n^{1/2} \ln n$ and 
$W_c^{\rm QD}(n) \gtrsim n^{3/4} \ln^{1/2} n$. This growth is in stark contrast to the behavior of 1D model, which exhibits $W_c(n)$ that is essentially independent of $n$, in agreement with analytic expectation $W_c^{\rm 1D}(n\to \infty)= {\rm const}$.  Our estimated value is $W_c^{\rm 1D}(n\to \infty) \approx 6.5$.

We also determine numerically the scaling of the transition width $\Delta W (n) / W_c(n)$,
see Fig.~\ref{fig:Relative_width_1D_QD_QREM}. The transition sharpens with increasing $n$; interestingly, it does it almost in the same way in all three models. However, this numerically observed scaling of the transition width is certainly not the asymptotic (large-$n$) one. For QREM, an analytical study implies that the asymptotic critical behavior, with a faster decay of $\Delta W (n) / W_c(n)$, will be approached for $n > n_*$, where $n_* \approx 22$. 
For 1D and QD models, we find, using the Harris bound \cite{chandran2015finite}, that a crossover to the asymptotic scaling behavior should happen at 
 $n < n_b$, where $n_b \sim 80$.   
While the corresponding values of $n$ are larger than those accessible to exact simulations on a classical computer, they are within the reach of quantum simulators (NISQ devices) that are already available or expected to be available in the near future.
Our results thus indicate, in consistency with existing experimental works \cite{schreiber2015observation,choi2016exploring,Yao2023}, feasibility of experimental studies of phase diagrams of the MBL transitions and associate scaling behavior in models of the type considered in our paper.

Advantages of considering the chosen class of models are discussed in Sec.~\ref{sec:models_def}. Furthermore, our results show that a finite-size drift in our 1D model is relatively small.
Importantly, one can also study models that interpolate between our 1D and QD models, by considering other spatial geometry (e.g., two-dimensional) and/or other distance dependence of interaction (e.g., power-law-decaying interaction). This can be done by keeping all key properties of the model, including the Fock-space graph (and thus the coordination number) and the distributions of diagonal and hopping Fock-space matrix elements. This will allow one to explore, in the clearest way, the effects of spatial dimensionality and the interaction character on the $n - W$ phase diagram and on the scaling of the MBL transition. Note that, at variance with the 1D model and in analogy with the QD model, the models with power-law interactions or in spatial dimensionality $d>1$ are predicted to exhibit critical disorder $W_c(n)$ that increases to infinity at $n\to \infty$. 

Another important direction for future research is to explore in more detail the properties of the MBL phase in the models studied in this paper. As discussed above, it is believed that proliferation of system-wide resonances 
\cite{villalonga2020, Ghosh2022resonance, Morningstar2022avalanches, 
Herre-thesis, Ha2023, long2023phenomenology, colbois2024statistics}
is responsible for the development of many-body delocalization when the critical disorder is approached from the MBL side. However, a detailed quantitative understanding of manifestations of resonances of various types in various experimentally measurable observables (such as, e.g., imbalance) and in various models remains to be developed.  Also, connections with the avalanche theory \cite{roeck17, Thiery2017a, goremykina2019analytically, morningstar2019renormalization, morningstar2020a, doggen2020slow, Morningstar2022avalanches} need to be better understood. 

Another interesting question, which is controversially discussed in recent literature, is whether the numerically observed residual, ultraslow transport deeply in the MBL phase of one-dimensional models \cite{Weiner2019slow, Kiefer2020a, Kiefer2020b, Kiefer2021a, Evers2023internal, Ghosh2022resonance, chavez2023ultraslow} requires any modification of our understanding of the MBL phase. A related question concerns proposals in the literature, also based on numerical data, that the MBL phase in one-dimensional models may actually be subdivided into two phases that differ in the form of asymptotics of some correlation or distribution functions \cite{Weiner2019slow, colbois2024statistics}. Although we are not aware of clear-cut analytical support for these intriguing  proposals, they cannot be excluded in view of the absence of a fully controllable analytical theory of the MBL transition. Our numerical data do not provide any indication of a second transition within the MBL phase. However, this would not be in contradiction with such conjectures if the observables that we study are not affected by a fine distinction between the two proposed parts of the MBL phase.

\section*{Acknowledgements}

   We thank Markus M\"uller for useful discussions and, in particular, for providing an idea on improving the lower bound on $W_c^{\rm QD}(n)$, which is implemented in
  Appendix \ref{app:QD-Wc-lower-bound}. 
  We also thank Christopher Laumann for useful discussions, in particular of implications of the Harris criterion for the MBL transition, Ref.~\cite{chandran2015finite}, which motivated us to derive results presented in Appendix \ref{app:Harris-criterion}. We further thank Piotr Sierant for sharing his exact-diagonalization codes, which were partly used in the exact-diagonalization part of this work.

\appendix

\section{Localization transition on RRG}
\label{app:RRG}

In Fig.~\ref{fig:analytics_RRG_Gaussian_m20}, the analytical ``finite-size phase diagram'' is shown for ergodicity-to-localization transition in an Anderson model on an RRG. In this Appendix, we present definition of the model and briefly summarize relevant formulas, which have been derived in Refs.~\cite{Herre2023,Scoquart2024}. The coordination number of the graphs is denoted by $m+1$; in Fig.~\ref{fig:analytics_RRG_Gaussian_m20} we choose $m=20$. To make a closer contact to the spin models, we parametrize the number of vertices of an RRG as $N=2^n$. The diagonal matrix elements $E_\alpha$ and the (complex) hopping matrix elements $T_{\alpha\beta}$ are independent random variables; we choose them to be Gaussian distributed in analogy with our spin models. Specifically, the distribution $\gamma(E)$ of on-site energies is $\mathcal{N}(0, W^2/2\pi)$, where $W$ is the disorder strength, and the hoppings $T_{\alpha\beta}$ are distributed as $\mathcal{N}(0, 2/\pi)+i \mathcal{N}(0, 2/\pi)$.
Specific choice of the numerical coefficients of order unity in these formulas are not essential; modifying them amounts simply to rescaling $W$. We choose these coefficients such that $\langle |T| \rangle \gamma(0)= 1/W$ in the notations of Refs.~\cite{Scoquart2024}.
The thermodynamic limit ($N \to \infty$) critical disorder $W_c^{\rm RRG}(m,\infty)$ 
is given by a solution of the equation
\begin{align}
4m \langle |T|\rangle \gamma(0) \ln \frac{1}{\langle |T|\rangle \gamma(0)} = 1\,
\label{eq:self_cons_critical_disorder_Herre}
\end{align}
with respect to $W$. 
For the above choice of the distributions of matrix elements, this equation becomes $W=4m\ln W$. 
We note that the accuracy of derivation is not sufficient to fix a numerical coefficient of order unity in the argument of logarithm in Eq.~\eqref{eq:self_cons_critical_disorder_Herre}; it is set to unity in Eq.~\eqref{eq:self_cons_critical_disorder_Herre}. Modifying this coefficient does not affect the large-$m$ asymptotics of 
$W_c^{\rm RRG}(m,\infty)$ but would lead to a subleading correction somewhat shifting $W_c^{\rm RRG}(m,\infty)$ at finite $m$.
 The finite-size critical disorder $W_c^\text{RRG}(m,2^n)$ is obtained by solving the equation 
 \begin{equation}
 N_\xi(W) = 2^n \,,
 \label{eq:RRG-transition-Nxi-N}
 \end{equation}
 where $N_\xi(W)$ is the correlation volume given by Eqs.~(48) and (49) of \cite{Scoquart2024}. 

For disorder $W$ close to the thermodynamic-limit transition, i.e., for
 \begin{equation}
\frac{W^{\rm RRG}_c(m,\infty)}{2} < W < W^{\rm RRG}_c(m,\infty),
\label{eq:RRG-critical regime}
\end{equation}
the correlation volume exhibits the following critical behavior:
\begin{equation}
\ln N_\xi (W) = \pi \ln m \left[\frac{3}{2} \left( 1 - \frac{W}{W_c^{\rm RRG}(m,\infty)}\right) \right]^{-1/2} .
\label{eq:RRG-N-xi-critical}
\end{equation}
Consequently, for sufficiently large system size, $n > n_*$, one has
\begin{equation}
\frac{W_c^{\rm RRG}(m,\infty) - W_c^{\rm RRG}(m,2^n)}{W_c^{\rm RRG}(m,\infty)} \propto n^{-1/\nu}
\label{eq:Wc-RRG-crit}
\end{equation}
with $\nu = 1/2$. The lower border $n_*$ of this (critical) regime increases logarithmically with $m$;  for $m=20$ chosen in Fig.~\ref{fig:analytics_RRG_Gaussian_m20}, we get $n_* \simeq 21$. The critical behavior \eqref{eq:Wc-RRG-crit} of $W_c^{\rm RRG}(m,2^n)$ is well seen in Fig.~\ref{fig:analytics_RRG_Gaussian_m20} below the dotted line corresponding to $n_*$.

For a large coordination number, when $W_c^{\rm RRG}(m,\infty) \simeq 4m\ln m$, there is also a parametrically defined pre-critical regime
\begin{equation}
m < W < \frac{W_c^{\rm RRG}(m,\infty)}{2} \,,
\label{eq:W-precrit-window}
\end{equation}
with the correlation volume $N_\xi(W)$ given by
\begin{equation}
N_\xi (W) \simeq \frac{W^2}{m} \exp\left( \frac{W}{2m} \right).
\label{eq:RRG-N-xi-precritical}
\end{equation}
Consequently, for not too large system sizes, $n < n_*$, the finite-size transition point $W_c^{\rm RRG}(m,2^n)$ exhibits an approximately linear drift with $n$. The overall magnitude of this drift increases logarithmically with $m$, as is seen from 
Eq.~\eqref{eq:W-precrit-window}. We have chosen in Fig.~\ref{fig:analytics_RRG_Gaussian_m20} a moderately large value of $m$ to illustrate both regimes of $W_c^{\rm RRG}(m,2^n)$: the critical and the pre-critical one, with a crossover at $n \simeq n_*$. 

For any finite $n$, the localization transition is in fact a crossover, i.e., it has a finite width $\Delta W$ around the disorder $W_c^{\rm RRG}(m,2^n)$. To estimate $\Delta W$, we use the fact that the relevant observables follow the volumic scaling (i.e., are functions of $N/N_\xi(W)$) at $W < W_c^{\rm RRG}(m,\infty)$
\cite{tikhonov2016anderson,garcia-mata17,biroli2018,tikhonov19statistics,tikhonov2021from,garcia-mata2022critical}. Thus, the interval of disorder $[W_-,W_+]$ in which the  transition takes place can be estimated from the equations
\begin{equation}
N_\xi(W_-) =  2^n b_- \,, \qquad
N_\xi(W_+) =  2^n b_+ \,,
\label{eq:RRG-finite-size-W-plus-minus}
\end{equation}
where $b_-$ and $b_+$ are numbers of order unity, $b_- < 1 < b_+$. In the critical regime
\eqref{eq:RRG-critical regime}, this yields the following scaling of the transition width $\Delta W = W_+ - W_-$:
\begin{equation}
\frac{\Delta W}{W^{\rm RRG}_c(m,2^n)} \propto n^{-1-1/\nu} \,,
\label{eq:RRG-transition-width}
\end{equation}
with $\nu=1/2$. Note that the transition width on RRG shrinks faster than the distance to the thermodynamic-limit critical value, Eq.~\eqref{eq:Wc-RRG-crit}. 

\section{Numerical methods}
\label{app:numerics_details}

To study the time-resolved dynamics of $\mathcal{I}^{(\alpha)}(t)$ up to a desired time $T$, we require an accurate numerical estimate of the intermediate states $\ket{\psi(t_i)} = \exp (-i \hat{H} t_i)\ket{\alpha}$, with a certain time grid $t_i \in [t_0,\dots,T]$. Thus, starting from $\ket{\psi(t=0)} = \ket{\alpha}$, we evolve the state from one time point to the next by numerically evaluating 
\begin{align}
\ket{\psi(t_{i+1})} = e^{-i\hat{H}(t_{i+1}-t_i)}\ket{\psi(t_i)}.    
\end{align}
The Hamiltonian $\hat{H}$ is a large sparse complex matrix, and the matrix exponential can be conveniently evaluated using the EXPOKIT algorithm, implemented in the C++ library SLEPC. To ensure that the algorithm converges to a desired precision $\epsilon$ at time $T$, each intermediate step is computed with required precision $\epsilon_i = \epsilon/(t_{i+1}-t_{i})$. Upon non-convergence of EXPOKIT over $1000$ iterations, the time step is further split into the minimal number of substeps required to achieve convergence. The memory-efficient compression of the matrices and vectors is done using the library PETSC. All of these libraries have a built-in MPI parallel support, which brings a significant improvement in time efficiency for sufficiently large matrices (typically when $n \geq 14$). The number of disorder realizations in our  time-evolution computations is given in the table below, for all three models and for systems sizes from $n=8$ to $n=16$: 

\vspace*{0.2cm}
\begin{centering}
\begin{tabular}{c|c|c|c|c|c|c|} \toprule
    & {} & {$n=8$} & {$n=10$} & {$n=12$} & {$n=14$} & {$n=16$}  \\ \hline
    & 1D  & 3200 & 800 & 400 & 200 & 100 \\
    \hline
    & QD  & 3200 & 1600 & 800 & 200 & 100  \\
    \hline
    & QREM  & 3200 & 800 & 400 & 200 & 100  \\
    \hline
\end{tabular}
\end{centering}

\vspace*{0.2cm}

The full exact-diagonalization data presented in this paper was obtained using the multithreaded version of the high-performance C++ scientific library Armadillo \cite{Sanderson2025}. 

 \section{Localization transition in QREM}
 \label{app:QREM}

Analytical predictions for the localization transition in QREM \cite{Scoquart2024} are obtained by exploting the 
fact that short loops are suppressed on QREM graph for large $n$, yielding a connection to the RRG model,
\begin{equation}
W_c^{\rm QREM} (n) = W_c^{\rm RRG} (n, 2^n) \,.
\label{eq:QREM-maintext-relation-RRG}
\end{equation}
It is used here that the coordination number (equal to $m+1$ for the RRG model) is given by $n$ in QREM and that the system volume $N$ is related to $n$ via $N=2^n$. In the large-$n$ limit, we have $\lim_{n\to \infty} W_c^{\rm QREM} (n) / W_c^{\rm RRG} (n, \infty) = 1$. For a finite $n$, this ratio is less than unity, approaching unity with increasing $n$.

 For our QREM, $\langle |T| \rangle = \sqrt{\pi}/2$ and $\gamma(0)= (\sigma\sqrt{2\pi})^{-1}$, where $\sigma^2 = nW^2$, see Eq.~\eqref{eq:distrib_energies_hoppings_1D}. 
Here we have neglected the second term in the variance of energies $E_\alpha$ in Eq.~\eqref{eq:distrib_energies_hoppings_1D} since it  is much smaller than the first term in the range of disorder $W \gg 1$ that is of interest for QREM. Thus, the product 
$\gamma(0) \langle | T | \rangle$ entering
Eq.~\eqref{eq:self_cons_critical_disorder_Herre} is
\begin{equation}
\gamma(0) \langle | T | \rangle = (3/2n)^{1/2} W^{-1} \,,
\label{eq:RRG-maintext-rescaling-factor}
\end{equation}
 so that 
Eq.~\eqref{eq:self_cons_critical_disorder_Herre}
for $W_c^{\rm RRG}(n,\infty)$ takes the form
\begin{equation}
W_c = 4 \sqrt{\frac{3}{2}} \: n^{1/2} \ln\left[ 
(2n/3)^{1/2} W_c \right].
\label{eq:RRG-maintext-Wc-rescaled-equation}
\end{equation} 
The leading large-$n$ asymptotics of the solution of this equation is 
\begin{equation}
W^{\rm RRG}_c (n,\infty) \simeq 4 \sqrt{\frac{3}{2}} \: n^{1/2} \ln n \,,
\label{eq:RRG-Wc-asympt}
\end{equation}
with a relative correction decaying slowly as $\ln \ln n / \ln n$.

In analogy with the RRG model, for a sufficiently large system size, $n > n_*$, we have $W_c^{\rm QREM}(n) > (1/2) W_c^{\rm RRG}(n,\infty)$, referred to as critical regime. The estimated lower border of this regime is $n_* \approx 22$. In this regime, one finds, in analogy with 
Eq.~\eqref{eq:Wc-RRG-crit}, 
\begin{equation}
\frac{W_c^{\rm RRG}(n,\infty) - W_c^{\rm QREM}(n)}{W_c^{\rm RRG}(n,\infty)} \simeq \frac{2\pi^2}{3 \ln^2 2} \: \frac{\ln^2 n}{n^2} \,.
\label{eq:QREM-maintext-Wc-finite-size-shift}
\end{equation}
The scaling of the transition width can be estimated using Eq.~\eqref{eq:RRG-finite-size-W-plus-minus}, which yields for the width $\Delta W^{\rm QREM}(n) = W_+(n,N) - W_-(n,N)$ in the critical regime ($n > n_*$)
\begin{equation}
\frac{\Delta W^{\rm QREM}(n)}{W_c^{\rm QREM}(n)} =
\frac{4\pi^2}{3\ln^2 2} \: \ln(b_+/b_-) \: \frac{\ln^2 n}{n^3}\,,
\label{eq:QREM-maintext-Wc-finite-size-width}
\end{equation}
in analogy with Eq.~\eqref{eq:RRG-transition-width}.

The values of $n$ accessible in exact-diagonalization simulations are smaller than (although not too far from) $n_* \approx 22$. The finite-size transition then happens in the pre-critical regime, 
$n^{1/2} < W_c^{\rm QREM}(n) < (1/2) W_c^{\rm RRG}(n,\infty)$, cf. Eq.~\eqref{eq:W-precrit-window}. The corresponding formula for the correlation volume $N_\xi(W)$ is then obtained from Eq.~\eqref{eq:RRG-N-xi-precritical} by substituting $m \mapsto n$ and $W\mapsto (2n/3)^{1/2}W$, which yields 
\begin{equation}
N_\xi \sim W^2 \exp\left[\frac{W}{(6n)^{1/2}} \right].
\label{eq:QREM-precrit-N-xi}
\end{equation}
From this formula, 
the finite-size transition point $W_c(n)$ can be obtained via Eq.~\eqref{eq:RRG-transition-Nxi-N}.

\section{Localized phase of 1D model }
\label{app:1D-loc}

This Appendix contains details of the calculation of observables in the localized phase of our 1D model; the corresponding results are presented in Sec.~\ref{sec:analytics-1D-loc}.
The Hamming-distance operator $\hat{x}$, Eq.~\eqref{eq:x-operator-def}, has the form 
$\hat{x} = \sum_{i=1}^n \hat{x}_i$, where 
the operator $\hat{x}_i = 
 \frac{1}{2} (1 - s_i^{(\alpha)} \sigma^z_i)
$ acts on the subspace of spin $i$, such that
\begin{align}
\bra{\beta}x_i\ket{\gamma} = (1-\delta_{\alpha_i \beta_i})\delta_{\beta\gamma} \,.
\end{align}
Summing these matrix elements over $i$, 
we recover the Hamming distance to the initial state:
\begin{align}
\bra{\beta}\hat{x}\ket{\gamma} = \bra{\beta}\sum_{i=1}^n \hat{x}_i\ket{\gamma} = \sum_{i=1}^n(1-\delta_{\alpha_i\beta_i})\delta_{\beta\gamma} = r(\alpha,\beta)\delta_{\beta\gamma}.
\end{align}

The representation of $\hat{x}$ as a sum of local spin operators is very helpful at strong disorder, because it aligns with the usual $l$-bit picture of the Hamiltonian in the localized phase \cite{Serbyn2013, Ros2015a, imbrie16a, Imbrie2016JSP, abanin2019colloquium}. At strong disorder, $W \gg 1$, almost all spins $i$ 
can be well approximated (up to parametrically small corrections $\sim 1/W$) by eigenstates of $\hat{S}_i^z$. Only a small ($\sim 1/W$) fraction of spins will be resonant, with the corersponding single-spin Hamiltonian
\begin{equation}
\hat{H}_i = {\tilde{\epsilon}_i} \, \text{sgn}(\epsilon_i) \, \hat{\tau}_i^z \,,
\label{eq:1D-loc-H_i}
\end{equation}
where
\begin{align}
\hat{\tau}_i^z  = \frac{1}{\tilde{\epsilon}_i}\left(\epsilon_i\hat{S}_i^z + h_i^x \hat{S}_i^x +h_i^y \hat{S}_i^y \right)\text{sgn}(\epsilon_i) \,,
\label{tauiz}
\end{align}
\begin{align}
h_i^{a} = 2\left(V^{a}_{i-1,i}{S}_{i-1}^z + V^{a}_{i+1,i}{S}_{i+1}^z\right), \qquad a=x,y \,,
\end{align}
and
\begin{align}
\tilde{\epsilon}_i = \sqrt{\epsilon_i^2 + (h_i^x)^2 +(h_i^y)^2} \,. 
\label{tilde-epsilon}
\end{align}
Here, $\hat{\tau}_i^z$ is essentially an approximate form of the $l$-bit. Note that
${S}_{i\pm 1}^z$ are treated here as numbers since neighbors of a resonant spin are almost always non-resonant; this approximation discards corrections of the next order in $1/W$. Also, we have omitted a correction $2(V_{i,i+1}^z S^z_{i+1} + V_{i-1,i}^z S^z_{i-1})$ to energy $\epsilon_i$, which does not affect the result in the leading order in $1/W$.

The eigenstates $\ket{I}$ of our 1D Hamiltonian \eqref{eq:1D-splittting} are now obtained as product states of single-spin Hamiltonians \eqref{eq:1D-loc-H_i}, i.e., they are diagonal
in the $\hat{\tau}_i^z$  basis. This approximation neglects disorder realizations with two (or more) adjacent resonant spins
but these are corrections of higher orders in $1/W$ as explained above.

The eigenstates of $\hat{\tau}_i^z$ are 
\begin{align}
\ket{\tau_i}^\pm =\frac{1}{\sqrt{2\tilde{\epsilon}_i(\tilde{\epsilon}_i\pm \epsilon_i)}} \big[(\tilde{\epsilon_i} \pm \epsilon_i)\ket{\uparrow}_i + (h_i^x+i h_i^y)\ket{\downarrow}_i\big].
\end{align}
The overlaps of these states with the basis states (for a given spin $i$) are
\begin{align}
p_i^+&= \big|\bra{\alpha_i = +1}\ket{\tau_i = +1}\big|^2 = \big|\bra{\alpha_i = -1}\ket{\tau_i = -1}\big|^2\nonumber\\
&= \frac{1}{2}\left(1+\frac{|\epsilon_i|}{\tilde{\epsilon}_i}\right),\\
p_i^-&= \big|\bra{\alpha_i = +1}\ket{\tau_i = -1}\big|^2 = \big|\bra{\alpha_i = -1}\ket{\tau_i = +1}\big|^2\nonumber\\
&= \frac{1}{2}\left(1-\frac{|\epsilon_i|}{\tilde{\epsilon}_i}\right).
\end{align}

We turn now to the calculation of average imbalance and imbalance fluctuations.
As a first step, we compute $\langle x_i \rangle \equiv \langle x_i^2 \rangle$ (this is an exact equality since the operator $\hat{x}_i$ has eigenvalues 0 and 1 only):
\begin{align}
\langle x_i \rangle = \langle x_i^2 \rangle &= \sum_{\beta_i} \sum_{\tau_i}|\bra{\alpha_i}\ket{\tau_i}|^2|\bra{\beta_i}\ket{\tau_i}|^2(1-\delta_{\alpha_i \beta_i})
\notag \\
&= \sum_{\tau_i}|\bra{\alpha_i}\ket{\tau_i}|^2|\bra{\tilde{\alpha}_i}\ket{\tau_i}|^2 = 2p_i^+ p_i^- \,,
\end{align}
where $\ket{\tilde{\alpha}_i}$ denotes the basis state orthogonal to $\ket{\alpha_i}$, representing the opposite direction of spin $i$.
Using $\hat{x}=\sum_i \hat{x}_i$ in combination with $\langle x_i x_j \rangle = \langle x_i \rangle \langle x_j \rangle$ for $i\ne j$, we then get 
\begin{align}
\langle x \rangle &= \sum_{i=1}^n \langle x_i \rangle = 2\sum_{i=1}^n 
 p_i^+ p_i^-,
 \end{align}
 \begin{align}
 \langle x^2 \rangle &= 
 \sum_{i,j=1}^n \langle x_i x_j \rangle
 = \sum_{i=1}^n \langle x_i^2\rangle + \sum_{\substack{i,j=1\\i\neq j}}^n \langle x_i x_j \rangle\notag \\
 &= \langle x \rangle + 4\sum_{\substack{i,j=1\\i\neq j}}^n p_i^+ p_i^- p_j^+ p_j^-,
 \end{align}
 \begin{align}
  \langle x \rangle^2 &= \sum_{i,j=1}^n \langle x_i \rangle \langle x_j \rangle \notag\\
 &= 4\sum_{\substack{i=1}}^n (p_i^+ p_i^-)^2 + 4\sum_{\substack{i,j=1\\i\neq j}}^n p_i^+ p_i^- p_j^+ p_j^-.
 \label{eq:interm_variance_1D}
\end{align}
To evaluate the ensemble average of $\langle x \rangle$ as well as quantum and mesoscopic fluctuations of $x$, we use now statistical independence and equivalence of Hamiltonians (and thus of $p_i^\pm$) corresponding to different spin indices $i$. This yields
\begin{eqnarray}
\overline{\langle x \rangle} &=& 2n \: \overline{p_i^+ p_i^-} \,, 
\label{eq:1D-loc-average-1}
\\
\overline{\langle x^2 \rangle - \langle x \rangle^2} &=& 2n \: \overline{\left[p_i^+ p_i^-
- 2 (p_i^+ p_i^-)^2 \right]} \,, 
\label{eq:1D-loc-average-2q}
\\
\overline{\langle x \rangle^2} - \overline{\langle x \rangle}^2 & = & 
4n \: \left[\overline{(p_i^+ p_i^-)^2} - 
\overline{p_i^+ p_i^-}^2 \right] .
\label{eq:1D-loc-average-2m}
\end{eqnarray}

Two single-spin averages entering these expressions, $\overline{p_i^+ p_i^-}$ and  
$\overline{(p_i^+ p_i^-)^2}$ can be straightforwardly computed. For the first of them, we get
\begin{align}
\mathcal{C}_1(W) &\equiv \overline{p_i^+ p_i^-}
\notag \\
& = \frac{1}{4}\int\limits_{-W}^W \frac{{\rm d}\epsilon}{2W}\int\limits_{-\infty}^{\infty} \frac{{\rm d}h_x {\rm d}h_y}{2\pi} \frac{h_x^2+h_y^2}{\epsilon^2+h_x^2+h_y^2}
e^{-\frac{h_x^2+h_y^2}{2}}\notag 
\\
& \simeq \frac{1}{8W}\int\limits_{-\infty}^\infty {\rm d}\epsilon \int\limits_{0}^{\infty} {\rm d}h \,h\, \frac{h^2}{\epsilon^2+h^2}\, e^{-h^2/2}
\notag \\
& = \frac{\pi^{3/2}}{8\sqrt{2}}\frac{1}{W}.
\end{align}
Here, we have  extended the integration limits in the $\epsilon$-integral to $\pm \infty$ to obtain the leading asymptotics of $\mathcal{C}_1(W)$ at $W\gg 1$.
Likewise, we compute 
\begin{align}
\mathcal{C}_2(W) &\equiv \overline{p_i^+ p_i^- p_i^+ p_i^-}
\notag
\\
& = \frac{1}{32W}\int\limits_{-W}^W {\rm d}\epsilon \int\limits_{0}^{\infty} {\rm d}h \,h\, 
\left(\frac{h^2}{\epsilon^2+h^2}\right)^2\, e^{-h^2/2} \notag \\
& \simeq \frac{\pi^{3/2}}{64\sqrt{2}}\frac{1}{W} \,.
\end{align}
Substituting these results into 
Eq.~\eqref{eq:1D-loc-average-1}, \eqref{eq:1D-loc-average-2q}, and \eqref{eq:1D-loc-average-2m}, we obtain, to order $\mathcal{O}(W^{-1})$,
\begin{eqnarray}
\overline{\langle x \rangle} & \simeq & 
2 n\, \mathcal{C}_1(W)\simeq  \frac{\pi^{3/2}}{4\sqrt{2}}\frac{n}{W} \,,
\label{eq:1D-loc-average-result-1}
\\
\overline{\langle x^2 \rangle - \langle x \rangle^2} & \simeq & 2 n\, [\mathcal{C}_1(W) - 2 \mathcal{C}_2(W)] \simeq 
\frac{3 \pi^{3/2}}{16\sqrt{2}}\frac{n}{W} \,,
\nonumber \\ 
& & \label{eq:1D-loc-average-result-2q}
\\
\overline{\langle x \rangle^2} - \overline{\langle x \rangle}^2 & \simeq & 
4n\, \mathcal{C}_2(W) \simeq 
\frac{ \pi^{3/2}}{16\sqrt{2}}\frac{n}{W} \,,
\label{eq:1D-loc-average-result-2m}
\end{eqnarray}
which yields Eqs.~\eqref{1D-Iinfty}, \eqref{eq94}, and \eqref{eq95} for the average imbalance and its fluctuations.

In the same way, we also compute the large-$W$ asymptotics of the IPR $P_2$ and of the associated fractal exponent. We have for the IPR of an eigenstate $\ket{I}$:
\begin{align}
P_2 &= \sum_\beta |\bra{\beta}\ket{I}|^4
\notag
\\
&= \prod_{i=1}^n \left[\sum_{\beta_i = \pm} |\bra{\beta_i}\ket{\tau_i}|^4\right]= \prod_{i=1}^n \left[(p_i^+)^2 + (p_i^-)^2\right]
\notag
\\
&= \prod_{i=1}^n \left[\frac{1}{4}\left(1+\frac{|\epsilon_i|}{\tilde{\epsilon}_i}\right)^2 + \frac{1}{4}\left(1-\frac{|\epsilon_i|}{\tilde{\epsilon}_i}\right)^2\right].
\end{align}
To calculate the disorder-averaged IPR, we thus need the single-spin average
\begin{align}
\mathcal{C}_3(W) &\equiv \overline{(p_i^+)^2+(p_i^-)^2} = 1-2\mathcal{C}_1(W)
\notag\\
&\simeq  1- \frac{\pi^{3/2}}{4\sqrt{2}}\frac{1}{W} \,.
\end{align}
This yields for the average IPR
\begin{align}
\overline{P_2} &=
\left[\mathcal{C}_3(W)\right]^n 
\simeq \exp\left(- \frac{\pi^{3/2} n}{4\sqrt{2} W}\right) =  N^{- \displaystyle \tilde{\tau}_2}(W) \,,
\end{align}
where $N=2^n$ is the many-body Hilbert-space volume, and 
\begin{equation}
\tilde{\tau}_2(W) = \frac{\pi^{3/2}}{4\sqrt{2} \ln\!2 \: W} \simeq \frac{1.420}{W} \,.
\label{eq:1D-loc-tau-2-tilde}
\end{equation}

Finally, we turn to the fractal exponent
\begin{equation} 
\tau_2 = - \frac{\ln P_2}{\ln N} \,.
\label{eq:1D-loc-tau-2-def}
\end{equation}
Averaging this expression, we get
\begin{align}
\tau_2(W) &=  -\frac{1}{\ln N}\, \overline{\ln P_2}  \notag \\
& = -\frac{1}{n\ln 2}\sum_{i=1}^n \overline{\ln\left[(p_i^+)^2 + (p_i^-)^2\right]}.
\end{align}
 We thus need the following single-spin average
\begin{align}
\mathcal{C}_4(W)&=\overline{\ln\left[(p_i^+)^2 + (p_i^-)^2\right]} = \overline{\ln\left[\frac{1}{2}\left(1+\frac{\epsilon_i^2}{\tilde{\epsilon}_i^2}\right)\right]} \notag 
\\
&= \int\limits_{-W}^W \frac{{\rm d}\epsilon}{2W}\int\limits_{0}^{\infty} {\rm d}h \, h\ln\left[\frac{1}{2}\left(1+\frac{\epsilon^2}{\epsilon^2 + h^2}\right)\right] e^{-h^2/2} \notag \\
& \simeq -\frac{\pi^{3/2}(\sqrt{2}-1)}{2W}  \,,
\end{align}
where we have again replaced $\pm W$ by $\pm \infty$ in the integration limits for the $\epsilon$-integral to find the leading asymptotics.
This results in the following asymptotic behavior of the fractal exponent:
\begin{align}
\tau_2(W) &=-\frac{1}{ \ln 2} \, \mathcal{C}_4(W)\simeq \frac{\pi^{3/2}(\sqrt{2}-1)}{2\ln \!2 \, W} \simeq \frac{1.664}{W} \,,
\label{eq:1D-loc-tau-2}
\end{align}
which is Eq.~\eqref{eq:1D-loc-main-tau-2-final} of the main text.
It is worth noting a difference (although not too large) in numerical coefficients in 
Eqs.~\eqref{eq:1D-loc-tau-2-tilde} and \eqref{eq:1D-loc-tau-2}. In the first case, we averaged the IPR, while in the second case we averaged  its logarithm, i.e., essentially calculated the typical IPR. Thus, the scaling of average IPR is determined by rare events. At the same time, we emphasize that the fractal dimension $\tau_2$ defined by Eq.~\eqref{eq:1D-loc-tau-2-def} is self-averaging in the large-$n$ limit (since it is proportional to $\ln P_2$).

\section{Lower bound on critical disorder in QD model}
\label{app:QD-Wc-lower-bound}

In this Appendix, we present a derivation of the lower bound on the critical disorder $W_c^{\rm QD}$ in the QD model,
see Eq.~\eqref{sec:analytics-QD-Wc-bounds} of Sec.~\ref{sec:analytics-QD}.
This bound improves the result obtained in Ref.~\cite{Scoquart2024} (see Appendix C of that paper). 
We start by singling out the terms in the Hamiltonian $\hat{H}^\text{QD}$, Eqs.~(\ref{eq:QD})-(\ref{eq:QD-H1}), that involve spin $i$ (ignoring the self-interaction terms with $j=i$):
\begin{align}
    \hat{H}^\text{QD}_i& \equiv  \Bigg(\epsilon_i + \frac{2}{\sqrt{n}}\sum\limits_{j \neq i} V_{ij}^{z} \hat{S}_j^z+\frac{2}{\sqrt{n}}\smashoperator{\sum}\limits_{\substack{j\neq i\\ a \in \{x,y\}}} V_{ij}^{a}\hat{S}_j^a\Bigg)\hat{S}_i^z\notag 
    \\&+ \frac{2}{\sqrt{n}}\smashoperator{\sum}\limits_{\substack{j \neq i\\ a \in \{x,y\}}}  V_{ji}^{a}\hat{S}_j^z \hat{S}_i^a \,.
\end{align}
Following Ref.~\cite{Scoquart2024} and the consideration in Appendix~\ref{app:1D-loc} above, we describe spin $\hat{\mathbf{S}}_i$ by the Hamiltonian  
\begin{equation}
H^{(0)}_i \approx (\epsilon_i+ h^z_i) \hat{S}_i^z
+ h^x_i \hat{S}_i^x + h^y_i \hat{S}_i^y.
\label{eq:appendix-QD-H0i}
\end{equation}
This representation corresponds to a randomly oriented effective magnetic field of strength $\tilde{\epsilon}_i$ [cf. Eq.~\eqref{tilde-epsilon}], which includes interactions with all other spins,
\begin{align}
h^a_i\sim \frac{1}{\sqrt{n}}\sum_j V^a_{ij} S_j^z\sim 1, \quad a=x,y.
\end{align}
By a unitary rotation of operators $\hat{S}_i^a$, we rewrite Eq.~\eqref{eq:appendix-QD-H0i} in the form of Eq.~\eqref{eq:1D-loc-H_i}: 
$H^{(0)}_i = \tilde{\epsilon}_i \hat{\tau}_i^z $,
where $\hat{\tau}_i^z$ is the $z$-component of the spin operator in the new basis. 

In Ref.~\cite{Scoquart2024}, the bound $W_c^{\rm QD}(n) \gtrsim n^{3/4} (\ln n)^{-1/4}$ was obtained for the QD model by focusing on resonant spins, for which $\epsilon_i\sim 1$. Here, at variance with the approach of Ref.~\cite{Scoquart2024}, we analyze the role of non-resonant spins with $\epsilon_i\gg 1$, which may form two-spin resonances (cf. Refs.~\cite{gornyi2017spectral} and \cite{gutman2016energy}). As in Appendix~\ref{app:1D-loc}, we neglect $h_i^z$  in comparison with $\epsilon_i$. For simplicity, here, we discard the $y$-component of the random field, setting $h_i^y=0$
(this will not affect the resulting scaling of $W_c$ with $n$; we will restore the $y$-components later on). The components of the spin operator in the rotated basis are then given by (for $\epsilon_i>0$)
\begin{align}
    \hat{\tau}_i^z  &= \frac{\epsilon_i}{\tilde{\epsilon}_i}\hat{S}_i^z +  \frac{h_i^x}{\tilde{\epsilon}_i} \hat{S}_i^x,
    \\
    \hat{\tau}_i^x  &= -\frac{h_i^x}{\tilde{\epsilon}_i}\hat{S}_i^z +  \frac{\epsilon_i}{\tilde{\epsilon}_i} \hat{S}_i^x. 
\end{align}
With this rotation, we generate the XX-couplings in the full Hamiltonian, which are responsible for the flip-flop  processes (two-spin flips) in the new basis:
\begin{align}
   \sum\limits_{i,j}  V_{ij}^{x}\hat{S}_i^z \hat{S}_j^x
    \ \rightarrow \
    \sum\limits_{i,j} \frac{V_{ij}^{x}}{\tilde{\epsilon}_i\tilde{\epsilon}_j}
    \left(\epsilon_i\hat{\tau}_i^z -  h_i^x \hat{\tau}_i^x\right) \left(h_j^x\hat{\tau}_j^z +  \epsilon_j\hat{\tau}_j^x\right).
\end{align}
Here, the generated XZ-terms sum up to zero by construction of the rotation.  
The XX-terms in the Hamiltonian,
\begin{align}
    \frac{1}{\sqrt{n}} \sum\limits_{i,j}V_{ij}^{x}\,
    \frac{h_i^x \epsilon_j}{\tilde{\epsilon}_i\tilde{\epsilon}_j}\,
    \hat{\tau}_i^x \hat{\tau}_j^x
    \simeq \frac{1}{\sqrt{n}} \sum\limits_{i,j}V_{ij}^{x}\,
    \frac{h_i^x}{\epsilon_i}\,
    \hat{\tau}_i^x \hat{\tau}_j^x,
\end{align}
are responsible for the two-spin-flip resonances that occur when the energy distance $|\epsilon_i-\epsilon_j|$ is comparable to the flip-flop matrix element. Restoring the $y$-components of the random fields, we generate the XY- and YY-terms in the same way. The characteristic strength of the matrix element in the flip-flop terms is 
$$ M(\epsilon_i)\sim \frac{1}{\sqrt{n}}V_{ij}^{a}\,
   \frac{h_i^a}{\epsilon_i}\sim  \frac{1}{\sqrt{n} \epsilon_i},$$
since $h^a_i\sim V_{ij}^a\sim 1$.
We thus have the following approximation for the full Hamiltonian written in the $\tau$-basis:
\begin{align}
   \hat{H}^\text{QD}&\approx \sum\limits_{i = 1}^{n} \epsilon_i\hat{\tau}_i^z + \frac{1}{\sqrt{n}}\sum\limits_{i,j = 1}^{n} V_{ij}^{z}\,\hat{\tau}_i^z \hat{\tau}_j^z \notag
    \\
&+\frac{1}{\sqrt{n}}\smashoperator{\sum}\limits_{\substack{ j\neq i\\ a,b \in \{x,y\}}}\! \frac{1}{\epsilon_i} \, \hat{\tau}_i^a \hat{\tau}_j^b ,
\label{HQDtau}
\end{align}
where we suppressed the numerical factors of order unity.
Here, we retain the ZZ-terms that will be responsible for spectral diffusion.

Let us now explore the two-spin-flip resonances in the model described by Eq.~(\ref{HQDtau}). Consider spins $i$ with energies $\epsilon_i$ belonging to the energy window $[E,E+\delta E]$ with $1<E<W$, their number is $n_i(\delta E) \sim n\: \delta E /W$. The level spacing for the states resulting from a flip of a pair of spins $(i,j)$ including such spin $i$ and another spin $j$ with arbitrary $|\epsilon_j|<W$ is 
\begin{align}
    \Delta_2(\delta E)\sim \frac{W}{n_{ij}(\delta E)} \sim \frac{W^2}{\delta E\,  n^2},
\end{align}  
where $n_{ij}(\delta E)\sim n_i(\delta E) n =(n \,\delta E/W) n$ is the number of such pairs. Among these pairs, there are about 
\begin{align}
\frac{M(E)}{\Delta_2(\delta E)}\sim \frac{1/(E n^{1/2})}{W^2/(\delta E\, n^2)} = \frac{n^{3/2}}{W^2} \frac{\delta E}{E}
\end{align}
resonant pairs, defined by the condition $M\gtrsim \Delta_2$.
The total number of two-spin resonances is given
by the integral over energy slices,
\begin{align}
    n_{{\rm res}, 2} \sim \int_1^W \frac{dE}{E}\,  \frac{n^{3/2}}{W^2}  \sim \frac{n^{3/2}}{W^2} \ln W.
    \label{eq:app-QD-lower-bound-nres}
\end{align}
With lowering $W$, the system enters the regime with $n_{{\rm res}, 2}\gtrsim 1$
at
\begin{align}
    W\lesssim n^{3/4}\, \ln^{1/2} n \,.
    \label{W2}
\end{align}

As discussed in Ref.~\cite{gornyi2017spectral}, once we have resonances in a typical state, $n_{{\rm res}, 2}\gtrsim 1$, 
spectral diffusion may lead to full thermalization by activating further resonances. Let us check whether this is the case in our QD model. 
For resonances with $E<\epsilon_i <2E$, $m$ steps of spectral diffusion due to the ZZ-term in Eq.~(\ref{HQDtau}) shift the levels by $\sim \sqrt{m} V_{ij}^z\sim  \sqrt{m}/\sqrt{n}$. 
The condition that $m$ steps of spectral diffusion can be realized for the energy slice $[E,2E]$ is given by
\begin{align}
\frac{\sqrt{m}}{\sqrt{n}}>\Delta_2(E) m = \frac{W^2 m}{E n^2}.
\end{align}
Setting here $m=n$ (which means that mixing of states due to resonances proceeds over the entire Fock space), we find the condition (cf. Ref.~\cite{gornyi2017spectral})
\begin{align}
\frac{W^2}{E n}<1.
\end{align}
For the largest $W$ at which resonances start to appear, Eq.~\eqref{W2}, we find that the spectral diffusion continues to $m=n$ steps for (up to a logarithmic factor)
\begin{align}
    E>n^{1/2},
    \label{eq:app-QD-lower-bound-E_res}
\end{align}
which satisfies $E<W$. 
Thus, when resonances with such energies are present, they will trigger full thermalization involving all $n$ spins. 
Adjusting the lower limit of integral in Eq.~\eqref{eq:app-QD-lower-bound-nres} to $E$ modifies only the logarithmic factor and does not affect the resulting estimate \eqref{W2} of disorder $W$ at which two-spin-flip resonances 
[of spins with energies satisfying \eqref{eq:app-QD-lower-bound-E_res}]
emerge. 

Therefore, the condition \eqref{W2} of the appearance of the first resonance yields the lower bound for the critical strength of disorder,
\begin{equation}
  W_c^{\rm QD}(n) \gtrsim n^{3/4} (\ln n)^{1/2}\,,
\label{eq:app-QD-lower-bound-final}
\end{equation}
which is given 
in Eq.~(\ref{sec:analytics-QD-Wc-bounds}) of the main text. The leading (power-law) $n^{3/4}$ factor in this bound is identical to that
in the lower bound derived in Ref.~\cite{Scoquart2024}; however, the subleading (logarithmic) factor is improved by $\ln^{3/4}n$. 
It is plausible that the obtained bound 
\eqref{eq:app-QD-lower-bound-final}
yields in fact the actual scaling of $W_c^{\rm QD}(n)$ or is at least very close to it. 
This is further supported by our numerical results, see Fig.~\ref{fig:Wc_vs_n_1D_QD_QREM}.

\section{Lower bound on transition width for 1D and QD models}
\label{app:Harris-criterion}

In this Appendix, we present details of the derivation of the lower bound on the transition width in 1D and QD models, Sec.~\ref{sec:transition}.
The calculation is based on a bound on a derivative of an observable derived in Ref.~\cite{chandran2015finite}. 

Let $X$ to be an observable that exhibits, in the $n\to \infty$ limit, a jump of magnitude $(\Delta X)_t$ at the transition. Examples of such an observable are the average imbalance and mean adjacent gap ratio of level statistics. By a linear transformation of $X$, we make it to jump from $X=-1$ to $X=+1$, with $(\Delta X)_t=2$. 
To determine the width of the transition, we consider the average  $\overline{X}(W)$ and select a window of the width $(\Delta \overline{X})_w$ within the interval $[-1;1]$, such as, e.g., $[-1/2;1/2]$. This defines the 
disorder window $[W_-(n);W_+(n)]$ associated with the transition, and thus the transition width $\ln(W_+(n) / W_-(n)) \simeq \Delta W(n) / W_c(n)$. 
We use now a Lemma, Eq. (2) of Ref.~\cite{chandran2015finite} that bounds the derivative $d\overline{X}(W)/dW$ (see also
a related Lemma 2.3 in Ref.~\cite{Chayes1989correlation}):
\begin{equation}
\left  | \frac{d\overline{X}(W)}{dW}   \right | \le \alpha n^{1/2} \,,
\label{eq:app-Harris-1}
\end{equation}
where 
\begin{equation}
\alpha^2 = \int d\epsilon \, \frac{1}{p_W(\epsilon)} \left( \frac{dp_W(\epsilon)}{d\epsilon}\right)^2 \,.
\label{eq:app-Harris-2}
\end{equation}
Here $p_W(\epsilon)$ is the distribution of diagonal energies $\epsilon_i$ corresponding to the disorder $W$. We choose this distribution to be Gaussian with a variance $\sigma^2 = c W^2$  (where $c$ is a numerical coefficient) for the purpose of calculating $\alpha^2$; it is expected that the transition width should not strongly depend on this choice. We find then
$$
\frac{dp_W(\epsilon)}{d\epsilon} = \frac{1}{W} \, p_W(\epsilon)
\left( - 1 +  \frac{\epsilon^2}{\sigma^2}
\right),
$$
yielding
\begin{equation}
\alpha^2 = \frac{1}{W^2} \int d \epsilon \:  p_W(\epsilon) \left( - 1 +  \frac{\epsilon^2}{\sigma^2}
\right)^2 = \frac{2}{W^2} \,, 
\label{eq:app-Harris-3}
\end{equation}
and thus
\begin{equation}
W \left  | \frac{d\overline{X}(W)}{dW}   \right | \le \sqrt{2n} \,.
\label{eq:app-Harris-4}
\end{equation}
This upper bound on the derivative leads to the lower bound of the transition width: 
\begin{equation}
\ln \frac{W_+(n)}{W_-(n)} \ge \frac {(\Delta \overline{X})_w}{\sqrt{2n}} \,.
\label{eq:app-Harris-5a}
\end{equation}
For large $n$, the transition sharpens, and $\ln(W_+ / W_-) \simeq \Delta W (n) / W_c(n) \ll 1$. The inequality \eqref{eq:app-Harris-5a} tells us, however, that it cannot sharpen too fast.

We recall that we have assumed above that a linear transformation of $X$ has been made such that $(\Delta X)_t =2$. To get rid of this assumption, we rewrite Eq.~\eqref{eq:app-Harris-5a} in the form 
\begin{equation}
\delta(n)  \equiv R \, \ln \frac{W_+(n)}{W_-(n)} \ge \frac {1}{\sqrt{2n}} \,,
\label{eq:app-Harris-5}
\end{equation}
where
\begin{equation}
R = \frac{(\Delta X)_t}{2 (\Delta \overline{X})_w } \,.
\label{eq:app-Harris-6}
\end{equation}
Clearly, $R$ is invariant with respect to a linear transformation of $X$, so that  the bound in the form \eqref{eq:app-Harris-5}, \eqref{eq:app-Harris-6} holds independently of the value of $(\Delta X)_t$.  Equation \eqref{eq:app-Harris-5}
is Eq.~\eqref{eq:transition-width-bound}
of the main text. 

Let us emphasize that, in the above derivation, we did not make any assumptions concerning the scaling of the critical disorder $W_c(n)$ or concerning the spatial geometry. The derivation thus holds equally for the 1D and QD models. It also applicable to various extensions of these models, such as, e.g., systems of dimensionality $d>1$ or systems with a non-trivial spatial dependence of interaction. 

We discuss now implications of the bound on the form of scaling near the transition for an observable $\overline{X}$ of the type considered above. In this part, we will focus on the 1D model, for which $W_c(\infty)={\rm const}$, which simplifies the analysis. 
We will further make the following assumptions, which are supported by phenomenological theories and by numerical simulations on various 1D models: (i) $W_c(n)$ increases monotonically at sufficiently large $n$, so that $W_c(n) < W_c(\infty)$, (ii) in the thermodynamic limit, the system on the ergodic side of the transition, $W < W_c(\infty)$, is characterized by a  correlation length $\xi$ that diverges at $W\to W_c(\infty)$; (iii) the condition for the transition in a finite-size system is $\xi \simeq n$. 
We note that all these conditions hold for the RRG model, see Appendix \ref{app:RRG}
and Fig.~\ref{fig:analytics_RRG_Gaussian_m20}.
Let $\nu$ be the exponent characterizing divergence of the correlation length, $\xi \sim w^{-\nu}$, where $w= W_c(\infty)-W > 0$.
(A faster-than-power-law divergence of $\xi$ corresponds to $\nu=\infty$.) 

Very generally, one can write possible scaling 
of $\overline{X}$ at $w > 0$ in the form
\begin{equation}
\overline{X} = F (f(n) - f(\xi)) \equiv
G\left( \frac{\exp f(n)}{\exp f(\xi)} \right),
\label{eq:app-Harris-7}
\end{equation}
where $f(x)$ is a monotonically increasing function, $f(x\to \infty) = \infty$.  Two types of scaling that have been discussed in the literature in the context of MBL transition in 1D systems are the ``linear'' scaling $\overline{X} = G(n/\xi)$ and the 
``volumic'' scaling $\overline{X} = G(N/N_\xi)$, where $N=2^n$ is the full Fock-space volume and $N_\xi = 2^\xi$ is the correlation volume. Equation \eqref{eq:app-Harris-7} yields linear scaling for $f(x) \sim \ln x$ and volumic scaling for $f(x) \sim x$. 
One can also imagine an intermediate scaling form, $f(x) \sim x^\mu$ with $0 < \mu < 1$, although we are not aware of any analytical motivations for it. As pointed out in Appendix \ref{app:RRG}, the volumic scaling holds for the RRG model (for which $\nu=1/2$)
\cite{tikhonov2016anderson,garcia-mata17,biroli2018,tikhonov19statistics,tikhonov2021from,garcia-mata2022critical}. 

We are now going to analyze implications of the transition-width bound \eqref{eq:app-Harris-5a} for allowed scaling form in 1D model at large $n$. According to Eq.~\ref{eq:app-Harris-7}, we have
\begin{equation}
(\Delta \overline{X})_w \simeq F'(0) \, f'(\xi) \: \xi'(w) \, \Delta w \,,
\end{equation}
where primes denote the derivatives. As above, we take  $(\Delta \overline{X})_w \sim 1$, which implies, according to Eq.~\eqref{eq:app-Harris-5a}, $\Delta w  \gtrsim n^{-1/2} \approx \xi^{-1/2}$. In this analysis, we discard numerical coefficients (since we are only interested in the form of scaling), which is indicated by the symbols $\sim$ and $\gtrsim$. Since $F(y)$ should be a smooth scaling function, we have $F'(0) \sim 1$. 
Further, $\xi'(w) \sim w^{-\nu-1} \sim \xi^{(\nu+1)/\nu}$.
Thus,
\begin{equation}
f'(\xi) \sim \frac{1}{\xi'(w) \: \Delta w} \lesssim \xi^{-\frac{1}{2}-\nu^{-1}} \,.
\label{eq:app-Harris-8}
\end{equation}
Since by assumption $f(\xi) \to \infty$ at $\xi\to\infty$, we obtain from this inequality the Harris bound for the exponent $\nu$:
\begin{equation}
\nu \ge 2 \,,
\label{eq:app-Harris-9}
\end{equation}
in consistency with Ref.~\cite{chandran2015finite} (which considered only the linear scaling).
For $\nu = 2$, Eq.~\eqref{eq:app-Harris-8} becomes $f'(\xi) \lesssim 1/\xi$, leaving essentially the only option of linear scaling, $f(\xi) \sim \ln \xi$. For $\nu > 2$, in additional to the linear scaling, also a form of scaling that is intermediate between linear and volumic is consistent with Eq.~\eqref{eq:app-Harris-8}:
\begin{equation}
f(\xi) \sim \xi^\mu \,, \qquad 0<\mu \le \frac{1}{2} - \nu^{-1} \,.
\end{equation}
Importantly, the volumic scaling, $\mu=1$, is not allowed irrespective of the value of $\nu$.  

We reiterate that these conclusions (concerning the exponent $\nu$ and the type of scaling) refer to the asymptotic (large-$n$) critical behavior. Numerical results of this paper show that our 1D model is not in this asymptotic regime for values of $n$ that we were able to explore, see
Sec.~\ref{sec:transition}. To our knowledge, this applies also to all existent numerical studies of the MBL transition in various 1D systems. The effective exponent $\nu$ and the effective scaling form in such ``pre-critical'' regime may violate the above bounds. (This is at variance with the strict bound \eqref{eq:app-Harris-5}.)
As discussed in Sec.~\ref{sec:summary}, the ``true'' critical behavior may be accessible for quantum simulations of the MBL transition.

\bibliography{rrg}

\end{document}